\documentclass{aa}

\usepackage{graphicx}
\usepackage{subcaption}
\usepackage{txfonts}
\usepackage{multirow}
\begin{document}

   \title{BGM FASt: Besan\c{c}on Galaxy Model for big data}

   \subtitle{Simultaneous inference of the IMF, SFH, and density in the solar neighbourhood }

  \author{
R.\,Mor\inst{1} 
\and A.C.\, Robin\inst{2}
\and F.\, Figueras \inst{1}  
\and T.\, Antoja \inst{1} 
          }

 \institute{
Dept. F\'isica Qu\`antica i Astrof\'isica, Institut de Ci\`encies del Cosmos, Universitat de Barcelona (IEEC-UB), Mart\'i Franqu\`es 1, E08028 Barcelona, Spain.
\email{rmor@fqa.ub.edu}
\and
Institut Utinam, CNRS UMR6213, Universit\'e de Bourgogne Franche-Comt\'e, OSU THETA , Observatoire de Besan\c{c}on, BP 1615, 25010 Besan\c{c}on Cedex, France
}

   \date{Received Month dd, yyyy; accepted Month dd, yyyy}
 
\abstract
   {}
   {We develop a new theoretical framework to generate Besan\c{c}on Galaxy Model Fast Approximate Simulations (BGM FASt) to address fundamental questions of the Galactic structure and evolution performing multi-parameter inference. As a first application of our strategy we simultaneously infer the initial-mass function (IMF), the star formation history and the stellar mass density in the solar neighbourhood.}
   {The BGM FASt strategy is based on a reweighing scheme, that uses a specific pre-sampled simulation, and on the assumption that the distribution function of the generated stars in the Galaxy can be described by an analytical expression. To evaluate the performance of our strategy we execute a set of validation tests. Finally, we use BGM FASt together with an approximate Bayesian computation algorithm to obtain the posterior probability distribution function of the inferred parameters, by automatically comparing synthetic versus Tycho-2 colour-magnitude diagrams.}
   {The validation tests show a very good agreement between equivalent simulations performed with BGM FASt and the standard BGM code, with BGM FASt being $\sim 10^4$ times faster. From the analysis of the Tycho-2 data we obtain a thin-disc star formation history decreasing in time and a present rate of $1.2 \pm 0.2 M_\odot/yr$. The resulting total stellar volume mass density in the solar neighbourhood is $0.051_{-0.005}^{+0.002} M_\odot/pc^3$ and the local dark matter density is $0.012 \pm 0.001 M_\odot/pc^3$. For the composite IMF, we obtain a slope of $\alpha_2={2.1}_{-0.3}^{+0.1}$ in the mass range between $0.5 M_\odot$ and $1.53M_\odot$. The results of the slope at the high-mass range are trustable up to $4M_\odot$ and highly dependent on the choice of extinction map (obtaining  $\alpha_3={2.9}_{-0.2}^{+0.2}$ and $\alpha_3={3.7}_{-0.2}^{+0.2}$ , respectively, for two different extinction maps). Systematic uncertainties coming from model assumptions are not included.}
   {The good performance of BGM FASt demonstrates that it is a very valuable tool to perform multi-parameter inference using Gaia data releases.}
\titlerunning {BGM FASt for big data }

   \maketitle
%

\section{Introduction}\label{introduction} 
Recently, the astrophysics community has successfully carried out important ground-based and space missions generating very large data sets. Large sky surveys with photometric and astrometric data, such as Gaia data release 2 \citep{GaiaDR2}, among others, represent a challenge for Galaxy modelling in terms of both the new types of data and the  large amounts of data created. At the same time, new statistical techniques are having a significant effect on modern astronomy. The use of Bayesian statistics for the exploration of large parameter spaces, together with Monte Carlo Markov Chains (MCMC) or approximate Bayesian computation (ABC), among others, are in rapid development. 

Several attempts have been done to generate fast Milky Way simulations (e.g. \citealt{Girardi2005}, \citealt{Juric2008}, \citealt{Sharma2011} or \citealt{Pasetto2016}). It is demonstrated that the Galaxy models from \cite{Sharma2011} and  \cite{Pasetto2016} can be used to explore large parameters spaces under machine-learning algorithms, MCMC, and ABC (e.g. \citealt{Rybizki2015} and \citealt{Pasetto2016}). The Galaxia code is able to work in two different modes: it can simulate the Milky Way from a Galaxy model based on \cite{Robin2003}, or from N-body simulations. The fast performance of Galaxia relies on its sampling technique and its clever strategy adopted to avoid the simulation of unnecessary stars. The strategy of \cite{Pasetto2016}
to perform fast simulations of the Milky Way is based on the use of distribution functions. It constructs colour-magnitude diagrams (CMDs) of single stellar populations from N-body simulations, with low computational cost \citep{Pasetto2012}.

The Besan\c{c}on Galaxy Model (BGM; \citealt{Robin2003}) is also a stellar population synthesis model for the Milky Way. It is a very powerful and versatile tool for the statistical analysis of the structure and evolution of the Milky Way. Moreover, it is a valuable tool for the preparation and validation of catalogues for space- and ground-based observational instruments and surveys. Recently, BGM was used to  study the kinematics of the local disc from the RAVE survey and the Gaia first data release (DR1) \citep{Robin2017}, to evaluate the evolution  of the Milky Way's disc shape over time \citep{Amores2017}, to constrain Galactic and stellar physics \citep{Lagarde2017}, to constrain the local initial-mass function (IMF) using Galactic Cepheids and Tycho-2 data \citep{Mor2017}, and to study microlensing events in the Galactic bulge  \citep{Awiphan2016}. Furthermore, BGM has also been useful, together with other Milky Way models, to study the bulge bar \citep{simion2017} and for the validation of the Gaia DR1 \citep{Arenou2017}. Nowadays, a BGM standard (BGM Std) simulation (e.g. \citealt{Czekaj2014}) has a computational cost that is not adapted to exploring large parameter spaces using modern Bayesian iterative methods that require a very large number of simulations. To overcome this handicap we have developed a theoretical framework to generate very fast Milky Way approximate simulations based on BGM. This framework allows us to explore, among others, the parameter spaces of the IMF, the star formation history (SFH), and the density laws using ABC. The flexibility of the strategy presented here allows for the generation of fast approximate simulations for different Milky Way components, such as thin disc, thick disc, halo, and bulge. Our full strategy is codified to run on Apache Spark \footnote{https://spark.apache.org/} \citep{ApacheSpark} and  Apache Hadoop \footnote{http://hadoop.apache.org/}, which are engines coming from business science suited to deal with large surveys. Thanks to its codification, BGM FASt is implemented in the big data infrastructure known as Gaia Data Analytics Framework (GDAF, e.g. \citealt{Tapiador2017}). As a first application of this complex strategy we use ABC algorithms, BGM, and Tycho-2 data to constrain the IMF, the local SFH, the local stellar mass density, and the thin-disc density laws.

In Sect. \ref{BGM} we describe the BGM, and in Sect. \ref{FASt} we present the framework to generate the BGM fast approximate simulations (BGM FASt). In Sect. \ref{ldse} we describe the treatment of the local dynamical statistical equilibrium in the context of BGM FASt. In Sect. \ref{ABC} we briefly describe the approximate Bayesian computation technique applied to explore the parameter space of the fundamental functions of the Milky Way. In Sect. \ref{SolarN} we present an evaluation of the BGM FASt performance in the solar neighbourhood. Results are presented in Sect. \ref{Results} while a discussion and conclusions are presented in Sects. \ref{Discussion} and \ref{Conclusions}.

\section{The Besan\c{c}on Galaxy Model}\label{BGM}

In the present paper we use the following versions of the Galactic components of BGM chosen from a compromise between recent and stable updates. 

For the stellar halo component we use the model from \cite{Robin2014} and for the bulge-bar region we use the model described in \cite{Robin2012}. For the thick disc component we use the model from the best fit obtained in \cite{Robin2014} which is a thick disc with two main star-formation episodes at 10 and 12 Gyr. For the thin disc component we use the model described in \cite{Czekaj2014} with the updates on the parameters introduced in \cite{Mor2017}. The local dynamical statistical equilibrium of BGM is ensured by dynamical constraints based on \cite{Bienayme1987}. The last dynamics and kinematics updates from \cite{Bienayme2015} and \cite{Robin2017} are not considered in the present paper and will be incorporated in the near future.

\subsection{BGM star-generation strategy}\label{generation}

The BGM has two main working modes to compute the generation of the stars in the Galaxy. The traditional approach relies on using a precomputed Hess diagram \citep{Robin2003}, while the more updated approach is able to generate the stars from a given set of fundamental functions (e.g. IMF, SFH, age-metallicity among others), making them evolve using a desired set of stellar evolutionary models \citep{Czekaj2014}. For each Galactic component we can choose whether we want to simulate it using the Hess diagram or the updated strategy. Alternatively, from the updated star-generation strategy we can build a Hess diagram from a given set of fundamental functions, and ingest it into BGM code afterwards to be used in a traditional way.

In this section we summarise the stellar generation strategy described in \cite{Czekaj2014}, henceforth referred to as our standard strategy. Initially this strategy was developed for the thin-disc component but nowadays it can be used for other Galactic components.

In the BGM Std strategy, to generate stars born $\tau$ years ago for a given Galactic i-component (e.g. thin disc, thick disc, halo and bulge-bar), we start from a given total surface mass density at the position of the Sun $(\Sigma^i_\odot)$. We then use the SFH ($\psi^i_\odot(\tau)$) to distribute the surface mass density along $\tau$ as follows:

\begin{equation}\label{SigmaSol}
\Sigma^ i_\odot (\tau) \approx \Sigma^ i_\odot \cdot \psi^i_\odot(\tau)
.\end{equation}

For simplicity, the current version neglects the radial migration. We setup the model so that the stars are born in the plane. We then redistribute them in the process of secular evolution by using the surface-to-volume mass density ratio at the position of the Sun $(\mathcal{H}(\tau,\bar{x}_\odot)={\Sigma_\odot(\tau)}/{\rho_\odot(\tau)})$ to compute the volume stellar mass density from $\Sigma^ i_\odot$ , as follows.

\begin{equation}\label{RSV}
\rho_i(\tau,x_\odot,y_\odot,z_\odot)=\frac{\Sigma^i_\odot (\tau)}{\mathcal{H}_i(\tau,x_\odot,y_\odot,z_\odot)} = \frac{\Sigma^i_\odot \cdot \psi^i_\odot(\tau)}{\mathcal{H}_i(\tau,x_\odot,y_\odot,z_\odot)}
,\end{equation}

\noindent where we have expressed the position $\bar{x}$ in Cartesian Galactic coordinates as $(x,y,z)$. The volume mass density is distributed throughout the Galaxy as

\begin{equation}\label{rhoCzTau}
\rho_i(\tau,x,y,z) =  \rho_i(\tau,{x}_\odot,{y}_\odot,{z}_\odot) \cdot \mathcal{R}_i(\tau,x,y,z) 
,\end{equation}

\noindent where $\mathcal{R}_i(\tau,x,y,z)$  are the density laws for the given i-component and $\mathcal{R}_i(\tau,x_\odot,y_\odot,z_\odot)=1$.  We can then  write $\mathcal{H}_i(\tau,x_\odot,y_\odot,z_\odot)$ as the integral of the density law along the vertical direction at the position of the Sun:

\begin{equation}\label{Hi}
\mathcal{H}_i(\tau,x_\odot,y_\odot,z_\odot)=\frac{\Sigma^i_\odot(\tau)}{\rho_i(\tau,x_\odot,y_\odot,z_\odot)}=\int_{\forall z}\mathcal{R}_i(\tau,x_\odot,y_\odot,z)  \cdot dz
.\end{equation}

\noindent Finally, from Equations \ref{RSV}, \ref{rhoCzTau} and \ref{Hi}, we can write the distribution of the volume mass density along position and age as follows.

\begin{equation}\label{rhoCzTau2}
\rho_i(\tau,\bar{x}) =  \frac{\Sigma^ i_\odot \cdot \psi^i_\odot(\tau)}{\mathcal{H}_i(\tau)} \cdot \mathcal{R}_i(\tau,\bar{x}) 
,\end{equation}

\noindent where for simplicity we call $\mathcal{H}_i(\tau)$ to $\mathcal{H}_i(\tau,x_\odot,y_\odot,z_\odot)$. As explained in \cite{Czekaj2014}, the IMF distributes this mass density in three mass ranges. The star generation process goes through all the volume elements in the Galaxy. First, in a given volume element, for a given age sub-population the age of the star is drawn uniformly within the age limits. Afterwards, the mass of the star is drawn from the IMF. Next, the metallicity is assigned depending on the age and position of the stars. In the most updated versions, the $\alpha$-elements-to-iron abundance ($[\alpha/Fe]$) is assigned to each star with a given probability \citep{Lagarde2017}. The evolutionary stage is then assigned to the star by interpolating the stellar evolutionary tracks. In a following step, the process assigns to the generated star a given probability to be the primary component of a stellar multiple system. This probability is assigned following the guidelines of \cite{Arenou2011}. Finally, if the star is flagged as a primary component of a stellar multiple system, the standard strategy generates a secondary star with a mass drawn from the probability distributions described in \cite{Arenou2011}.

For consideration in the following sections we define a BGM Std simulation   to be one that works using the standard stellar generation strategy or a fixed Hess diagram built from the standard stellar generation strategy. 

\subsection{Standard thin disc component}\label{stdthin}

The thin disc component is described in Czekaj et al. (2014). The
stars are generated as described in Sect. \ref{generation}. The thin disc population is divided in seven age
sub-populations. Usually the chosen age intervals are those described in \cite{Bienayme1987} but for the two youngest populations we use the age limits described in \cite{Mor2017}. The density distribution of each sub-population
of the thin disc is assumed to follow an Einasto density profile
as described in \cite{Robin2012}, except for the
youngest sub-population which follows the expression described
in \cite{Robin2003}. These profiles are characterised by the eccentricities of the ellipsoid
(i.e. the axis ratio), the radial scale length of the disc ($h_{R}$), and
the radial scale length of the disc hole ($h_{Rh}$). A velocity dispersion as a function of age is adopted and the dynamical statistical equilibrium is ensured by using the strategy described in \cite{Bienayme1987}. Stellar evolutionary tracks
and model atmosphere, combined with an age-metallicity relation, allow us to go from masses, ages, and metallicities to the space of the observables. In this process a three-dimensional (3D) interstellar extinction map is adopted.

\section{Framework for the Besan\c{c}on Galaxy Model Fast Approximate Simulation}\label{FASt}

A BGM Std simulation has a computational cost of $\sim 432$ hours of CPU time for a simulation of $10^6$ stars, excluding the use of iterative methods like ABC or MCMC to explore large parameter spaces. Hence, we have developed a new method, called BGM FASt, which is able to robustly simulate the Galaxy with a computational cost of $\sim 240$ seconds of CPU time for a simulation of $10^6$ stars. Thanks to the use of Apache Hadoop and Apache Spark environments \citep{ApacheSpark} the computational cost should not scale with the number of stars as is the case in standard environments (Julbe, F. private comm.).

\subsection{The BGM FASt concept}\label{FASt1}

 The BGM FASt is a population-synthesis simulation of the Milky Way, obtained from a clever modification of a BGM Std simulation. The BGM FASt development is based on the distribution function of the generated stars ($\mathcal{D}_i$) . The $\mathcal{D}_i$  carries on the information about the generation of the stars in the i-component of the Galaxy (e.g. thin disc, thick disc, halo, bulge-bar) throughout the life of the given component up to the present day. This distribution function contains the chemo-dynamical information that is classically expressed by fundamental functions such as the IMF, the SFH, density distribution, the age-metallicity relation, and the radial metallicity gradient, among others. The $\mathcal{D}_i$ is defined in a N dimensional space $(\P^i$) for each of the i-components of the Galaxy. This N dimensional space contains all the parameters that can be involved in a distribution function of the generated stars in the Galaxy. Let us introduce the parameter space as follows:

\begin{equation}\label{Pi}
\P^i \equiv \tau \times M \times  Z  \times \bar{x} \times \bar{v}  \times \bar{p}
,\end{equation}

\noindent where  $\tau$ is the present age of the stellar object, M and $Z$ are its initial  mass and metallicity, and $\bar{x}$ and $\bar{v}$ are position and velocity, respectively. $\bar{p}$ accounts for other independent parameters that, for some specific purposes, would be interesting to have explicitly introduced in the distribution function.  The $\alpha$-elements-to-iron abundance ratio ($\left [ \alpha/Fe \right ]$) is an example of one of the possible $\bar{p}$ parameters and we treat this in the following section.
 
The strategy to generate a BGM FASt begins with the choice of a specific Mother Simulation with an imposed set of fundamental functions. We use Mother Simulation to refer to a BGM Std simulation used as a seed to generate a BGM FASt simulation. This Mother Simulation is used as a main constituent to generate one or several BGM FASt simulations with different assumptions for the fundamental functions. The idea behind the BGM FASt strategy is that the number of stars generated in a given interval of the parameter space ($\mathcal{N}_i(\Delta \P)$) is proportional to the mass dedicated to generate stars for that given interval ($\mathcal{M}_i(\Delta \P)$):  

\begin{equation}\label{prop}
\mathcal{N}_i(\Delta \P) \propto \mathcal{M}_i(\Delta \P)
,\end{equation}

\noindent where $\Delta \P \equiv (\Delta \tau, \Delta M, \Delta Z,\Delta \bar{x},\Delta \bar{v},\Delta \bar{p})$. Equation (\ref{prop}) is valid for both the Mother Simulation and the BGM FASt simulation. If  $\Delta \P$ is small enough,  we can write a proportion between the number of stars and the masses relating both the Mother Simulation and the BGM FASt simulation:

\begin{equation}\label{prop2}
\frac{\mathcal{N}^{FASt}_i(\Delta \P)}{\mathcal{N}^{MSt}_i(\Delta \P)} \propto \frac{\mathcal{M}^{FASt}_i(\Delta \P)}{\mathcal{M}^{MSt}_i(\Delta \P)}
.\end{equation}

\noindent Then we can approximate the number of stars for a given interval for a BGM FASt simulation as follows:

\begin{equation}\label{propN}
\mathcal{N}^{FASt}_i(\Delta \P) \approx \frac{\mathcal{M}^{FASt}_i(\Delta \P)}{\mathcal{M}^{MSt}_i(\Delta \P)} \cdot \mathcal{N}^{MSt}_i(\Delta \P)
.\end{equation}

\noindent Let us call weight to the mass ratio of equation (\ref{propN}):

\begin{equation}\label{iniW}
{w}_i = \frac{\mathcal{M}^{FASt}_i(\Delta \tau, \Delta M, \Delta Z,\Delta \bar{x},\Delta \bar{v},\Delta \bar{p})}{\mathcal{M}^{MSt}_i(\Delta \tau, \Delta M, \Delta Z,\Delta \bar{x},\Delta \bar{v},\Delta \bar{p})}
,\end{equation}

\noindent where we compute the mass dedicated to generate stars, in a given interval, from the distribution function of the generated stars that we present in following sections. 

In practice, we generate a BGM FASt simulation by applying a weight to each star of the Mother Simulation. This is done according to the parameters of the star, such as mass, age, position, and   distance, among others. Thus, the resulting simulation is an approximation of the BGM Std simulation that would be obtained with the standard BGM star generation process.   

We present the theoretical framework and the practical implementation for the generation of a BGM FASt as follows.

First, in Sect. \ref{GSDf}, we describe the  distribution function of the generated stars $\mathcal{D}_i$ in the most generic context and its relation with the masses involved in Equation (\ref{iniW}). This allows us to introduce the classical fundamental functions, such as the IMF and the SFH, and functions describing more complex scenarios. In a following step we consider a set of assumptions and approximations to reach a $\mathcal{D}_i$ function compatible with the $\mathcal{D}_i$ implicitly involved in a BGM Std simulation. 

Next, in Sect. \ref{bin}, we discuss the treatment of stellar multiple systems as modelled in a BGM. We introduce the probability of obtaining a binary system at birth in our approximated $\mathcal{D}_i$. The obtained expressions are useful for both the process ensuring the local dynamical statistical equilibrium and the computation of the surface mass stellar density at the position of the Sun. Finally after describing a generalizable weight expression, in Sect. \ref{FAStweight}, we constrain it to the BGM context including stellar multiple systems.

\subsection{The distribution function of the generated stars }\label{GSDf}

\subsubsection{Generic context and fundamental functions}\label{GSDf1}

In this section we describe the distribution function of the generated stars in its most generic context. Under the given definition of the parameter space ($\P^i$)  we can precisely define some of the parameters belonging to $\bar{p}$. It is convenient for future purposes to write the distribution function $\mathcal{D}_i(\tau, M, Z,\bar{x},\bar{v}, \bar{p})$ accounting explicitly for the ratio $[\alpha/Fe]$. Subsequently, $[\alpha/Fe]$ is considered as one of the $\bar{p}$ parameters and we can write $\mathcal{D}_i(\tau, M, Z,\bar{x},\bar{v}, [\alpha/Fe], \bar{p'})$ in the N dimensional space:

\begin{equation}\label{Pi2}
\P^i \equiv \tau \times M \times  Z  \times \bar{x} \times \bar{v} \times \bar{p} =  \tau \times M \times  Z  \times \bar{x} \times \bar{v} \times [\alpha/Fe] \times \bar{p'}
.\end{equation}

\noindent In our line of action, for the moment, we are interested in
explicitly representing age, mass, metallicity, position, velocity, and the ratio  $[\alpha/Fe]$ in the distribution function. Subsequently we marginalize $\mathcal{D}_i(\tau,  M,  Z,\bar{x},\bar{v}, [\alpha/Fe], \bar{p'})$ over the rest of the $\bar{p'}$ parameters:

\begin{equation}\label{GSr}
\mathcal{G}_i(\tau,M,Z,\bar{x},\bar{v}, [\alpha/Fe]) = \int_{ \forall \,  \bar{p'} \, \in \, \P^i, } \mathcal{D}_i(\tau, M, Z,\bar{x},\bar{v}, [\alpha/Fe],\bar{p'}) \cdot d\bar{p'} 
,\end{equation}

\noindent where $\mathcal{G}_i$ is the distribution function of the generated stars for the i-component in the reduced space $\P^i_{r}$:

\begin{equation}\label{Pi_rd}
\P^i_{r} \equiv \tau \times M \times  Z  \times \bar{x} \times \bar{v} \times [\alpha/Fe]
.\end{equation}

\noindent For simplicity let us henceforth refer to $[\alpha/Fe]$ using only $\alpha$.

The $\mathcal{G}_i(\tau,M,Z,\bar{x},\bar{v},\alpha)$ distribution function is such that the integral over all the parameters that belong to the parameter space is the total number of generated stars in the i-component of the Galaxy:

\begin{equation}\label{Ni}
\int_{\forall \, \tau,M,\bar{Z},\bar{x}, \bar{v},\alpha \,  \in \, \P^i_{r}} \mathcal{G}_i(\tau,M,Z,\bar{x},\bar{v},\alpha) \cdot d\tau dM \, dZ \, d\bar{x} \, d\bar{v} d\alpha= N_{i}
,\end{equation}

\noindent and if we multiply by the mass before the integration, we have the total mass of the generated stars for the i-component of the Galaxy:

\begin{equation}\label{Mi}
\int_{\forall \, \tau,M,\bar{Z},\bar{x}, \bar{v},\alpha \,  \in \, \P^i_{r}} \mathcal{G}_i(\tau,M,Z,\bar{x},\bar{v},\alpha) \cdot M \cdot d\tau dM \, dZ \, d\bar{x} \, d\bar{v} d\alpha= M_{i}
.\end{equation}

\noindent The mass of equation (\ref{prop}) can then be expressed as the integral of the distribution function for a given interval in the parameter space $(\Delta \P^i_{r})$ :

\begin{equation*}
\mathcal{M}_i(\Delta \tau,\Delta M,\Delta Z,\Delta \bar{x},\Delta \bar{v},\Delta \alpha)=
\end{equation*}

\begin{equation}\label{Mint}
= \int_{\forall \, \tau,M,\bar{Z},\bar{x}, \bar{v},\alpha \,  \in \,  \Delta \P^i_{r}} \mathcal{G}_i(\tau,M,Z,\bar{x},\bar{v},\alpha) \cdot M \cdot d\tau dM \, dZ \, d\bar{x} \, d\bar{v} d\alpha
,\end{equation}
 
\noindent and we can write equation (\ref{iniW}) as

\begin{equation}\label{WG}
 w_i=\frac{\int_{ \Delta \P^i_{r}} \mathcal{G}^{FASt}_i(\tau,M,Z,\bar{x},\bar{v},\alpha) \cdot M \cdot d\P_r}{\int_{ \Delta \P^i_{r}} \mathcal{G}^{MSt}_i(\tau,M,Z,\bar{x},\bar{v},\alpha) \cdot M \cdot d\P_r}
,\end{equation}

\noindent where $d\P_r \equiv d\tau dM \, dZ \, d\bar{x} \, d\bar{v} d\alpha$.

The true $\mathcal{G}_i(\tau,M,Z,\bar{x},\bar{v},\alpha)$ distribution function is unknown, but its marginalization over combinations of parameters results in deeply studied functions such as the IMF, the SFH, the age-metallicity relation, the radial metallicity gradient, and also functions carrying information about the density distribution of the Galaxy or dynamical and chemo-dynamical information. Let us exemplify mathematically how some of these fundamental functions can be treated related to the distribution function.

The marginalization over the parameters $\tau,Z,\bar{x},\bar{v}$ and $\alpha$ within the values that belong to the $\P^i_{r}$ space can be written as:

\begin{equation}\label{IMF}
\int_{\forall \, \tau,Z,\bar{x},\bar{v}, \alpha \, \in \, \P^i_{r}} \mathcal{G}_i(\tau,M,Z,\bar{x},\bar{v},\alpha) \, d\tau \, dZ \,  d\bar{x} \, d\bar{v} \, d\alpha = \xi_i(M)
,\end{equation}
 
\noindent where $\xi_i(M)$ is the composite IMF for each one of the i-components. Marginalizing now the $\mathcal{G}_i$ over $M,Z,\bar{x},\bar{v},\alpha$ within the values that belong to the $\P^i_{r}$ space we have the $\tau$ distribution of the generated stars that can be interpreted as the SFH of the whole i-component:

\begin{equation}\label{SFH}
\int_{\forall \, M,Z,\bar{x},\bar{v},\alpha \, \in  \, \P^i_{r}} \mathcal{G}_i(\tau,M,Z,\bar{x},\bar{v},\alpha)\, dM \, dZ \,  d\bar{x} \, d\bar{v} \, d\alpha= \Psi_i(\tau)
.\end{equation}

\noindent IF we are interested in studying how the $\tau$ distribution depends on the position, we can then perform a marginalization over M, Z, $\bar{v}$ and $\alpha,$ obtaining

\begin{equation}\label{sfh}
\int_{\forall \, M,Z,\bar{v},\alpha \,\in \, \P^i_{r}} \mathcal{G}_i(\tau,M,Z,\bar{x},\bar{v},\alpha) \,dM \, dZ \,   d\bar{v} \, d\alpha = f_i(\tau,\bar{x})
.\end{equation}

\noindent If in equation (\ref{sfh}) we set $\bar{x}=\bar{x}_\odot$, the position of the Sun, then the resulting function can be interpreted as the SFH at the position of the Sun.

The functions involving metallicity, such as the age-metallicity relation or the radial metallicity gradient, could also be considered in BGM FASt. If we marginalize $\mathcal{G}_i(\tau,M,Z,\bar{x},\bar{v},\alpha)$ over mass, $\alpha,$ and phase-space we get the Z distribution of stars formed $\tau$ years ago:

\begin{equation}\label{AM}
\int_{\forall \, M,\bar{x},\bar{v},\alpha \,  \in \, \P^i_{r}} \mathcal{G}_i(\tau,M,Z,\bar{x},\bar{v},\alpha) \, dM \, d\bar{x} \,   d\bar{v} \, d\alpha = \chi_i(\tau,Z)
.\end{equation}

The radial metallicity gradient can be deduced from a more complex expression obtained marginalizing $G_i$ over age, mass, velocity and $\alpha$:

\begin{equation}\label{eta}
\int_{\forall \, \tau,M,\bar{v},\alpha \, \in \, \P^i_{r}} \mathcal{G}_i(\tau,M,Z,\bar{x},\bar{v},\alpha) \, d\tau \,  dM \,  d\bar{v} \, d\alpha  = \eta_i(Z,\bar{x})
.\end{equation}

\noindent This latter expression is the position and metallicity distribution of the generated stars throughout the life of the i-component. 

Finally, information about chemo-dynamics and kinematics can be introduced with the following two equations.

\begin{equation}\label{MC}
\int_{\forall\, M,\bar{x}, \, \alpha \, \in \, \P^i_{r}} \mathcal{G}_i(\tau,M,Z,\bar{x},\bar{v},\alpha) \, dM \,d\bar{x} \, d\alpha = \mathcal{Q}_i(\tau,Z,\bar{v})
,\end{equation}

\begin{equation}\label{MG}
\int_{\forall \, M,Z, \, \alpha \, \in  \, \P^i_{r}} \mathcal{G}_i(\tau,M,Z,\bar{x},\bar{v},\alpha) \, dM \, dZ \, d\alpha  = \mathcal{K}_i(\tau,\bar{x},\bar{v})
.\end{equation}

The spatial distribution of the volume mass density ($\rho^i_{g}(\bar{x})$) that has been dedicated to generating stars for the Galactic i-component throughout its life can be written as follows.

\begin{equation}\label{densi}
\rho^i_{g}(\bar{x})= \int_{\forall \, \tau,M,\bar{Z},\bar{v},\alpha \,  \in \,\P^i_{r}} \mathcal{G}_i(\tau,M,Z,\bar{x},\bar{v},\alpha) \cdot M  \, d\tau \, dM \, dZ \, d\bar{v} \, d\alpha
.\end{equation}

\noindent In the following steps it is useful to have the equation of the mass density dedicated to generating stars born $\tau$ years ago in a position $\bar{x}$:

\begin{equation}\label{densii}
\rho^i_{g}(\tau,\bar{x})= \int_{\forall \, M,\bar{Z},\bar{v} \,  \in \,\P^i_{r}} \mathcal{G}_i(\tau,M,Z,\bar{x},\bar{v},\alpha)   \cdot M \, dM \, d\bar{Z} \, d\bar{v} \, d\alpha
.\end{equation}

Until now, we have described the distribution function of the generated stars in the Galaxy in a generic context. We have emphasised that our strategy can be generalizable and any of the fundamental equations described above (Equations \ref{IMF} to \ref{MG}) can be used if we are able to write an analytical expression for them.

\subsubsection{The approximate solution}\label{GSDf2}

In this section we find an approximation to the distribution function of the generated stars compatible with BGM. At the same time this approximation aims to be extensible to other models of the Galaxy that use similar star-generation strategies. As the exploration of the velocity spaces is not included in the present paper, for simplicity we do not consider the kinematic part here. Our goal is therefore to find an approximate solution to the integral $\int_{\forall \, \bar{v} \, \in  \, \P^i_{r}} \mathcal{G}_i(\tau,M,Z,\bar{x},\bar{v},\alpha) \cdot d\bar{v}$.

In this context the first assumption comes from a traditional strategy (e.g. \citealt{Tinsley1980}), assuming that mass and age distributions are separated. Splitting the mass function from the function of $\tau$ and $\bar{x}$ as follows.

\begin{equation}\label{split1}
\xi_i(M) \cdot \mathcal{F}_i(\tau,\bar{x})
.\end{equation}

Moreover, the BGM assumes a metallicity distribution that depends on position and age. We can therefore introduce, in the equations, the probability that a star of a given age in a given position has a metallicity Z:  $\mathcal{P}_i(Z|\tau,\bar{x})$. Furthermore, BGM has recently included the possibility to use $[\alpha/Fe]$, a parameter which affects the stellar evolutionary tracks \citep{Lagarde2017}, assuming, from observational surveys, a certain probability that a given star has a given $[\alpha/Fe]$. In general, for a given i-component, this probability depends on the age, the position, and the metallicity of the star and we denote it as $\mathcal{P}_i(\alpha|\tau,Z,\bar{x})$.

From Equation (\ref{split1}), and the metallicity and $[\alpha/Fe]$ distributions, we can write:

\begin{equation}\label{split}
\int_{\forall \, \bar{v} \, \in  \, \P^i_{r}} \mathcal{G}_i(\tau,M,Z,\bar{x},\bar{v},\alpha) \cdot d\bar{v}  \approx \xi_i(M) \cdot \mathcal{F}_i(\tau,\bar{x}) \cdot \mathcal{P}_i(Z|\tau,\bar{x}) \cdot \mathcal{P}_i(\alpha |\tau,Z,\bar{x})
,\end{equation}

\noindent where the assumptions and approximations behind the mathematical expression of the functions $\mathcal{F}_i(\tau,\bar{x})$, $\mathcal{P}_i(Z|\tau,\bar{x})$ and  $\mathcal{P}(\alpha|\tau,Z,\bar{x})$ are imposed to be compatible with the BGM. Equation (\ref{split}) assumes, from the statistical point of view, that the probability to generate a star with mass M and the probability to generate a star  $\tau$ years ago in a given position are conditionally independent. This means that the IMF is assumed to be independent of time and position. 

The standard star-generation strategy described in Sect. \ref{generation} guides us by using Equation (\ref{rhoCzTau2}) to approximate the function  $\mathcal{F}_i(\tau,\bar{x})$ as follows.

\begin{equation}\label{func}
\mathcal{F}_i(\tau,\bar{x}) \approx \frac{\Sigma^{i}_{\odot} \cdot \psi^{i}_\odot(\tau) }{\mathcal{H}_i(\tau)}\cdot \mathcal{R}_i(\tau,\bar{x})
,\end{equation}

\noindent where $\Sigma^{i}_\odot$ is the stellar surface mass density $(*/pc^2)$ of the generated stars at the position of the Sun for the  Galactic i-component, $\mathcal{H}_i(\tau)$ is the surface-to-volume-density ratio at the position of the Sun, $\mathcal{R}_i(\tau,\bar{x})$ is the density distribution, and $\psi^i_\odot$ is the SFH in the solar neighbourhood.

Taking into account the approximations implicitly or explicitly adopted in standard BGM  we present a solid solution for the following integral.

\begin{equation*}
\int_{\forall \,  \bar{v} \, \in \, \P^i_{r}} \mathcal{G}_i(\tau,M,Z,\bar{x},\bar{v},\alpha)    \, d\bar{v} \approx 
\end{equation*}
\begin{equation}\label{GSend}
 \approx \frac{\Sigma^{i}_{\odot} \cdot \psi^i_\odot(\tau) }{\mathcal{H}_i(\tau)}\cdot \mathcal{R}_i(\tau,\bar{x}) \cdot \xi_i(M) \cdot  \mathcal{P}_i(Z|\tau,\bar{x}) \cdot \mathcal{P}_i(\alpha |\tau,Z,\bar{x})
,\end{equation}

\noindent where the IMF is normalized as $\int_{\forall \, M \, \in \, \P^i_{r}} \xi(M) \cdot M dM=1$, the SFH in the solar neighbourhood is normalized as $\int_{\forall \, \tau \, \in \, \Pi_{r}} \psi_\odot(\tau) d\tau =1$, and by definition $\mathcal{P}_i(Z|\tau,\bar{x})$ and  $\mathcal{P}_i(\alpha|\tau,Z,\bar{x})$ are normalized to 1.  $\mathcal{R}_i(\tau,\bar{x})$ and $\mathcal{H}_i(\tau)$ are such that the integral over the whole parameter space of $\mathcal{G}_i(\tau,M,Z,\bar{x},\bar{v},\alpha)$ gives the total number of generated stars in the Galactic i-component.

For simplicity, as BGM Std does, we assume an axi-symmetric structure with no radial migration. Additionally, we assume that for a given volume element, $\frac{d\rho}{dt} \approx 0$ , where $\rho$ is the stellar volume mass density. These assumptions allow us to use the density distributions ($\mathcal{R}_i(\tau,\bar{x})$) described in \cite{Robin2003} and \cite{Robin2014}, derived to match the present density distribution of each Galactic i-component.

\subsection{Handling of the stellar multiple systems}\label{bin}

The probability of obtaining a multiple system at birth in a star formation event is unknown, very complex, and can depend on many parameters, such as the metallicity, the mass of the molecular cloud, the turbulence, mass segregation, mass competition, and mass accretion rate (e.g. \citealt{Bonnell2007}, \citealt{Kroupa2013}). For these reasons, the BGM approach (e.g. \citealt{Czekaj2014} and  \citealt{Robin2012}) is performed adopting an empirical law to match the observed present distribution function of multiple stellar systems \citep{Arenou2011}. This approach assigns a certain probability to a generated star  of being the primary component of a multiple system, depending on its mass and its luminosity class. Inside our framework this means a dependence on mass, age, and, through the stellar evolutionary models,  Z and $\alpha$. This probability is therefore $P(bin|\tau,M,Z,\alpha)$.  

Let us refer to the generated star susceptible to be flagged as either the single star or primary component of the binary system as the "primal star". Once a primal star is decided as being a primary component of a multiple system, a secondary star is generated with a mass m. The mass m of the secondary is assigned following a probability distribution function (PDF) that depends mainly on the mass M and the luminosity class of the primary component, leading to $P(m|\tau, M,Z,\alpha)$.

We introduce the stellar multiple systems in BGM FASt by simplifying their treatment under two assumptions: (1) The masses of primal stars (singles and primaries) are drawn from the IMF while the mass of the secondary follows the empirical laws described above, and (2) the age of the primary star follows the SFH, while the age of the secondary is assumed to be the same as the primary (i.e. both components were born together). With these two assumptions, the multiple stellar systems can be introduced in our analytical approach at low computational cost. We can compute first, at the position of the Sun, the surface mass density of secondary stars ($\Sigma^{i,\$}_\odot$) as a function of the surface mass density of the primal stars ($\Sigma^{i,\triangle}_\odot$) and afterwards compute $\Sigma^{i,\triangle}_\odot$ as a function of $\Sigma_\odot^{i}$. We begin by expressing $\Sigma^i_\odot$ as

\begin{equation}\label{sigmaall}
\Sigma_\odot^{i}=\Sigma^{i,\triangle}_\odot + \Sigma^{i,\$}_\odot
.\end{equation}

\noindent Subsequently, the stellar volume mass density of primary stars at the position of the sun for the i-component is given by

\begin{equation*}
 \rho_g^{i,\triangle}(\bar{x}_\odot) \approx \int_{\forall \, \tau, M, \,Z, \, \alpha \, \in  \, \P^i_{r}}   \frac{\Sigma^{i,\triangle}_{\odot} \cdot \psi^i_\odot(\tau) }{\mathcal{H}_i(\tau)}  \cdot \mathcal{R}(\tau,\bar{x}_\odot) \cdot \xi_i(M) \cdot M \cdot
\end{equation*} 
\begin{equation}\label{rhotrisolg}
 \cdot \mathcal{P}_i(Z|\tau,\bar{x}_\odot) \cdot \mathcal{P}_i(\alpha | \tau,Z,\bar{x}_\odot)  d\tau \, dM \, dZ \, d\alpha
,\end{equation}

\noindent and the stellar volume mass density of secondary stars at the position of the sun for the i-component is given by

\begin{equation*}
 \rho_g^{i,\$}(\bar{x}_\odot) \approx \int_{\forall \, \tau, M, \,Z, \, \alpha, \, \in  \, \P^i_{r}} \int^{m_{max}}_{m_{min}}  \frac{\Sigma^{i,\triangle}_{\odot} \cdot \psi^i_\odot(\tau) }{\mathcal{H}_i(\tau)}  \cdot \mathcal{R}(\tau,\bar{x}_\odot) \cdot \xi_i(M)  \cdot \mathcal{P}_i(Z|\tau,\bar{x}_\odot) \cdot 
\end{equation*}
\begin{equation}\label{rho$solg} 
\cdot \mathcal{P}_i(\alpha | \tau,Z,\bar{x}_\odot) \cdot P(bin|\tau,M,Z,\alpha) \cdot P(m|\tau,M,Z,\alpha) \cdot m \, dm  d\tau \, dM \, dZ \, d\alpha
.\end{equation}

\noindent We also want to compute the surface mass density; then, if we leave out $\mathcal{H}_i$ from the integrand of equation (\ref{rho$solg}), we get the surface mass density of secondary stars at the position of the Sun:

\begin{equation*}
\Sigma^{i,\$}_\odot \approx  \Sigma^{i,\triangle}_{\odot} \int_{\forall \, \tau, M, \,Z, \, \alpha \, \in  \, \P^i_{r}} \int^{m_{max}}_{m_{min}}    \psi^i_\odot(\tau)    \cdot \xi_i(M)  \cdot \mathcal{P}_i(Z|\tau,\bar{x}_\odot) \cdot
\end{equation*}
\begin{equation}\label{Sigma$solg} 
\cdot P_i(\alpha | \tau,Z,\bar{x}_\odot) \cdot P(bin|\tau,M,Z,\alpha) \cdot P(m|\tau,M,Z,\alpha) \cdot m \, dm  d\tau \, dM \, dZ \, d\alpha
.\end{equation}

\noindent Finally, $\Sigma^{i,\triangle}_\odot$ is computed from  $\Sigma^{i}_\odot$  using Equations (\ref{sigmaall}) and (\ref{Sigma$solg}). 

This result is very useful for both determining the weight function when considering multiple stellar systems as implemented in BGM (Sect. \ref{FAStweight}) and computing the local dynamical statistical equilibrium (Sect. \ref{ldse}). In Sect. \ref{ldse} we also particularize the computation of $\Sigma^{i,\triangle}_\odot$ when specifically using the thin disc component as described in \cite{Czekaj2014}.


\subsection{The weight}\label{FAStweight}

The strategy developed in previous sections allows us to give the analytical expression to compute the weights. These weights are able to transform the distribution of stars in the Mother Simulation, linked to a given set of Galactic fundamental functions, into the distribution of the stars linked to other Galactic fundamental functions adopted for a BGM FASt simulation. From (\ref{WG}) and (\ref{GSend}) we obtain an expression for the weights applicable to Galaxy models that sample the stars from a distribution function of the form of Equation (\ref{split}), considering the assumptions discussed in Sect. \ref{FASt}:

\begin{equation*}
w_i(\tau,  M, Z, \bar{x}, \alpha) \approx  \frac{ \int_{\Delta \P^i_r} \frac{\Sigma^{i,FASt}_{\odot} \cdot \psi^{i,FASt}_\odot }{\mathcal{H}_i^{FASt}(\tau)} \cdot \mathcal{R}_i^{FASt}(\tau,\bar{x})}          { \int_{\Delta \P^i_r}\frac{\Sigma^{i,MS}_{\odot} \cdot \psi^{i,MS}_\odot }{\mathcal{H}_i^{MSt}(\tau)}\cdot \mathcal{R}_i^{MSt}(\tau,\bar{x})} \cdot 
\end{equation*}
\begin{equation}\label{wii}
\cdot \frac{\xi_i^{FASt}(M) \cdot M \cdot \mathcal{P}^{FASt}_i(Z|\tau,\bar{x}) \cdot \mathcal{P}^{FASt}_i(\alpha|\tau,Z,\bar{x}) \cdot d\P_r}{\xi_i^{MSt}(M) \cdot M \cdot \mathcal{P}^{MSt}_i(Z|\tau,\bar{x}) \cdot \mathcal{P}^{MSt}_i(\alpha|\tau,Z,\bar{x}) \cdot d\P_r}
.\end{equation}

In the BGM context, including multiple stellar systems and taking into account the scenario described in Sect. \ref{bin}, the weight that we apply in practice to generate a BGM FASt simulation is the following.

\begin{equation*}
w_i( \tau, M, Z, \bar{x}, \alpha) \approx \frac{ \int_{\Delta \P^i_r}  \frac{\Sigma^{i,\triangle,FASt}_{\odot} \cdot \psi^{i,FASt}_\odot }{\mathcal{H}_i^{FASt}(\tau)}\cdot \mathcal{R}_i^{FASt}(\tau,\bar{x})}{ \int_{\Delta \P^i_r}  \frac{\Sigma^{i,\triangle,MS}_{\odot} \cdot \psi^{i,MS}_\odot }{\mathcal{H}_i^{MSt}(\tau)}\cdot \mathcal{R}_i^{MSt}(\tau,\bar{x})} \cdot
\end{equation*}
\begin{equation}\label{wii2}
\cdot \frac{\xi_i^{FASt}(M) \cdot M \cdot \mathcal{P}^{FASt}_i(Z|\tau,\bar{x}) \cdot P^{FASt}_i(\alpha|\tau,Z,\bar{x}) \cdot d\P_r}{\xi_i^{MSt}(M) \cdot M \cdot \mathcal{P}^{MSt}_i(Z|\tau,\bar{x}) \cdot \mathcal{P}^{MSt}_i(\alpha|\tau,Z,\bar{x}) \cdot d\P_r}
,\end{equation}

\noindent where we have substituted  $\Sigma^{i}_\odot$ with  $\Sigma^{i,\triangle}_\odot$ from Equation (\ref{wii}).

In our approach, the stellar evolutionary tracks are  the same for both the Mother Simulation and the BGM FASt simulation. For simplicity, the probabilities $\mathcal{P}_i(Z|\tau,\bar{x})$  and $\mathcal{P}_i(\alpha|\tau,Z,\bar{x})$ are imposed and we are not going to explore them;  they are equal for both the Mother Simulation and the BGM FASt simulation and we can marginalize the numerator and denominator over Z and $\alpha$ to obtain the final expression for the weights:

\begin{equation*}
w_i( \tau ,  M,\bar{x}) \approx \frac{ \int_{\Delta \tau, \, \Delta M, \Delta \bar{x}} \frac{\Sigma^{i,\triangle,FASt}_{\odot} \cdot \psi^{i,FASt}_\odot }{\mathcal{H}_i^{FASt}(\tau)}\cdot \mathcal{R}_i^{FASt}(\tau,\bar{x})}{\int_{\Delta \tau , \Delta M,\Delta \bar{x}} \frac{\Sigma^{i,\triangle,MS}_{\odot} \cdot \psi^{i,MS}_\odot }{\mathcal{H}_i^{MSt}(\tau)}\cdot \mathcal{R}_i^{MSt}(\tau,\bar{x})  } \cdot 
\end{equation*}
\begin{equation}\label{wiii}
 \cdot \frac{   \xi_i^{FASt}(M) \cdot M  \cdot  d\tau \, dM  \, d\bar{x}}{ \xi_i^{MSt}(M) \cdot M \cdot d\tau \, dM \, d\bar{x}}
.\end{equation}

In practice, the weight  in Equation (\ref{wiii}) is applied to each single star and each stellar system. This means that both the primary and the secondary components of the stellar system are weighted with the same value, according to the parameters of the primary component. We choose this way to apply the weights because in BGM Std the mass and age of the secondary star are drawn according to the mass and the age of the primary star. This choice ensures that the mass and age distributions of the primal stars in a BGM FASt simulation follow the pertinent IMF and SFH, and that the mass distribution of the secondary stars follows the empirical distributions described in \cite{Arenou2011}.

The choice of the intervals for the integrals must be discussed for each particular case. Generally they must be small enough to preserve Equations (\ref{prop2}) and (\ref{propN}). For tests and cases presented in the present paper, the choice of the mass interval is set to be very small ($0.025 M_\odot$). For a given star of age $\tau$ we set the limits of the integral to be the age limits of the age sub-populations to which the star belongs. Finally, as the BGM Std generation strategy assigns the density of its centre to the whole volume element,  in BGM FASt we do not need to perform a volume element integral.

\section{Local dynamical statistical equilibrium (LDSE)}\label{ldse}

By local dynamical statistical equilibrium we understand that, at the position of the Sun, the mass density distribution and the potential satisfy both the Poisson equation and the first-order moment of the collisionless Boltzmann equation for the vertical direction  \citep{Bienayme1987}. In this approach we consider axi-symmetry to solve both equations. To solve the first-order moment of the collisionless Boltzmann equation in the vertical direction, we also assume steady state, isothermal state, and decoupled radial and vertical motions. One possible methodology to ensure the LDSE is described in \cite{Czekaj2014} and summarised in Sect. \ref{fullLDSE}. As it requires a computational time that is not affordable for the methodology presented in this paper, we develop here analytical (Sect. \ref{AnalLDSE}) and approximate (Sect. \ref{AppLDSE}) methods that ensure LSDE,  significantly reducing  the computational cost.

\subsection{Full LDSE}\label{fullLDSE}

The iterative strategy described in \cite{Czekaj2014} performs, for a given set of Galactic fundamental functions, a local normalization and a sphere simulation around the Sun to compute at the position of the Sun: the surface mass density of the generated stars, $\Sigma_\odot$, the volume mass density of the stars generated $\tau$ years ago, that is $\rho_g(\tau,\bar{x}_\odot)$, and the volume mass density for stars generated $\tau$ years ago which are not stellar remnants at present, that is  $\rho(\tau,\bar{x}_\odot)$  \footnote{We note that the equivalent nomenclature for $\rho_g(\tau,\bar{x}_\odot)$ and  $\rho(\tau,\bar{x}_\odot)$ in \cite{Czekaj2014} is $\rho^{all}_\odot(i)$ and  $\rho^{obs}_\odot(i).$ }. We define the stars which are not remnant at present as those with $\tau \leq T_{lim}(M,Z,\alpha)$, where $T_{lim}(M,Z,\alpha)$ is the maximum age that a star of a given mass, metallicity, and $[\alpha/Fe]$ reaches without becoming a remnant object. The stellar mass density at the position of the Sun is fitted with the stars that are not remnants at present and the white dwarfs' density is added separately. The mass lost by the stars during their evolution and the interaction between components of a stellar multiple system are neglected. The total mass in stars, plus that of the interstellar medium,  the dark matter, and the central mass, allows us to compute the radial force. At this stage, the dark matter density distribution and the central mass is adjusted such that the model rotation curve fits the observations, the fit is done using the least-squares method in velocity. Finally, the Poisson and the first-order moment of the collisionless Boltzmann equation in the vertical direction are iteratively solved. The whole strategy is iterated from the beginning until convergence is reached.

\subsection{Analytical LDSE}\label{AnalLDSE}

In this analytical approach to ensure LDSE, we follow the process described in Sect. \ref{fullLDSE}, but instead of using a simulation of a sphere around the Sun, we analytically derive the surface mass density of the generated stars ($\Sigma_\odot$), the volume mass density of the stars generated $\tau$ years ago  ($\rho_g(\tau,\bar{x}_\odot)$), and the volume mass density for stars generated $\tau$ years ago which are not remnant at present ($\rho(\tau,\bar{x}_\odot)$). As this approach concerns only the thin-disc component, from now on we avoid the use of the index i.

Using Equation (\ref{sigmaall}) we can express $\Sigma_\odot$ as the sum of the surface density for primal stars $(\Sigma^{\triangle}_\odot)$ and the surface density for secondary stars $(\Sigma^{\$}_\odot)$. In an equivalent way, the $\rho_g(\tau,\bar{x}_\odot)$ for the generated stars can be expressed as

\begin{equation}\label{rhosuntotg}
 \rho_g(\tau,\bar{x}_\odot) = \rho^\triangle_g(\tau,\bar{x}_\odot) + \rho^\$_g(\tau,\bar{x}_\odot)
,\end{equation}

\noindent and the stellar volume mass density of the stars with $\tau \leq T_{lim}(M,Z,\alpha)$ can be expressed as

\begin{equation}\label{rhosuntot}
 \rho(\tau,\bar{x}_\odot) = \rho^\triangle(\tau,\bar{x}_\odot) + \rho^\$(\tau,\bar{x}_\odot)
.\end{equation}

\noindent Now we want to derive the stellar volume mass density of the primal stars at the position of the Sun $\rho^{\triangle}_g(\tau,Z,\bar{x}_\odot,\alpha)$. Therefore, from equation (\ref{rhotrisolg}) we obtain

\begin{equation*}
 \rho_g^{\triangle}(\tau,Z,\bar{x}_\odot,\alpha) \approx \int_{\forall \,  M \, \in  \, \P_{r}}   \frac{\Sigma^{\triangle}_{\odot} \cdot \psi_\odot(\tau) }{\mathcal{H}(\tau)}  \cdot \xi(M) \cdot M \cdot
\end{equation*} 
\begin{equation}\label{rhotrisolgn}
 \cdot \mathcal{P}(Z|\tau,\bar{x}_\odot) \cdot \mathcal{P}(\alpha | \tau,Z,\bar{x}_\odot)   dM 
.\end{equation}

 At this stage we assume that at the position of the Sun all the stars have solar metallicity $Z=Z_\odot$ and for the moment BGM assumes that the $\alpha$-elements-to-iron abundance for the thin disc is $\alpha=\alpha_\odot$. The probabilities for $Z$ and $\alpha$ then become

\begin{equation}\label{pz}
\mathcal{P}(Z|\tau,\bar{x}_\odot)
\begin{cases}
    1 ,& \text{if $Z=Z_\odot$} \\
    0,              & \text{otherwise}
\end{cases}
,\end{equation}

\noindent and

\begin{equation}\label{palpha}
\mathcal{P}(\alpha | \tau,Z,\bar{x}_\odot)
\begin{cases}
    1 ,& \text{if $\alpha=\alpha_\odot$} \\
    0,              & \text{otherwise}
\end{cases}
.\end{equation}

\noindent Combining Equations (\ref{rhotrisolgn}), (\ref{pz}), and (\ref{palpha}) we can approximate $\rho_g^{\triangle}(\tau,\bar{x}_\odot)$ by

\begin{equation}\label{rhotrisolend}
 \rho_g^{\triangle}(\tau,\bar{x}_\odot) \approx \int_{\forall \,  M \, \in  \, \P_{r}}   \frac{\Sigma^{\triangle}_{\odot} \cdot \psi_\odot(\tau) }{\mathcal{H}(\tau)}  \cdot \xi(M) \cdot M \cdot dM
.\end{equation}

As mentioned above, the fit with the observational value of the volume stellar mass density at the position of the Sun is done with the stars with $\tau \leq T_{lim}(M,Z,\alpha)$. Therefore, as a next step, we need to derive an analytical expression for the density of non-remnant primal stars at the position of the Sun. This is given by

\begin{equation*}
 \rho^{\triangle}(\tau,Z,\bar{x}_\odot,\alpha) \approx \int_{\forall \,  M \, \in  \, \P_{r}} \mathcal{O}(\tau,M,Z,\alpha)  \cdot \frac{\Sigma^{\triangle}_{\odot} \cdot \psi_\odot(\tau) }{\mathcal{H}(\tau)}  \cdot \xi(M) \cdot M \cdot
\end{equation*} 
\begin{equation}\label{rhotrisolgn3}
 \cdot \mathcal{P}(Z|\tau,\bar{x}_\odot) \cdot \mathcal{P}(\alpha | \tau,Z,\bar{x}_\odot)   dM 
,\end{equation}

\noindent where the function $\mathcal{O}(\tau,M,Z,\alpha)$ discussed in Sect. \ref{omega} is defined as

\begin{equation}\label{tlim}
\mathcal{O} (\tau,M,Z,\alpha)=
\begin{cases}
    0 ,&  \tau > T_{lim}(M,Z,\alpha) \\
   
    1,&  \tau \leq T_{lim}(M,Z,\alpha)

\end{cases}
.\end{equation}

As before, we assume all the stars at the position of the Sun have $Z=Z_\odot$ and $\alpha=\alpha_\odot$ (Equations (\ref{pz}) and (\ref{palpha}) ), thus

\begin{equation}\label{rhotrisolgn2}
 \rho^{\triangle}(\tau,\bar{x}_\odot) \approx \int_{\forall \,  M \, \in  \, \P_{r}} \mathcal{O}(\tau,M,Z_\odot,\alpha_\odot)  \cdot \frac{\Sigma^{\triangle}_{\odot} \cdot \psi_\odot(\tau) }{\mathcal{H}(\tau)}  \cdot \xi(M) \cdot M \cdot dM
.\end{equation} 

Up to now we have derived analytical expressions for the volume densities of the primal stars (\ref{rhotrisolend} and \ref{rhotrisolgn2}). To complete Equations (\ref{rhosuntotg}) and (\ref{rhosuntot}) we need to derive the analytical expressions for the mass density of secondary components. Considering Equation (\ref{rho$solg}) we can write $\rho^{\$}_g(\tau,Z,\bar{x}_\odot,\alpha)$:

\begin{equation*}
\rho^{\$}_g(\tau,Z,\bar{x}_\odot,\alpha) \approx \int_{\forall \,  M  \in  \, \P^i_{r}} \int^{m_{max}}_{m_{min}}  \frac{\Sigma^{\triangle}_{\odot} \cdot \psi_\odot(\tau) }{\mathcal{H}(\tau)}  \cdot \xi(M)  \cdot \mathcal{P}(Z|\tau,\bar{x}_\odot) \cdot 
\end{equation*}
\begin{equation}\label{rho$solg2} 
\cdot \mathcal{P}(\alpha | \tau,Z,\bar{x}_\odot) \cdot P(bin|\tau,M,Z,\alpha) \cdot P(m|\tau,M,Z,\alpha) \cdot m \, dm  \, dM 
.\end{equation}

\noindent Next, to continue reducing computational time, we assume that the mass of the secondary is only dependent on the mass of the primary component $P(m|\tau,M,Z,\alpha) \approx P(m|M)$ and we consider this probability is uniform and given by


\begin{equation}\label{pmM}
p(m| M)\approx 
\begin{cases}
    \frac{1}{M-M_{min}} ,& \text{if } M_{min} < m\leq M\\
    0,              & \text{otherwise}
\end{cases}
,\end{equation}

\noindent where $M_{min}$ is the minimum mass needed to generate a star. This expression states that the mass of the secondary star is always equal or less massive than the corresponding primary component. 

We introduce a second approximation assuming that for a given generated primal star the probability of being the primary component of a multiple system is given by

\begin{equation}\label{pbin*}
P(bin| \tau,M,Z,\alpha)\approx 
\begin{cases}
    0 ,& \tau > T_{lim} \\
    P(bin|M),              & 0 < \tau \leq T_{lim}
\end{cases}
.\end{equation}

This probability, as defined, is not null only for the non-remnant stars. Therefore, using Equation (\ref{tlim}), it can also be expressed as $p(bin| \tau,M,Z,\alpha) = \mathcal{O} (\tau,M,Z,\alpha) \cdot p(bin|M)$. In other words, this approach assumes that the probability of being a primary component of a binary system is null for those stars which are remnant at present. Introducing Equations (\ref{pmM}) and (\ref{pbin*}) in (\ref{rho$solg2}), we have

\begin{equation*}
\rho^{\$}_g(\tau,Z,\bar{x}_\odot,\alpha) \approx \int_{\forall \,  M  \in  \, \P^i_{r}} \int^{m_{max}}_{m_{min}}  \mathcal{O}(\tau,M,Z_\odot,\alpha_\odot) \cdot \frac{\Sigma^{\triangle}_{\odot} \cdot \psi_\odot (\tau) }{\mathcal {H}(\tau)}  \cdot \xi(M)  \cdot 
\end{equation*}
\begin{equation}\label{rho$solg5} 
\cdot \mathcal{P}(Z|\tau,\bar{x}_\odot) \cdot  \mathcal{P}(\alpha | \tau,Z,\bar{x}_\odot) \cdot P(bin|M) \cdot P(m|M) \cdot m \, dm  \, dM 
.\end{equation}

\noindent The above assumptions imply that secondary stars are never remnants, thus $\rho^{\$}(\tau,Z,\bar{x}_\odot,\alpha)=\rho^{\$}_g(\tau,Z,\bar{x}_\odot,\alpha)$.

As done for primal stars, we assume that all the stars at the position of the Sun  have solar metallicity $Z=Z_\odot$ and $\alpha=\alpha_\odot$. Therefore,

\begin{equation*}
\rho^{\$}(\tau,\bar{x}_\odot)=\rho^{\$}_g(\tau,\bar{x}_\odot) \approx \int^{M=M_{max}}_{M=M_{min}} \int^{m=M_{max}}_{m=M_{min}}   \mathcal{O}(\tau,M,Z_\odot,\alpha_\odot)  \cdot
\end{equation*}
\begin{equation}\label{rho$solg3} 
\cdot \frac{\Sigma^{\triangle}_{\odot} \cdot \psi_\odot (\tau) }{\mathcal {H}(\tau)}  \cdot \xi(M) \cdot P(bin|M) \cdot P(m|M) \cdot m \, dm  \, dM 
.\end{equation}

As done for primal stars, we leave out $\mathcal{H}(\tau)$ from (\ref{rho$solg3}) to reach the expression for $\Sigma^\$ _\odot$:

\begin{equation*}
\Sigma^\$ _\odot \approx \int^{\tau=\tau_{end}}_{\tau=0} \int^{M=M_{max}}_{M=M_{min}} \int^{m=M_{max}}_{m=M_{min}}   \mathcal{O}(\tau,M,Z_\odot,\alpha_\odot) \cdot \Sigma^{\triangle}_{\odot} \cdot \psi_\odot (\tau)  \cdot \xi(M) \cdot
\end{equation*}
\begin{equation}\label{Sigma$sol4} 
 \cdot P(bin|M) \cdot P(m|M) \cdot m \, dm  \, dM \, d\tau
.\end{equation}

At this point we have on hand all the analytical expressions to compute the $\Sigma_\odot$ that fits the observed stellar mass density at the position of the Sun ($\rho^\odot_{obs}$) for the thin disc stars  with $\tau \leq T_{lim}(M,Z,\alpha)$. To do so we use Equations (\ref{rho$solg3}) and (\ref{rhotrisolgn2}) in  (\ref{rhosuntot}) and set $ \rho(\tau,\bar{x}_\odot)= \rho^\odot_{obs}$, with $\rho^\odot_{obs}$  being the observed stellar mass volume density at the position of the Sun for stars which are not remnant at present. We then solve the resulting equation for $\Sigma^ \triangle_\odot$ and compute $\Sigma_\odot$ from (\ref{sigmaall}) and (\ref{Sigma$sol4}).

For the practical implementation of the analytical approach developed in this section we adopt $M_{min}=0.09 M_\odot$ and $M_{max}=120 M_\odot$, the mass range of the evolutionary models we are using at present (\citealt{Chabrier1997}, \citealt{Bertelli2008,Bertelli2009}). For $p(bin|M)$ in Equation (\ref{pbin*}) we follow the expression for main sequence stars in \cite{Arenou2011}.

\subsection{The non-remnant fraction $(\Omega)$}\label{omega}

The $\mathcal{O}$ function (45) is directly related with the stellar evolutionary tracks considered through the expressions of $T_{lim}(M,Z,\alpha)$. We define $T_{lim}(M,Z,\alpha)$  as the maximum age for which a star of a given mass, metallicity, and $\alpha$-elements-to-iron abundance is still not a stellar remnant. As discussed in the previous section,  for the $\mathcal{O}$ function we consider solar metallicity $Z=Z_\odot$ and solar $\alpha=\alpha_\odot$. We use, for the moment, the stellar evolutionary tracks \cite{Bertelli2008} and \cite{Chabrier1997} considered in \cite{Czekaj2014} which do not consider the $\alpha$-element abundance; therefore $T_{lim}(M,Z,\alpha)=T_{lim}(M,Z)$. From the mentioned stellar evolutionary tracks we derive  $T_{lim}(M,Z_\odot)$  fitting three truncated logarithmic expressions to the two-dimensional (2D) grid of mass and age limit for $Z=Z_\odot$. We end up with the following expressions

\noindent for $M/M_\odot \geq 7$.
\begin{equation}
T_{lim}(M,Z_\odot) \approx exp(-1.6 \cdot ln(M)+20.8)
\end{equation}
\noindent for $2.2<M/M_\odot<7.0$,
\begin{equation}
T_{lim}(M,Z_\odot) \approx exp(-2.7 \cdot ln(M)+23.0)
\end{equation}
\noindent for $2.0<M/M_\odot<2.2$,
\begin{equation}
T_{lim}(M,Z_\odot) \approx  exp(-2.7 \cdot ln(2.2)+23.0)
\end{equation}
\noindent for $M/M_\odot \leq 2.2$,
\begin{equation}
T_{lim}(M,Z_\odot) \approx exp(-3.5 \cdot ln(M)+23.3)
.\end{equation}

With these latter expressions we have everything we need for the execution of the analytical LDSE as described in Sect. \ref{AnalLDSE}. At this point, to take into account that the BGM Std strategy samples the age of the stars uniformly inside each age sub-population (see Sect. \ref{stdthin}) we substitute  the $\mathcal{O}$ function by the $\Omega$ function in the equations of Sect. \ref{AnalLDSE}. The $\Omega(\tau,M,Z_\odot)$ function is defined as

\begin{equation}\label{tlim2}
\Omega(\tau,M,Z_\odot) =
\begin{cases}
    1 ,&  T_{lim}(M,Z_\odot) > T^j_{end}(\tau) \\
    \frac{T_{lim}(M,Z_\odot)-T^j_{ini}(\tau)}{T^j_{end}(\tau) - T^j_{ini}(\tau)},              & T^j_{ini}(\tau) \leq T_{lim}(M,Z_\odot) \leq T^j_{end}(\tau) \\
    
    0,& T_{lim}(M,Z_\odot) < T^j_{ini}(\tau),

\end{cases}
\end{equation}

\noindent where $T^j_{ini}(\tau)$ and $T^j_{end}(\tau)$ are respectively the low and high age boundaries for each of the seven age sub-populations of the thin disc ( j=1...7). We note that $T^j_{ini}(\tau)$ and $T^j_{end}(\tau)$ obviously depend on $\tau$, because the age $\tau$ of the star establishes the j-sub-population to which the star belongs. We emphasise that the  $\Omega$ function defined above gives the fraction of stars which are not remnant at present and is constant with $\tau$ inside the limits of each age sub-population. Furthermore, when defining $\mathcal{O}$ and $\Omega$ functions we are neglecting both the entangled evolution of stars belonging to the multiple stellar systems and the mass lost by the star during its evolution. If in the future we want to implement the effect of the mass lost by the stars during its evolution, we can do this by introducing a new function, $\mathcal{L}(\tau,M,Z,\alpha),$ into the equations, accounting for the fraction of mass that a given star has lost during its evolution up to the present time. If at some point we are interested in accounting for the entangled evolution of a stellar multiple system then we need to implement the desired evolutionary model of stellar systems directly in the BGM Std strategy,  generate a Mother Simulation, and finally modify $T_{lim}$ accordingly.

\subsection{Approximate LDSE}\label{AppLDSE}

In Sects. \ref{AnalLDSE} and \ref{omega} we explained how, 
in the full process to ensure the LDSE (Sect. \ref{fullLDSE}), we can substitute the entire simulation of a sphere around the Sun by crafted analytical expressions which are computationally undemanding. Here we complete the construction of an approximate LDSE strategy by complementing Sects. \ref{AnalLDSE} and \ref{omega} with a final assumption to make the process even faster. As described in Sect. \ref{fullLDSE} in the full LDSE process, the central mass and the dark matter density distribution is adjusted such that the model rotation curve fits the observations. We can avoid the computational cost of this adjustment by assuming that the central mass density and the dark matter density distribution are invariant under variations of the Galactic fundamental functions \footnote{In practice we impose that the central mass density and the dark matter density distribution take the values of the Default Model Variant in \cite{Mor2017}.}. 

We shall test the performance of the approximate LDSE to evaluate the assumptions that we made and to constrain its range of validity. In the following section we present a set of tests comparing the results obtained when applying the full LDSE process against the results of our approximate LDSE.

\subsection{Validation of the approximate LDSE}\label{ldsetest}

We develop three tests to evaluate the goodness of our approximate method (Sects. \ref{AnalLDSE} to \ref{AppLDSE}). These tests  quantify the differences, when using the approximate instead of full method, in some of the key parameters resulting from the LDSE process. All the tests are performed using the seven model variants described in \cite{Mor2017} that were built with different assumptions of the IMF, the SFH, and the density laws. In Table \ref{variants1} we present their main parameters. These model variants cover the range of parameters that we want to explore in this paper well, and therefore they are a good set to perform the tests. The parameters for the thin disc density laws which are not listed in Table \ref{variants1} are adopted to be the same as used in \cite{Mor2017}, being the functional forms of the density laws listed in \cite{Robin2003}. The rest of the model ingredients which are not specifically indicated in Table \ref{variants1}, for example the atmosphere models, the stellar evolutionary tracks, the age-metallicity relation, and the age-velocity dispersion, are adopted to be the same as in "Model B" listed in Table 5 of \cite{Czekaj2014}.

\begin{table}
\caption{For the seven model variants adopted from \cite{Mor2017} we present: (1) The three slopes of the initial mass function ($\alpha_1$,$\alpha_2$ and $\alpha_3$) and the two mass limits ($x_1$ and $x_2$) between the three truncated power laws of a Kroupa-like type; (2) the stellar volume mass density at the position of the Sun $(\rho_\odot)$; and (3) The radial scale length of the old thin disc $(h_R)$.}             
\label{variants1}       
\centering
\resizebox{\columnwidth}{!}{
\begin{tabular}{|c c c c c c c c c |}     
\hline\hline       

 \multirow{2}{*}{ID} & SFH &\multicolumn{5}{|c|}{IMF} & \multicolumn{2}{c|}{Dens. law} \\ 
 & \multicolumn{1}{|c|}{$\gamma$ ($Gyr^{-1}$)} & \multicolumn{1}{|c}{$\alpha_1$} &  $\alpha_2$ & $\alpha_3$ & $x_1$ &$x_2$ & \multicolumn{1}{|c}{$\rho_\odot$ ($M_\odot/pc^3$)} & $h_R$ (pc)\\
\hline

DAV & 0.12 &1.3 & 1.8 & 3.2  & 0.5& 1.53 & 0.033& 2170 \\
  DM & 0.12 & 1.3 & 1.8 & 3.2  & 0.5& 1.53 & 0.033& 2530 \\
  
  DBV & 0.12 & 1.3 & 1.8 & 3.2  & 0.5& 1.53 & 0.039& 2530 \\
  DCV & 0.00 & 1.3 & 1.8 & 3.2  & 0.5& 1.53 & 0.033& 2530 \\
  
  HRV & 0.12 & 1.6 & 1.6 & 3.0  & 0.5& 1.00 & 0.033& 2530 \\
  
  HRVB & 0.12 &1.6 & 1.6 & 3.0  & 0.5& 1.00 & 0.039& 2530  \\
  
  SV & 0.12 & 2.35 & 2.35 & 2.35  & 0.5& 1.53 & 0.033& 2530 \\

\hline                  
\end{tabular}}
\end{table}

\begin{figure}
   \centering
   \includegraphics[width=\hsize]{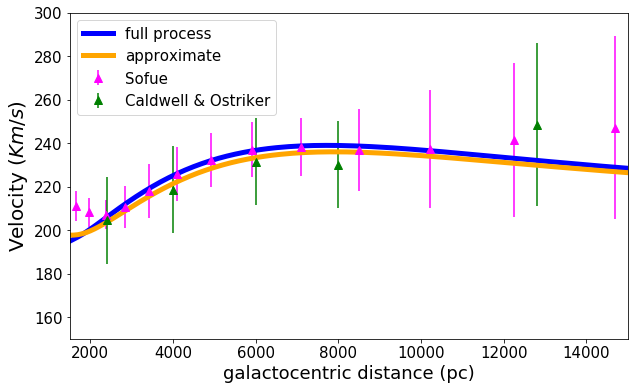}
   \includegraphics[width=\hsize]{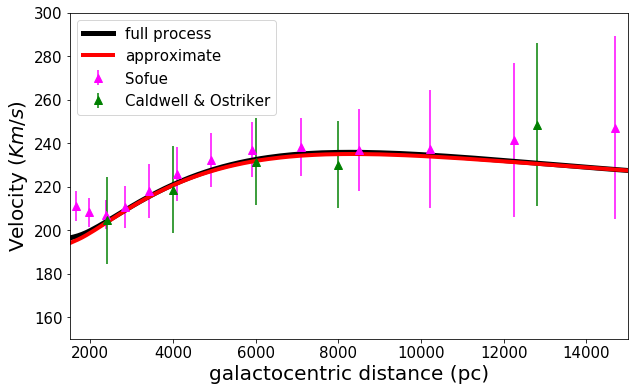}
      \caption{Two examples of the comparison between the rotation curve obtained with both the full process to ensure LDSE and the rotation curve obtained with the approximate LDSE. \textbf{Top:} Results for the DAV variant from \cite{Mor2017}. \textbf{Bottom}: Results for the HRBV variant from \cite{Mor2017}. The green triangles show the data points derived from the \cite{Caldwell1981} rotation curve assuming $R_0=8000$ pc and $V_0=230$ km/s for the Sun. The magenta triangles show the data points of the rotation curve from \cite{Sofue2015}. For Sofue data, error bars are provided by the author,  while we have estimated the errors of the data from   \cite{Caldwell1981} following the expressions and tables in their paper.}
         \label{RotCurveDAV} 
   \end{figure}

In the first test we show that our approximate method obtains a rotation curve compatible with the one resulting from the full method. For all model variants, we obtain differences of less than $2\%$ in the rotational velocity between a galactocentric radius from $3$ kpc to $14$ kpc. Moreover these differences are much smaller than discrepancies between the observational values (e.g. \citealt{Caldwell1981} and \citealt{Sofue2015}). In Fig. \ref{RotCurveDAV} we present the results for the two model variants where we found the highest discrepancies between the rotation curve obtained from both methods. The discrepancies along the curve from $3$ kpc to $14$ kpc  are always smaller than 5 km/s. Thus we consider that the approximated LDSE is valid within these galactocentric radii.

The second test compares for each model variant the eccentricities of the Einasto density profiles which are fitted inside the process that ensures the LDSE. The Einasto eccentricities are computed for each one of the seven age sub-populations of the thin disc. In all cases we find differences smaller than $0.7\%$.

Our third test compares the local volume stellar mass density of all the thin-disc age sub-populations. For each model variant we check our capability to recover the density values of Table 1 of \cite{Mor2017}. The discrepancies are within the error bars and are always smaller than $5\%$. 

We want to point out that these tests are the first empirical demonstration that our strategy for the BGM FASt simulation is well founded. Whereas it is true that they validate the equations behind the weight function, they do not evaluate the capabilities of the weight function itself to generate a BGM FASt simulation. To do so we need to directly test distributions of observable parameters obtained with BGM FASt simulations. Tests on age and mass distribution are also needed to deeper validate the BGM FASt strategy. These tests are presented in Sect. \ref{SolarN} and Appendix \ref{ApA1}.

\section{ABC to infer Galactic fundamental functions}\label{ABC}

\subsection{The ABC method}\label{ABCmethod}

For very complex models such as the BGM, to obtain the exact likelihood function is mathematically impossible or computationally prohibitive. In these cases the ABC algorithms allow us to compute an approximate posterior probability distribution function of the explored parameters. In this paper we use a sequential Monte Carlo approximate Bayesian computation algorithm (SMC-ABC) because of its balance between high acceptance rate and independence of the outcomes. The algorithm that we use is further described in \cite{Jennings2017}, and is available as a python package named astroABC. The SMC-ABC method is an improvement of basic ABC acceptance-rejection sampling (ARS-ABC) to optimise the acceptance rate. The basic ARS-ABC contains the theoretical bases of the SMC-ABC algorithm. This latter first generates a proposed set of parameters $\bar{\theta}$ from the prior PDF. Subsequently, the algorithm generates the simulated data using the proposed set of parameters in a given model, and finally the set of parameters is accepted as part of the posterior PDF if the simulated data ($\mathcal{D}_{simu}$) are equal to the real data ($\mathcal{D}$); otherwise $\bar{\theta}$ is rejected and we start the process again. This algorithm is very restrictive because it is almost impossible to find a model with a given set of parameters that  perfectly reproduces the data. Usually one uses the ARS-ABC, relaxing the condition $\mathcal{D}_{simu}=\mathcal{D}$ and approximating the method as follows.

\begin{enumerate}[1]
\item Generate $\bar{\theta}$ from the prior PDF. 
\item Simulate data $\mathcal{D}_{sim}$  from the model $\mathcal{M}$ with parameters $\bar{\theta.}$
\item Calculate the distance $\delta(\mathcal{D},\mathcal{D}_{sim})$ between $\mathcal{D}$ and $\mathcal{D}_{sim}$. 
\item Accept $\bar{\theta}$ if $\delta$ is smaller than a given threshold ($\upsilon$) $\delta \leq \upsilon $; return to 1. 
\end{enumerate}

This approximate algorithm needs to adopt an adequate distance metric $\delta$ and a threshold $\upsilon$. When $\upsilon \to 0,$ the algorithm is sampling exactly from the posterior PDF. Very small values of $\upsilon$ diminish the acceptance rate of the algorithm, and therefore the choice of $\upsilon$ must be a compromise between computational cost and accuracy of the posterior PDF. We want to emphasise that ARS-ABC samplers generate independent outcomes; each iteration is independent from the previous.

Other ABC methods, like the so-called free likelihood MCMC samplers (MCMC-ABC), are also built to optimise the acceptance rate. The MCMC-ABC increases the acceptance rate by modifying MCMC sampling algorithms (such as Metropolis-Hastings) to be able to work without the need of the likelihood \citep{Marjoram2003}, but the price paid in this case is that the outcomes are dependent \citep{Marjoram2003}. In a different way, the SMC-ABC improves the acceptance rate with a double entangled optimisation. For each iteration, the algorithm uses kernels to assign a higher sampling probability to the sets of parameters with better results. The resulting new sampling probability is used in the next iteration. In other words, regions of the parameter space with better results are visited more frequently. The application of kernels in the limits of the prior PDF can sometimes produce a posterior PDF slightly wider than the limits stated by the prior PDF. The use of the kernels is complemented with an adaptive threshold $\upsilon$, where the upper and lower limits have to be set. This allows to go from a more relaxed threshold in the first iterations down to a small threshold for the last ones, optimising the sampling of the posterior PDF in terms of computational time. We emphasise that for a given expression for the distance metric and a given $\upsilon,$ the outcomes are still independent in the SMC-ABC. 

Now we need to choose the data in accordance with the parameters that we want to infer. Observations can only provide a small subset of all the potential data defining the Milky Way system. This limitation forces us to search for summary statistics, which are of a lower dimension and are incomplete. If the chosen summary statistics, $\mathcal{S,}$ are statistically sufficient for $\mathcal{D}$ then the posterior PDF under the summary data $P(\theta|\mathcal{S})$ is equivalent to the posterior PDF under the full data $P(\theta|\mathcal{D})$. In practice it is very difficult, or impossible, to formally identify rigorous summary statistics sufficient for $\mathcal{D}$. As suggested in \cite{Marjoram2003} we use a more heuristic approach for our specific problem. In the following section we propose summary statistics $\mathcal{S}$ that capture information on the $\bar{\theta}$ essential parameters of the Milky Way. Once sufficient statistics are established, $\mathcal{D}$ is replaced by $\mathcal{S}$ in the algorithm as done in \cite{Marjoram2003}. The theoretical basis  for these algorithms can be found in \cite{Marin2011} , \cite{Beaumont2008} and \cite{Sisson2010}, for example.

\subsection{Bayesian inference in the solar neighbourhood}\label{suf}

For the appropriate use of the ABC algorithm described above, we need to define, for each of the specific scientific goals that we want to achieve, our sufficient statistics $\mathcal{S}$, a distance metric $\delta,$ and a threshold $\upsilon$. For both the evaluation of the BGM FASt performance (Sect. \ref{SolarN}) and the science demonstration cases (Sect. \ref{Results}), we consider as our sufficient statistics $\mathcal{S}$ the star counts in a binned four-dimensional space of position, apparent magnitude, and observed colour. This means that for the purposes of this paper we define our sufficient statistics as the number of stars in each bin of latitude, longitude, visual Tycho magnitude ($V_T$), and Tycho-2 colour $(B-V)_T$. Specifically, our $\mathcal{S}$ is the colour-magnitude diagram split in three latitude rangesù: $(|b|<10)$, $(10<|b|<30),$ and $(30<|b|<90)$. We choose the bin size of the colour-magnitude diagram to be large enough to allow for a robust statistical analysis to be performed, but small enough to avoid loosing information. As mentioned in the previous section, our choice of $\mathcal{S}$ is based on our previous experience when comparing BGM with observational data. We know that the observed colour-magnitude diagrams in different regions of the sky offer very valuable information to constrain the IMF, the SFH, the local stellar mass density, and the density laws (e.g. \citealt{Mor2017}, \citealt{Robin2014}, \citealt{Robin2012}). In some of these papers these colour-magnitude diagrams have already been used as sufficient statistics combined with ABC algorithms (e.g. \citealt{Robin2014}). Furthermore, in \cite{Czekaj2014} it was demonstrated that the division of the sky into the three above-mentioned  latitude ranges is very useful to analyse the Galactic fundamental functions by fitting BGM to Tycho-2 data. In our star counts analysis we always work inside the completeness limits of the observational catalogues and therefore our sufficient statistics when using Tycho-2 data are limited at $V_T=11$ where Tycho-2 is complete up to $99 \%$.

Once our sufficient statistics are defined, we need to choose the distance metric to quantify differences between the observed and simulated data, $\delta \left(\mathcal{S}_{obs},\mathcal{S}_{simu}\right)$. We use this to compare the simulated $\mathcal{S}_{simu}$ and the observed $\mathcal{S}_{obs}$ data. For this paper we use the following expression, which we call Poissonian distance:

\begin{equation}\label{Lr}
\delta_P \left(\mathcal{S}_{obs},\mathcal{S}_{simu}\right) = \left |\sum_{i=1}^N q_i \cdot \left(1-R_i+ln(R_i)\right )\right |
,\end{equation}

\noindent where $R_i$ is defined as the quotient $R_i = f_i / q_i$ and $q_i$ and $f_i$ are the number of stars in the data and the model, respectively. Additionally, we penalise the cases where there are no stars observed in the bin but the model predicts stars to be present by replacing $q_i$  and $f_i$ in the formula with
$q_i+1$ and  $f_i+1$, respectively. The defined distance metric becomes zero when the simulation and the observations have the same number of stars in each bin. The smaller the value of the distance metric, the closer $\mathcal{S}_{simu}$ is to $\mathcal{S}_{obs}$. From \cite{Kendall1973} and \cite{Bienayme1987} we know this formula is a good choice for the comparison between observed and simulated colour-magnitude diagrams in terms of star counts. Furthermore, in \cite{Mor2017} we already introduced the idea that expression (\ref{Lr}) can be understood as a distance metric. 

Finally, we choose an upper limit and a lower limit for the threshold $\upsilon$ by experimenting with different values to have a good balance between computational cost and accuracy of the posterior PDF.

\section{Evaluating BGM FASt at the solar neighbourhood}\label{SolarN}
 
In this section we evaluate the behaviour of BGM FASt in the solar neighbourhood and we demonstrate that the performance of the BGM FASt strategy is independent of the Mother Simulation. In Sect. \ref{fastvsstd} we use both BGM FASt and BGM Std to simulate the solar neighbourhood. We then analyse the resulting samples limited in apparent magnitude by comparing colour distributions and CMDs. Additionally, we extend the comparison to mass and age distributions for a deeper analysis of the BGM FASt performance. We do these comparisons in the completeness regime of Tycho-2 catalogue, and therefore we use samples limited in Tycho visual apparent magnitude ($V_T$) up to 11, where Tycho-2 is complete at $99\%$.

In Sect. \ref{imposedSFH} we present a test for the BGM FASt framework together with the ABC algorithm, consisting in exploring the parameter space of the SFH, demonstrating that we are able to correctly recover an imposed SFH in the solar neighbourhood.

\subsection{BGM FASt versus BGM Std}\label{fastvsstd}

\begin{table*}
\caption{Summary of the results of the  BGM FASt vs. BGM Std  tests and the parameters for the SFH,  the IMF, and the density laws of the model variants. $x_1=0.5$ for all the model variants. The units for this table are the same as in Table \ref{variants1}. The variants marked with an asterisk are those which are less affected by the sampling noise discussed in Appendix B. The final column indicates the assumed 3D extinction map in each case; it can be \cite{Drimmel2001} (D) or \cite{Marshall2006} (M).}             
\label{table:2}      
\centering  
\centering
\resizebox{\textwidth}{!}{
\begin{tabular}{|c  c c c c c c ||c  c c c c c  c c|c c|c|}     
\hline\hline       
  \multicolumn{7}{|c||}{\textbf{Parameters of the Mother Simulation}} & \multicolumn{8}{c|}{\textbf{Parameters of BGM FASt and BGM Std. simulations}} & \multicolumn{2}{c|}{\textbf{FASt vs. Std}} & \multicolumn{1}{c|}{Ext.} \\ 
\multicolumn{1}{|c}{} & \multicolumn{4}{c}{} & \multicolumn{2}{c||}{} &  \multirow{ 2}{*}{} & \multicolumn{1}{c}{} & \multicolumn{4}{c}{} & \multicolumn{2}{c|}{} & \multirow{ 2}{*}{} & \multirow{ 2}{*}{} &\\
  \multicolumn{1}{|c|}{SFH} & \multicolumn{4}{c|}{IMF} & \multicolumn{2}{c||}{Dens. law} &  \multirow{ 2}{*}{ID} & \multicolumn{1}{|c|}{SFH} & \multicolumn{4}{c|}{IMF} & \multicolumn{2}{c|}{Dens. law} & \multirow{ 2}{*}{$\delta_{p}$} & \multirow{ 2}{*}{$\%$} & \\ 
\multicolumn{1}{|c|}{$\gamma$} & $\alpha_1$ & $\alpha_2$ & $\alpha_3$ & $x_2$ & \multicolumn{1}{|c}{$\rho_\odot$} & $h_R$ & & \multicolumn{1}{|c|}{$\gamma$} & $\alpha_1$ & $\alpha_2$ & $\alpha_3$  & $x_2$ & \multicolumn{1}{|c}{$\rho_\odot$} & $h_R$ &  & &\\
\hline
    \multirow{ 8}{*}{0.12} & \multirow{8}{*}{1.3}  & \multirow{8}{*}{1.8} & \multirow{ 8}{*}{3.2}  & \multirow{8}{*}{1.53}  &  \multirow{8}{*}{0.033} & \multirow{ 8}{*}{2170}   & \multirow{ 2}{*}{DBV$^*$} & \multirow{ 2}{*}{0.12} & \multirow{ 2}{*}{1.3} & \multirow{ 2}{*}{1.8} & \multirow{ 2}{*}{3.2}  & \multirow{ 2}{*}{1.53} & \multirow{ 2}{*} {\textbf{0.039}}& \multirow{ 2}{*}{\textbf{2530}} & 1178 &+2.38 & D\\
  &    &  &   &   &  &   &  &     &  &  &   &  &  & & 685 &+0.57 & M \\
  &    &  &   &  &  &   & \multirow{ 2}{*}{DCV$^*$} & \multirow{ 2}{*}{\textbf{0.00}} & \multirow{ 2}{*}{1.3} & \multirow{ 2}{*}{1.8} & \multirow{ 2}{*}{3.2}  & \multirow{ 2}{*}{1.53} &  \multirow{ 2}{*}{0.033}& \multirow{ 2}{*}{\textbf{2530}} & 1623 &+3.7& D\\  
   &  &  &   & &  &  &  &  & & &   &  &  & & 932 & +2.00& M \\

   &    & \multicolumn{3}{c}{\multirow{4}{*}{\Large{(DAV)}}}   &  &   & \multirow{ 2}{*}{HRV} & \multirow{ 2}{*}{0.12} & \multirow{ 2}{*}{\textbf{1.6}} & \multirow{ 2}{*}{\textbf{1.6}} & \multirow{ 2}{*}{\textbf{3.0}}  & \multirow{ 2}{*}{\textbf{1.00}} &  \multirow{ 2}{*}{0.033}& \multirow{ 2}{*}{\textbf{2530}} &1722 &+1.49& D\\
     &    &  &  &  &  &     &  &  & \ & \ &  &  &   & & 1450&-0.55& M \\

      &    &  &  &  &  &  & \multirow{ 2}{*}{SV} & \multirow{ 2}{*}{0.12} & \multirow{ 2}{*}{\textbf{2.35}} & \multirow{ 2}{*}{\textbf{2.35}} & \multirow{ 2}{*}{\textbf{2.35}}  & \multirow{ 2}{*}{1.53} &  \multirow{ 2}{*}{0.033} & \multirow{ 2}{*}{\textbf{2530}} & 2284 &-5.22 & D\\
      &    &  &  &  &   &  &  &  &  &  &   &  &  & & 2642 &-7.01& M \\

\hline
    \multirow{ 8}{*}{0.00} & \multirow{8}{*}{1.3}  & \multirow{8}{*}{1.8} & \multirow{ 8}{*}{3.2}    & \multirow{8}{*}{1.53}  &  \multirow{8}{*}{0.033} & \multirow{8}{*}{2530}   & \multirow{ 2}{*}{DAV$^*$} & \multirow{ 2}{*}{\textbf{0.12}} & \multirow{ 2}{*}{1.3} & \multirow{ 2}{*}{1.8} & \multirow{ 2}{*}{3.2}  & \multirow{ 2}{*}{1.53} &  \multirow{ 2}{*}{0.033}& \multirow{ 2}{*}{\textbf{2170}}& 1893 &-1.13& D\\
  &    &  &  &   &  &   &  &  &  &  &   &  &  & & 1683 &+0.77& M\\
  &    &  &   &  &  &   & \multirow{ 2}{*}{DBV$^*$} & \multirow{ 2}{*}{\textbf{0.12}} & \multirow{ 2}{*}{1.3} & \multirow{ 2}{*}{1.8} & \multirow{ 2}{*}{3.2}  & \multirow{ 2}{*}{1.53} & \multirow{ 2}{*} {\textbf{0.039}}& \multirow{ 2}{*}{2530} & 1698 &+1.43& D\\
   &  &  &   &  &  &  &  &  & & &   &  &  & & 1866 &+1.60& M\\

   &    &   \multicolumn{3}{c}{\multirow{4}{*}{\Large{(DCV)}}}     &  &   & \multirow{ 2}{*}{HRV} & \multirow{ 2}{*}{\textbf{0.12}} & \multirow{ 2}{*}{\textbf{1.6}} & \multirow{ 2}{*}{\textbf{1.6}} & \multirow{ 2}{*}{\textbf{3.0}}  & \multirow{ 2}{*}{\textbf{1.00}} &  \multirow{ 2}{*}{0.033}& \multirow{ 2}{*}{2530} & 1748 &+0.47& D\\
     &    &  &   &  &  &     &  &  & \ & \ &  &  &   & & 1939 &+0.48& M\\

      &    &  & &  &  &  & \multirow{ 2}{*}{SV} & \multirow{ 2}{*}{\textbf{0.12}} & \multirow{ 2}{*}{\textbf{2.35}} & \multirow{ 2}{*}{\textbf{2.35}} & \multirow{ 2}{*}{\textbf{2.35}}  & \multirow{ 2}{*}{1.53} &  \multirow{ 2}{*}{0.033} & \multirow{ 2}{*}{2530} & 2993 &-6.26& D\\
      &    &  &  &  &   &  &  &  &  &  &   &  &  & & 3144 &-8.87& M\\

\hline            
                
\end{tabular}}
\end{table*} 
 
To evaluate the performance of the BGM FASt we compare simulations generated with the same sets of fundamental functions obtained from both the BGM FASt and BGM Std strategy. We have selected five model variants from \cite{Mor2017}: DAV, DBV, DCV, HRV and SV. All of them constitute a good framework to analyse the BGM FASt behaviour in the solar neighbourhood as they adequately cover the parameter space that we want to explore with BGM FASt in Sect. \ref{Results}. For each model variant we perform the tests using both the \cite{Drimmel2001} and \cite{Marshall2006} extinction maps. \footnote{The \cite{Marshall2006} extinction map covers the longitude ranges $-100<l<100$ and the latitude ranges $|b|<10$. Therefore, when simulating with the Marshall map we use the Drimmel map for the rest of the sky.} The main set of comparisons aims to evaluate the behaviour of BGM FASt when using as Mother Simulation the best fit variant from \cite{Mor2017} (the DAV variant) obtained using Galactic Cepheids and Tycho-2 data. The DAV variant is our choice for the Mother Simulation when exploring a six-dimensional (6D) parameter space in Sect. \ref{CaseB}. In the tests we evaluate the effects of changing  one or more parameters of the fundamental functions when generating BGM FASt simulations according to the model variants DBV, DCV, HRV and SV. 

Additionally we repeat the comparisons but use the DCV variant (constant SFH) as the Mother Simulation. These additional tests help us to evaluate the BGM FASt performance in several aspects. First, they allow us to demonstrate whether or not the performance of the BGM FASt strategy is independent of the Mother Simulation. Second, we are able to analyse possible dependencies on the total number of stars of the Mother Simulation. Third, they allow us to study simultaneous changes of the IMF and the SFH. Finally, they provide us with more data, which can be used to better interpret the obtained results.

In Table \ref{table:2} we summarise the parameters of the IMF, the SFH, and the density laws for the variants involved in the tests. The functional form for the SFH is assumed to be an exponential law, the IMF is assumed to be a three-times-truncated power law and the density profiles are assumed to be of Einasto shape. In the first column we show the parameters that we use as Mother Simulation: the DAV and DCV. In the second column we show the parameters for the BGM FASt simulations. The same parameters are also used to perform BGM Std simulations. We mark in bold text the parameters that have been modified from the Mother Simulation to generate the BGM FASt simulation. In the third column (FASt vs. Std) we present a summary of the global results comparing BGM FASt with BGM Std simulations, $\delta_p$ is for the Poissonian distance from equation (\ref{Lr}) and $\%$ refers to the discrepancies of the total number of stars between BGM FASt and BGM Std simulations, that is (($\#$ Stars in BGM FASt - $\#$ Stars in BGM Std)/$\#$ Stars in BGM Std)$*100$). We note that for all cases except the SV variant the discrepancies in the total number of stars is smaller than $4\%$. When looking to the Poissonian distance metric we note that all the values are below 2000. Except, again, for the SV variant. The Poissonian distance between BGM FASt and BGM Std simulations is one order of magnitude smaller than the difference between the best fit variant from \cite{Mor2017} and Tycho-2 data. 

When comparing colour, age, and mass distributions between BGM FASt and BGM Std, for the cases of Table \ref{table:2}, we note that in general the differences in relative counts per bin are smaller than $5\%$ (see Figs. \ref{A1}, \ref{A2}, and \ref{A3} in the Appendix A for details). In some cases, such as when using the IMF (SV variant) from  \cite{Salpeter1955}  or when changing the mass limits of the IMF (HRV variant), the difference in the youngest population of the thin disc and for the high mass range can reach about $10\%$. These differences are also reflected in the colour distributions. We demonstrate in Appendix B that these discrepancies ($>10\%$) come from the sampling noise involved in the BGM Std generation strategy. Occasionally, mostly for flat values of the IMF at high masses, the mass distribution of the youngest massive stars is influenced by the distribution of sampling noise in very low-mass reservoirs \footnote{As described in \cite{Czekaj2014} equation (1) in each age bin, the mass available to be spent on star production in a given volume element, is the mass reservoir.}. The sampling noise in very-low-mass reservoirs produces, on average, more stars than predicted by the imposed distribution function for the mass range between $1.53 M_\odot$ and $4M_\odot$. From Appendix B, we deduce that we must use a Mother Simulation such that the combination of the imposed fundamental function minimises the effects of the noise in the very small mass reservoir; this is the case of the DAV variant,  for example.  

After the analysis presented in this section and also in Sect. \ref{ldsetest} we conclude that BGM FASt is performing correctly in the solar neighbourhood. 

\subsection{Recovering an imposed SFH}\label{imposedSFH}

  \begin{figure}
   \centering
   \includegraphics[width=\hsize]{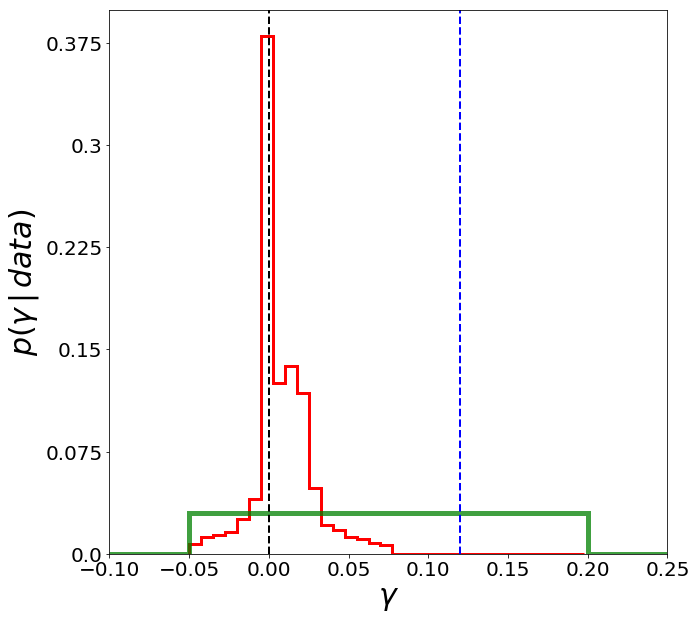}
      \caption{ Probability distribution function for the inverse of the characteristic time scale ($\gamma$ parameter) of the thin disc SFH, assuming a decreasing exponential shape. The right vertical blue dotted line indicates the  $\gamma$ value of the Mother Simulation used in the test. The left black dotted vertical line ($\gamma=0$) indicates the value to be recovered by the test. In green we show the prior probability distribution function assumed for the test (uniform distribution between $-0.05$ and $0.20$). In red we show the resulting approximate posterior probability distribution function $P(\gamma|data)$. }
         \label{sfhtestlast}
   \end{figure}

To show that BGM FASt together with the ABC algorithm is capable of inferring a given parameter, we perform several tests trying to recover an imposed SFH of the thin disc. We assume the SFH to be a decreasing exponential function:

\begin{equation}
\psi=K_\psi \cdot e^{\gamma\tau} 
,\end{equation}

\noindent where $K_\psi$ is the normalization constant, $\gamma$ is the inverse of the characteristic timescale (e.g. \citealt{Snaith2015} \footnote{We want to emphasise that we use different nomenclature from \cite{Snaith2015}}) and $\tau$ is the time. The coordinate origin of the $\tau$ space is at present ($\tau=0$) and the values are positive in the direction backwards in time, to the past, until the age of the thin disc (in this case we impose $\tau=10Gyr$). We use one BGM Std simulation, with an imposed value of $\gamma$, playing the role of the observations, and we use one BGM Std simulation, with an imposed $\gamma$, as a Mother Simulation to generate BGM FASt simulations while mapping the parameter space. 

We use Equation (\ref{Lr}) as a distance metric and the sufficient statistics $\mathcal{S}$ described in Sect. \ref{suf}. The threshold $\upsilon$ is set to be small enough to obtain statistically significant results. The simulations are covering the full sky with a limiting apparent magnitude of $V_T=11$ to mimic the completeness selection function of Tycho-2. Moreover we added the Tycho-2 photometric errors to the simulated observables (\citealt{Czekaj2014} and \citealt{Mor2017}). The first test consists in using a BGM Std simulation with $\gamma=0.12Gyr^{-1}$ as the observational data and as the Mother Simulation. For the second test, we use a BGM Std simulation with $\gamma=0.12Gyr^{-1}$ as observational data and another BGM Std simulation with $\gamma=0.00Gyr^{-1}$ as the Mother Simulation. Our third test consists in using  a BGM Std simulation with $\gamma=0.00Gyr^{-1}$ as observational data and another BGM Std simulation with $\gamma=0.12Gyr^{-1}$ as the Mother Simulation, that is, the inverse of the second test. We find that in the three tests presented here our strategy succeeded in recovering the imposed $\gamma$ value with a narrow posterior probability distribution function. The posterior probability distribution for the $\gamma$ parameter resulting from the last test is plotted in Fig. \ref{sfhtestlast}. In green we show the prior PDF while in red we plot the posterior PDF. It is interesting to see how from a given prior PDF we end up with a posterior PDF of a very different shape. This is expected as according to the Bayesian statistics theory,  the better the information coming from the data, the  smaller the dependency of the posterior PDF on the prior PDF.

\section{BGM FASt science demonstration cases}\label{Results}

To show the capabilities and the strength of BGM FASt simulations, we present two science demonstration cases using Tycho-2 data. We want to prove that our BGM FASt strategy together with the ABC algorithm obtain consistent results by analysing data from the solar neighbourhood. The solar vicinity is a good region to start with, allowing comparisons with previous results. In both cases we use Tycho-2 data with $V_T < 11$. We transform the photometry from Johnson to Tycho as done in \cite{Mor2017}. Following \cite{Czekaj2014} we adopt a spacial resolution of 0.8 arcsec, according to the Tycho-2 catalogue, to decide if the binary stellar systems are resolved or unresolved. In addition, we add photometric errors to the simulations to mimic Tycho-2 data. In both cases, for the ABC,  we use the sufficient statistics defined in Sect. \ref{suf} and the distance metric from Equation (\ref{Lr}). The lower limit of the threshold is chosen in each case to be large enough to ensure the existence of sets of parameters ($\bar{\theta}$) able to fulfil the condition $\delta_P<\upsilon$  ( that is: $\exists \hspace{2pt} \bar{\theta} \hspace{2pt}|\hspace{2pt} \delta_P(\bar{\theta})<\upsilon$) and at the same time to be small enough to achieve an accurate posterior PDF.

\subsection{Case A: The SFH in the solar neighbourhood}\label{CaseA}

The goal of case A is twofold. On one hand, we evaluate 12 sets of parameters under variations of the thin disc SFH to analyse which one best fits the observational data, obtaining a posterior PDF for the SFH under the given prior. On the other hand, the obtained results allow us to decide which Mother Simulation and parameter space are the best to design case B (see Sect. \ref{CaseB}). 

For case A we work in a 2D space. The first dimension is the inverse of the characteristic time scale ($\gamma$) for a SFH modelled with an exponential law (see Eq. 59). The second dimension is the projection of all the other parameters involved in the thin disc simulation, which we call the Variant dimension. For the $\gamma$ parameter of the SFH we choose a uniform prior within $\gamma=0$ (constant SFH) and $\gamma=1/3$. We set these limits to be the minimum and maximum value of the inverse of the characteristic timescale compatible with the data in \cite{Snaith2015} (their Fig. 5), when fitting a SFH with an exponential shape. To build the prior for the variant dimension we select six model variants (DAV, DM, DBV, HRV, HRVB and SV) whose details are shown in Table \ref{variants1} and Sect. \ref{ldsetest}. These model variants, using two different extinction maps, one from \cite{Drimmel2001} and  the other from \cite{Marshall2006}, constitute 12 sets of parameters belonging to our second dimensional space. We then assign a prior probability of $\frac{1}{12}$ to each set. This is a very restrictive prior and limits our exploration to 12 slices of the full parameter space.

\begin{figure}
   \centering
   \includegraphics[width=\hsize]{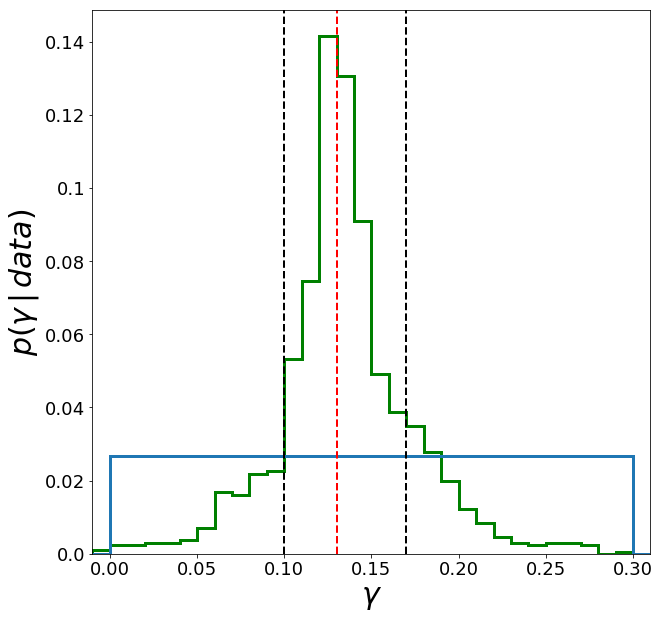}
      \caption{ $P(\gamma | data)$ (in green) Approximate posterior probability distribution function for the $\gamma$ parameter resulting from case A, given the adopted prior (see text). The vertical red dotted line indicates the mode of the distribution $\gamma = 0.13$ with an uncertainty in the range indicated by the two black vertical lines corresponding to the quantiles 0.16 and 0.84, respectively ($\gamma = 0.13^{+0.04}_{-0.03} Gyr^{-1}$). We obtain these results when using as observational data the stars in Tycho-2 catalogue with visual apparent magnitude $V_T < 11$. In blue we show the adopted prior probability distribution function.}
        \label{gammaposterior}
   \end{figure}

\begin{figure}
   \centering
   \includegraphics[width=\hsize]{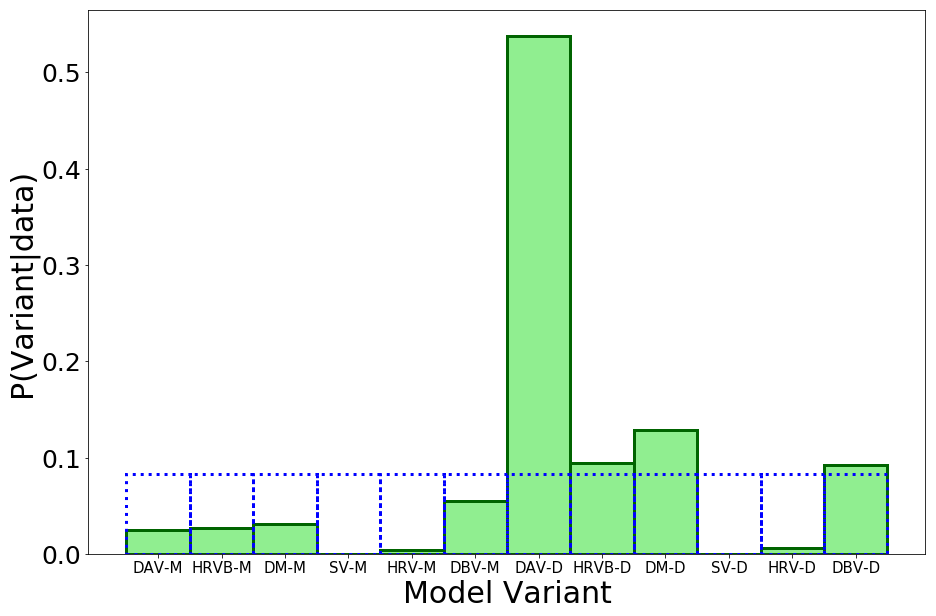}
      \caption{$P(Model | data)$: Approximate posterior probability distribution function for the variants  (see Table \ref{table:2}) given the adopted prior (see text). We obtain these results when using as observational data the stars in Tycho-2 catalogue with visual apparent magnitude $V_T < 11$. Variants whose names end with "-D" use the extinction maps from \cite{Drimmel2001}  while names ending with "-M" use those from \cite{Marshall2006}. With a dotted blue line we show the adopted prior probability distribution function.}
         \label{variantposterior}
   \end{figure}

In Fig. \ref{gammaposterior} we present the projection to the $\gamma$ space of the approximate posterior probability distribution function. The obtained value for the $\gamma$ parameter is $\gamma = 0.13 ^{+0.04}_{-0.03} Gyr^{-1}$. In Fig. \ref{variantposterior} we present the projection to the Variant space of the approximate PDF. We show here that the DAV variant with Drimmel extinction map is the most probable result. We also show that variants using the Drimmel extinction map are carrying more than $80\%$ of the probability. Moreover, the variants using Salpeter IMF has almost null probability.

\begin{figure}
   \centering
   \includegraphics[width=\hsize]{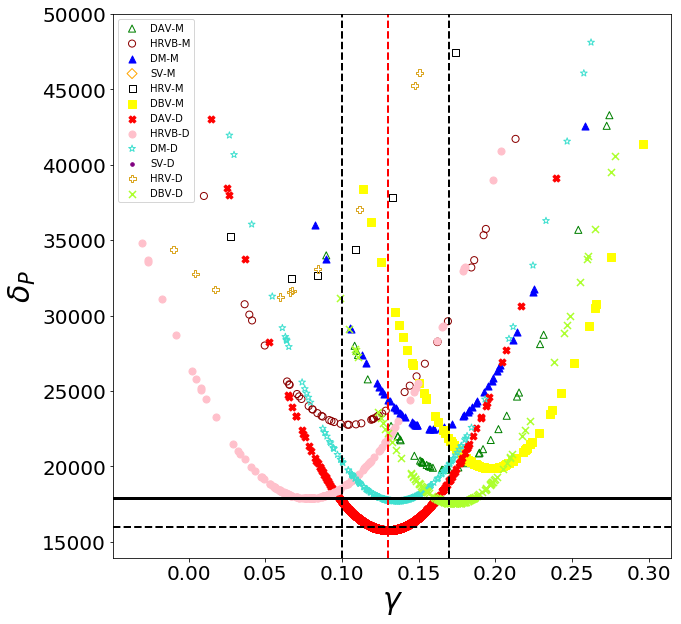}
    \caption{This plot is built from the results of Case A shown in Figs. \ref{posterior6D} and \ref{posterior6D4}. Each point in the plot represents a set of parameters accepted by the ABC algorithm as part of the posterior probability distribution function. For each accepted set we plot the Poissonian distance ($\delta_P$) as a function of the inverse of the characteristic timescale ($\gamma$ parameter) of an assumed decreasing exponential SFH. $\delta_P$ is computed using the expression of Equation (\ref{Lr}); the smaller the value of $\delta_P$ , the better its agreement with the data. Dotted horizontal line gives the $\delta_P$ value around the best fit ($\delta_P \approx 16000$) and the solid horizontal line is to emphasise the degeneracy for three different sets of parameters of the model variants (see Table \ref{table:2}). The vertical dashed black lines denote the quantile 0.16 (left) and quantile 0.84 (right) for the distribution of Fig. \ref{gammaposterior}.  The vertical red dashed line is for the mode of the $\gamma$ distribution in Fig. \ref{gammaposterior}. We obtain these results when using as observational data the stars in the Tycho-2 catalogue with visual apparent magnitude $V_T < 11$. Variant/extinction map combinations are as described in Fig.4.}
         \label{lrdistance}
   \end{figure}

In Fig. \ref{lrdistance} we present, for the accepted sets of parameters in the ABC algorithm, the Poissonian distance ($\delta_P$) as a function of the $\gamma$ parameter for the 12 model variants in the Variant space. The variant giving the best fit (smaller $\delta_p$) is the DAV variant using the Drimmel extinction. Furthermore, we want to emphasise that this variant gives a better fit in a large range of $\gamma$ values, that is, from $\gamma \approx 0.10$ to $\gamma \approx 0.16$. As discussed for Fig. \ref{variantposterior}, the four variants closest to the data (pink, cyan, green and red) use the Drimmel extinction map. We show using the horizontal solid line how three variants (HRVB, DM and DBV) with different combinations of parameters could result in the same Poissonian distance ($\delta_P$) to Tycho-2 data. We note that within $1\sigma$ (black dotted vertical lines show the 0.16 and 0.84 quantiles) in Fig. \ref{lrdistance} , the DAV variant can have a $\delta_P$ that is compatible with the $\delta_P$ of HRVB, DM and DBV variants.

By analysing the obtained approximate posterior PDF we can state that when modelling the SFH as a simple exponential law (allowing constant SFH when $\gamma=0$) our results point to a decreasing SFH in the solar neighbourhood with $\gamma = 0.13 ^{+0.04}_{-0.03} Gyr^{-1}$. However, we must keep in mind that in our parameter space the dimension of the SFH is continuously covered while the second dimension is discrete. Therefore, non-considered combinations of the full parameter space could lead to a better fit with Tycho-2 data (see case B).  

As mentioned above, the results obtained from case A guide us for the design of case B. We want to choose a Mother Simulation for case B as close to the observational data as possible. This choice minimises the effects of the approximations taken in BGM FASt and also the effects due to noise explained in Appendix C. From Fig. \ref{variantposterior} it is clear that the Mother Simulation for case B should be the DAV variant. 
   
Finally, to design the parameter space of case B we must try to include both the most important parameters as well as the ones that produce degeneracies in the results. From Figs. \ref{variantposterior} and \ref{lrdistance} it is clear that the IMF and the SFH produce important variations on the results. In Fig. \ref{lrdistance} we note that the stellar volume mass density at the position of the Sun is causing degeneracy among three models differing by the IMF and the SFH. Therefore, the parameter space of case B should contemplate the IMF, the SFH, and the density laws.

\subsection{Case B: Simultaneous inference of the SFH, the IMF, and the density laws}\label{CaseB}

In this second science demonstration case we explore a 6D parameter space of the IMF, the SFH, and the density laws for the thin disc component. As in case A, the SFH is assumed to be an exponential law. Therefore, one of the dimensions is the inverse of the characteristic timescale $\gamma$. The IMF is assumed to be a Kroupa-like IMF; this is a three truncated power law with three slopes $\alpha_1$, $\alpha_2,$ and $\alpha_3$ and two mass limits $x_1$ and $x_2$. For simplicity in this case we fix $x_1=0.5 M_\odot$ and $x_2=1.53 M_\odot$ which are the values used in the best-fit models from \cite{Czekaj2014}. Finally, for the thin disc density laws we use as main parameters the scale length ($h_R$) and the thin disc stellar volume mass density at the position of the Sun ($\rho_\odot$). As the youngest sub-population is too young to be considered isothermal, the dynamical constraints are not applied on the population with age less than $0.10 Gyr$ and its $h_R$ is fixed to be the one assumed in \cite{Robin2003}, that is $h_R=5000pc$. We explore the scale-length value for the rest of the age sub-populations assuming that it is independent of the age.

The Mother Simulation that we use in this case B is the DAV variant (see Table \ref{table:2}). We use the 3D extinction map from \cite{Drimmel2001} and additionally an alternative 3D extinction map called Stilism. This new map was constructed using the method of \cite{Capitanio2017} but setting the \cite{Marshall2006} map as a prior at a distance of 1.5 kpc. The full process is explained in \cite{Rosine2018} (their Sect. 5). This map is still at the testing stage. However, we find it interesting to include this preliminary version in our study because it is the most updated 3D extinction map built specially for the BGM. 

To sum up, we explore a  6D space that includes $\gamma$, $\alpha_1$, $\alpha_2$, $\alpha_3$, $h_R$ and $\rho_\odot$. For all of these parameters we consider a uniform prior PDF within the values of Table \ref{priors}. In the Bayesian strategy for the parameter inference the assumption of the prior is of marginal importance if the data are good enough to constrain the explored parameters. Our choice for the low and high limits aims to cover the range of values reported in the literature for the explored parameters.

\begin{table}
\caption{Lower and upper limits of the imposed initial uniform Prior probability distribution function for each of the explored parameters. Units are as in Table \ref{variants1}.}
\label{priors}      
\centering                          
\begin{tabular}{c c c}        
\hline\hline                 
parameter & lower limit & upper limit \\    
\hline                        
   $\gamma$ & 0.00 & $0.30$  \\      
   $\alpha_1$ & 0.5 & 2.0    \\
   $\alpha_2$ & 0.5 & 3.0      \\
   $\alpha_3$ & 0.5 & 4.0     \\
   $h_R$ &2000 & 2600    \\ 
      $\rho_\odot$ & 0.030 & 0.040    \\ 
\hline                                   
\end{tabular}
\end{table}

\begin{figure*}
   \centering
   \includegraphics[width=\hsize]{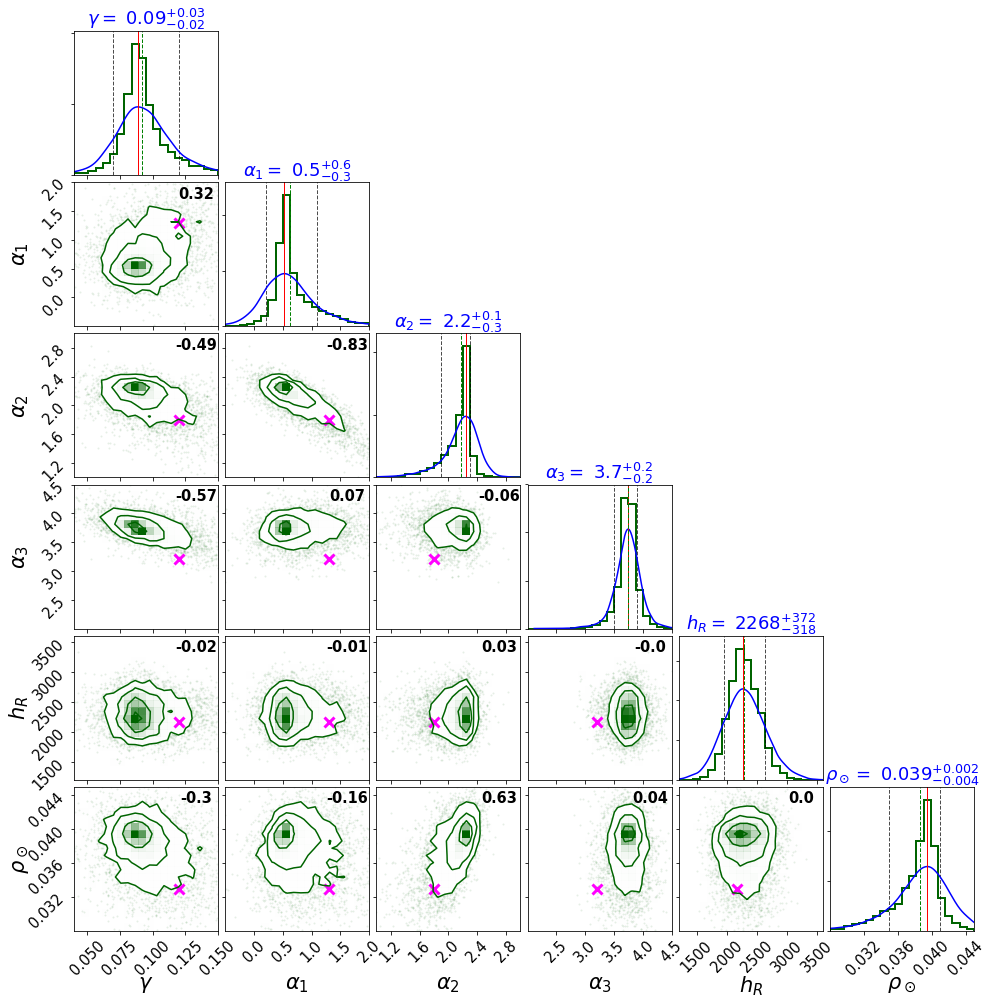}
    \caption{Corner plot with the projection of the approximate posterior probability distribution function obtained from the exploration of the six parameters of the thin disc component, including the IMF, the SFH, and the density laws. The vertical solid red lines and the vertical green dashed lines indicate the mode and the median for each parameter, respectively. For $\alpha_3,$ the median and the mode are superimposed. Green 2D contours and 1D distributions are for the distributions obtained directly from the ABC algorithm. The blue 1D posterior probability distribution functions are built by accounting, in the posterior PDF, for the differences between BGM Std and BGM full simulations (see text).  The vertical black dashed lines indicate the quantiles 0.16 and 0.84 of this distribution. On top of each 1D histogram the mode of the distribution and the interval of the 0.16 and 0.84 quantiles for the blue distributions are indicated. In black at the top right of each one of the 2D panels we show the Pearson's correlation coefficient. The magenta cross indicates the parameters of the adopted Mother Simulation. $\gamma$ is the inverse of the characteristic timescale when assuming an exponentially decreasing SFH. $\alpha_1$, $\alpha_2$ and $\alpha_3$ are the slopes of the IMF assumed to be a three-times truncated power law with the mass limits fixed at $x_1=0.5 M_\odot$ and $x_2=1.53 M_\odot$. $\rho_\odot$ is the thin disc stellar volume mass density at the position of the Sun and $h_R$ is the radial scale length of the thin-disc density profile assumed to be an Einasto shape (see units in Table \ref{variants1}). We obtain these results assuming the \cite{Drimmel2001} extinction map and using as observational data the stars in Tycho-2 catalogue with visual apparent magnitude $V_T < 11$.}
         \label{posterior6D}
   \end{figure*}

\begin{figure*}
   \centering
   \includegraphics[width=\hsize]{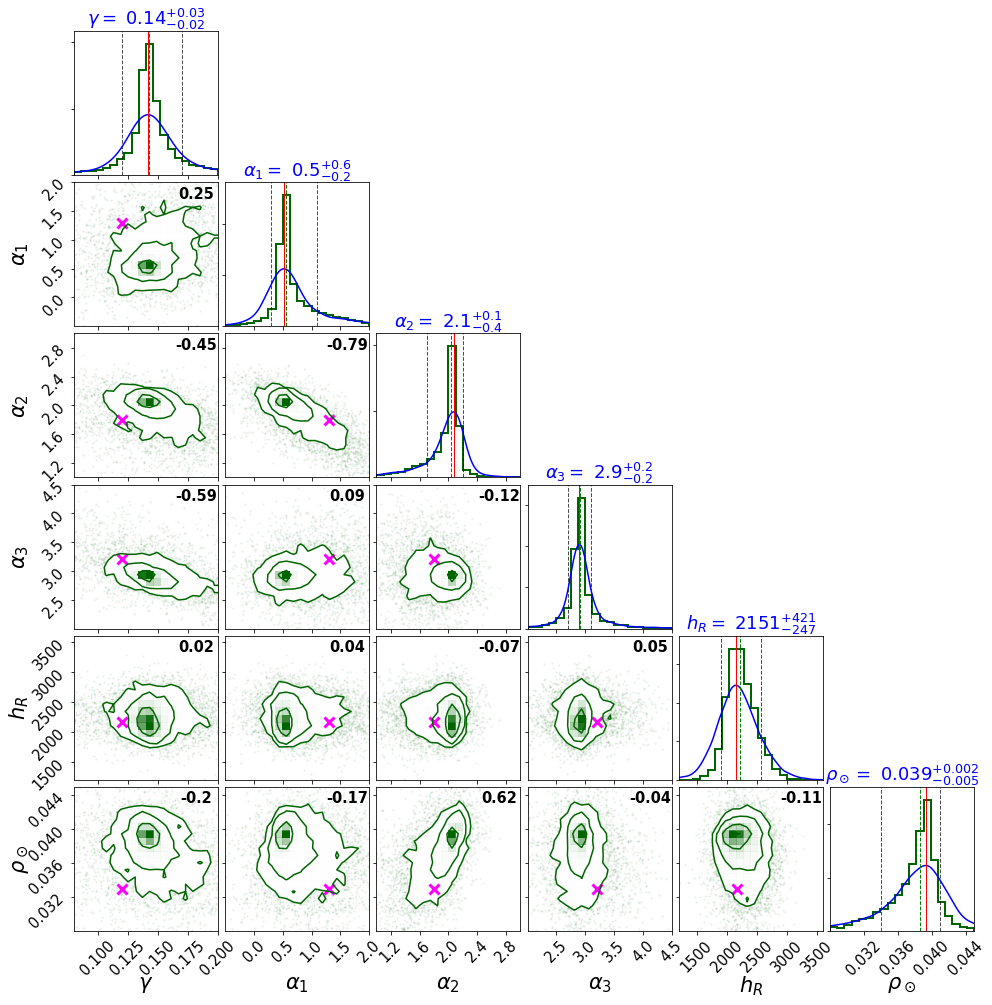}
    \caption{As in Fig. \ref{posterior6D} but using the Stilism extinction map.}
         \label{posterior6D4}
   \end{figure*}

In Figs. \ref{posterior6D} and \ref{posterior6D4} we show the corner plot resulting from the 6D exploration using BGM FASt and the ABC algorithm, using the Drimmel and Stilism 3D extinction maps, respectively. The green histograms are the 1D marginal posterior PDFs obtained directly from the ABC algorithm. The blue solid lines are the posterior PDFs of the six explored parameters accounting explicitly for the differences between BGM FASt and BGM Std simulations. To build these PDFs we first assume that the computed Poissonian distance between simulations and observations has a Gaussian error, with $\sigma$ equal to the mean Poisson distance \footnote{The mean is computed from Table \ref{table:2} considering only the 8 tests (ID name with *) whose results are marginally affected by the noise at low mass reservoir described in appendix B.} ($\delta_P$) between BGM FASt and BGM Std simulations ($\sigma=1445$). Then, for each set of parameters accepted by the SMC-ABC algorithm, we transform their single distance $\delta_p$ to a Gaussian distribution centred on the pertinent $\delta_p$ and with $\sigma=1445$. After this transformation we recompute the 1D distributions and finally, normalizing accordingly, we get the blue PDFs. 

The results point towards a decreasing SFH with  $\gamma={0.09}_{-0.02}^{+0.03} Gyr^{-1}$ (using Drimmel) and $\gamma={0.14}_{-0.02}^{+0.03} Gyr^{-1}$ (using Stilism). We obtain almost the same value of the slope of the IMF at the low mass range of $\alpha_1={0.5}_{-0.3}^{+0.6}$ when using the Drimmel map and $\alpha_1={0.5}_{-0.2}^{+0.6}$ when using the Stilism map. The same occurs for the IMF slope in the mass range between $0.5 M_\odot$ and $1.53 M_\odot$ where we obtain values of $\alpha_2={2.2}_{-0.3}^{+0.1}$ and $\alpha_2={2.1}_{-0.4}^{+0.1}$ when using Drimmel and Stilism maps, respectively. For masses greater than $1.53 M_\odot$, we find a very steep IMF with a slope value of $\alpha_3={3.7}_{-0.2}^{+0.2}$  for the Drimmel map and a more common slope for Stilism ($\alpha_3={2.9}_{-0.2}^{+0.2}$). The value obtained for the thin disc stellar mass volume density at the position of the Sun is $\rho_\odot={0.039}_{-0.004}^{+0.002}M_\odot/pc^3$ and $\rho_\odot={0.039}_{-0.005}^{+0.002}M_\odot/pc^3$ for Drimmel and Stilism, respectively. The radial scale length for the thin disc component has large uncertainty, being $h_R={2268}_{-318}^{+372} pc$ and  $h_R={2151}_{-421}^{+247} pc$ for Drimmel and Stilism, respectively.  

In both corner plots (Figs. \ref{posterior6D} and \ref{posterior6D4}) we can see in the 2D projection of the PDF the correlations (degeneracies) between the explored parameters. The $\gamma$ parameter of the SFH is clearly  correlated with the $\alpha_2$ and $\alpha_3$ slopes of the IMF. Steeper values of the SFH correspond to flatter slopes of the IMF. This effect is largely discussed in the literature (e.g. \citealt{Haywood1997} and \citealt{Aumer&Binney2009}). Steeper values of the SFH lead to less young stars and hence less massive stars alive at present. Flatter IMFs are therefore needed to compensate for this effect and reproduce the observations. On the contrary, flatter values of the SFH produce more young stars and hence more massive stars alive at present. Steeper IMFs are therefore needed to compensate this effect and reproduce the observations. We also note the correlation between the density $\rho_\odot$ and the second slope of the IMF ($\alpha_2$). A flatter slope corresponds to smaller density and a steeper slope to higher density. One possible explanation for this is based on two facts. First, about $60\%$ of our sample belongs to the mass range of $\alpha_2$. Second, in general for fixed values of $\alpha_1$ and $\alpha_3$, the smaller the value of $\alpha_2$, the higher the amount of mass dedicated to generate stars in its mass range. As a consequence, flatter slopes need smaller values of $\rho_\odot$ to fit the data. The last correlation that can be seen in the figure is between $\alpha_1$ and $\alpha_2$ because of the continuity at the IMF mass limit.

 \begin{figure*}
   \centering
   \includegraphics[width=\hsize]{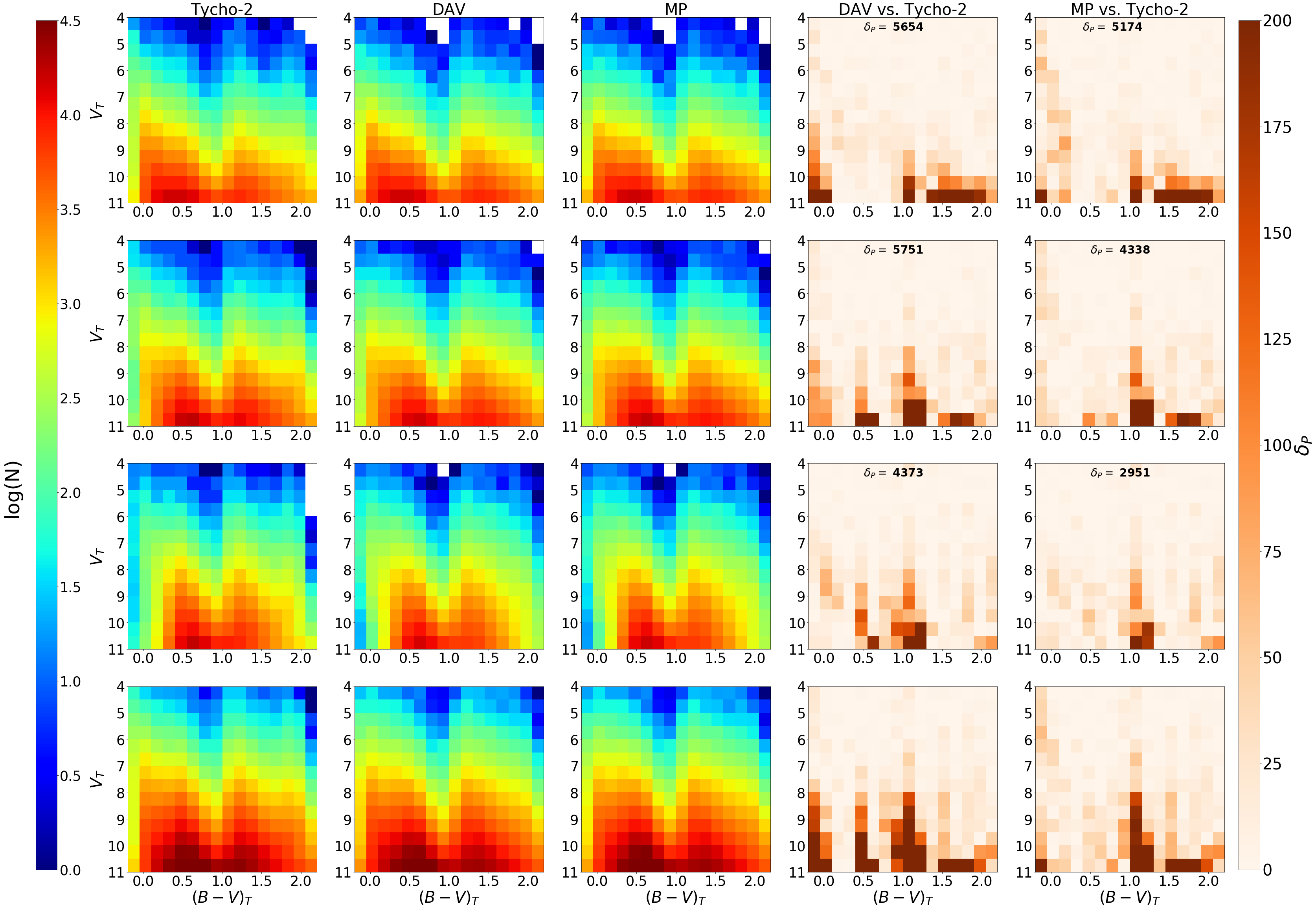}
    \caption{Colour-magnitude diagram ( apparent visual magnitude Tycho ($V_T$) vs. Tycho $(B-V)_T$ colour) divided into three latitude ranges: \textbf{first row:} $|b|<10$, \textbf{second row:} $10<|b|<30$ \textbf{third row} $30<|b|<90$ and for the whole sky (\textbf{bottom row}). The colour-map of the first, second, and third columns shows the logarithm of the star counts in each bin. \textbf{First column} is Tycho-2 data, \textbf{second} is the best-fit model variant from \cite{Mor2017} (DAV) obtained using Galactic classical Cepheid and Tycho-2 data, and the \textbf{third} is the MP variant combination of the most probable value for six parameters explored in Sect. \ref{CaseB}. \textbf{The BGM simulations performed for this figure use the Drimmel extinction map}. The colour map of the fourth and fifth rows represents the Poissonian distance computed for each bin. The total distance indicated in each plot is the Poissonian distance ($\delta_P$) computed using Equation \ref{Lr}. The smaller the value of $\delta_P$, the better the agreement. \textbf{The fourth column} contains Poissonian distance between DAV and Tycho-2, and the  \textbf{fifth column} is the Poissonian distance between MP and Tycho-2. Observational data and simulations are samples limited in visual apparent magnitude considering the stars with $V_T < 11$. }
         \label{Bestfitlat3}
   \end{figure*}

   \begin{figure}
   \centering
   \includegraphics[width=\hsize]{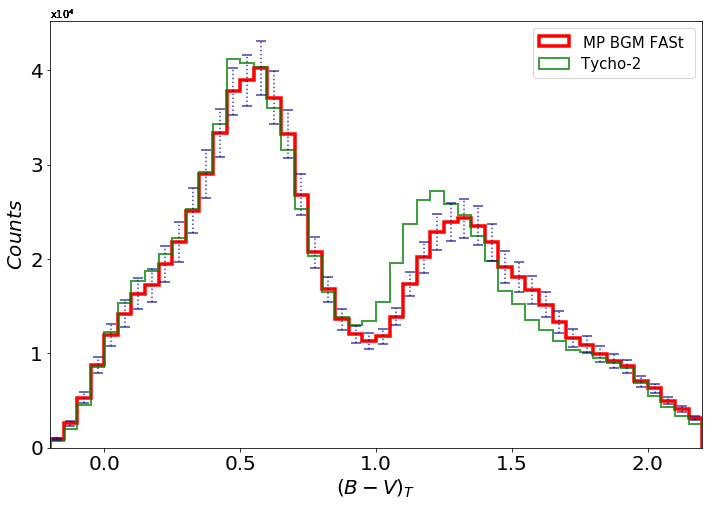}
      \caption{Colour distribution of MP variant and Tycho-2 with a limit in magnitude of $V_T=11$. Our Bayesian approach allows us to compute the error bars (dotted blue) from the simulations computed using the parameters inside the $1\sigma$ level of the posterior probability distribution functions shown in Fig. \ref{posterior6D}.}
         \label{bestfit1}
   \end{figure}

Using the most probable values shown in Figs. \ref{posterior6D} and \ref{posterior6D4}, we built two new model variants, the Most Probable (MP) variant for Drimmel and Stilism extinction maps, respectively. In Fig. \ref{Bestfitlat3}, we present the apparent visual magnitude Tycho ($V_T$) versus Tycho ${(B-V)}_T$ colour divided in three latitude ranges, $|b|<10$, $10<|b|<30,$ and $30<|b|<90$ and for the whole sky. The colour-map shows the Poissonian distance ($\delta_{Lr}$) between model and Tycho-2 data computed for each bin. This visualisation allows us to identify differences between model and data in the CMDs. The fifth column is for the MP model variant using the Drimmel extinction map versus Tycho-2. In the fourth column, for comparison, we show the DAV model variant versus Tycho-2, in order to show the improvement that the MP variant represents with respect to the best fit model obtained in \cite{Mor2017}. The improvements are significant in the three latitude ranges, with a clear impact on the bluest half ($(B-V)_T < 1.0 $) of the diagrams  while the reddest half ($(B-V)_T > 1.0) $ behaves almost equally for the DAV and MP variants. 

Previous works fitting the BGM thin disc component with observational data result in an excess of blue stars in the simulations at all latitudes (\citealt{Czekaj2014} and \citealt{Mor2017}). As can be seen in Fig. \ref{Bestfitlat3}, the MP variant is improving the fit on the blue side, where young stars dominate, in all latitude ranges, even in the high latitudes where the impact of the assumption of a 3D extinction map is expected to be small.

In Fig. \ref{bestfit1} we present the colour distribution of Tycho-2 data (green)  and the MP variant (red) using the Drimmel extinction map. Our Bayesian approach allows for the first time to introduce error bars in this comparison that account for the $1\sigma$ level. The agreement of MP with Tycho-2 is very good in the blue peak while, as already reported in \cite{Czekaj2014} and \cite{Mor2017}, the red peak is shifted by about $0.05$ magnitudes to the right. 

\begin{table*}
\caption{Local densities $M_\odot / pc^3$. Contribution to the total dynamical mass of the population components from BGM Std performed with the obtained most probable values (MP variant). Columns with (D) present the values obtained when using the Drimmel extinction map while columns with (S) present the values obtained when using the Stilism map. Additionally, we present the eccentricities of the Einasto density profiles for the age sub-populations of the thin disc component. The components marked with an asterisk are fixed and not derived in this paper.}  
     
\centering          
\begin{tabular}{r c c c c c  }     
\hline\hline       
Component & Age (Gyr)& $\rho_\odot$ (D) & $\epsilon (D)$ & $\rho_\odot$ (S)&    $\epsilon$ (S) \\ 
\hline  
\vspace{5pt}
   thin disc 1  & 0-0.10 & ${0.0015}^{+0.0001}_{-0.0002}$ & 0.0140& ${0.0013}^{+0.0001}_{-0.0002}$ & 0.0140  \\
\vspace{5pt}
   2  & 0.10 -1 &${0.0069}^{+0.0003}_{-0.0007}$ & 0.0210 &${0.0060}^{+0.0003}_{-0.0007}$ & 0.0209\\
\vspace{5pt}
   3  & 1-2 & ${0.0055}^{+0.0003}_{-0.0006}$  &  0.0299& ${0.0048}^{+0.0003}_{-0.0006}$  &  0.0298  \\
\vspace{5pt}
   4 & 2-3 & ${0.0037}^{+0.0002}_{-0.0004}$ & 0.0451& ${0.0033}^{+0.0002}_{-0.0004}$ & 0.0448 \\
\vspace{5pt}   
   5  & 3-5 &${0.0061}^{+0.0003}_{-0.0006}$ & 0.0577 &${0.0059}^{+0.0003}_{-0.0006}$ & 0.0574\\
   \vspace{5pt} 
   6  &5-7& ${0.0060}^{+0.0003}_{-0.0006}$ & 0.0655& ${0.0063}^{+0.0003}_{-0.0006}$ & 0.0652\\
   \vspace{5pt} 
   7  & 7-10  &${0.0103}^{+0.0005}_{-0.001}$ & 0.0660 &${0.0123}^{+0.0005}_{-0.001}$ & 0.0657 \\
   \vspace{5pt}  
  brown and white dwarfs* &  & 0.0071 &  & 0.0071 &\\
   \vspace{5pt} 
   total thin disc  & 0-10 &${0.047}^{+0.002}_{-0.004}$ &  & ${0.047}^{+0.002}_{-0.005}$ \\
   \vspace{5pt} 
   young thick disc*  &  10 &0.0036 &  &  0.0036 &\\
   \vspace{5pt} 
   old thick disc*  & 12 &0.0005  &&0.0005  & \\
   \vspace{5pt} 
   stellar halo*          & 14 &  4.1e-05&&  4.1e-05& \\
   \vspace{5pt} 
   total stellar component & & ${0.051}^{+0.002}_{-0.004}$& &${0.051}^{+0.002}_{-0.005}$ \\
   \vspace{5pt} 
   ISM* & & 0.05&& 0.05&\\
   \vspace{5pt} 
   dark matter halo & & $0.012^{+0.001}_{-0.001}$& &$0.012^{+0.001}_{-0.001}$&\\
\hline
\vspace{5pt}
 Total & & $0.113^{+0.002}_{-0.004}$& & $0.113^{+0.002}_{-0.005}$&\\
\hline                  
\end{tabular}
\label{densvalues}
\end{table*}

Finally, we ran a BGM Std simulation of the MP variants, using both Drimmel and Stilism extinction maps, and then we have compared this with our BGM FASt simulation confirming that the differences between them are within the $5\%$ reported in Sects. \ref{ldse} and \ref{SolarN} (see detailed tests in Appendix \ref{ApA2}). To perform these two BGM Std simulations we previously ran the full process to ensure the local dynamical statistical equilibrium (see Sect. \ref{ldse}). With the obtained results we have the local densities of the Milky Way components for the MP variant and the eccentricities of the Einasto density profiles ($\epsilon$) for the age sub-populations of the thin disc. The results are shown in Table \ref{densvalues}, the components tagged with an asterisk are imposed while the others are derived in our strategy. We obtain a total volume mass density in the solar neighbourhood of $0.113^{+0.002}_{-0.005} M_\odot/pc^3$ and a local dark matter density of $0.012 \pm 0.001 M_\odot/pc^3$. For the total stellar volume mass density of the thin disc we obtain $0.047^{+0.002}_{-0.005}M_\odot/pc^3$ (see the Table \ref{densvalues} for details on the density of each thin disc age range). Finally, for the total stellar volume mass density in the solar neighbourhood we obtain $0.051^{+0.002}_{-0.005}M_\odot/pc^3$.

\subsection{The local star-formation scene}\label{localSF}

In this section we discuss the results obtained in Sect. \ref{CaseB}. The whole IMF that we have obtained for the thin disc component, when using the Stilism 3D extinction map, is, within $1\sigma$ level, compatible for all mass ranges with the Galactic-field IMF given in \cite{Kroupa2013} (their Eq. 59). For the very-low-mass range, our results ($\alpha_1={0.5}_{-0.3}^{+0.6}$) are compatible with a flat function and have large uncertainties. We have to take into account that the number of stars with $M/M_\odot < 0.5$ in our simulations is smaller than $0.02\%$ of the total sample; this is about $100$ stars with very low weight in the comparison of synthetic versus observed CMDs. The integral over the whole mass range from $0.09M_\odot$ to $0.5M_\odot$ is in this case more important than the slope itself. The value obtained for the second slope of the IMF, in the mass range from $0.5M_\odot$ to $1.53 M_\odot$, is $\alpha_2={2.2}_{-0.3}^{+0.1}$ and it is very close to the value reported in \cite{Kroupa2008} ($\alpha_2= 2.2 \pm 0.5$) and not far from the Salpeter IMF ($\alpha=2.35$). We stress that about  $60\%$ of our simulated sample belongs to stars in this second mass range. For the high-mass range we found an IMF slope not compatible with Salpeter IMF  $(\alpha=2.35)$ at  $\sim 3\sigma$ level. We obtain $\alpha_3={3.7}_{-0.2}^{+0.2}$ (with Drimmel 3D extinction map) and $\alpha_3={2.9}_{-0.2}^{+0.2}$ (with Stilism 3D extinction map). We want to emphasise that our slope of the IMF at high mass range is only valid up to $4M_\odot$ as we estimate that in our samples just $1\%$ of the stars have masses larger that $4M_\odot$. As largely discussed in \cite{Mor2017}, the IMF considered in BGM is a composite IMF (or Integrated Galactic IMF; IGIMF). Moreover, the Tycho-2 data are a mixture of stars in the field and clusters. The abundant low-mass clusters do not have massive stars while the rare massive clusters do. This leads to a steepening of the composite IMF ($\alpha_{field} > \alpha_{cluster}$), which is a sum of all the IMFs in all the clusters that  spawn the Galactic population (\citealt{Kroupa2003} and \citealt{Kroupa2013}). Additionally, the deduced shape of the IMF, when studying field stars, could be influenced by the dynamical ejection of OB stars from dynamically unstable cores of young clusters \citep{Elmegreen2006}. This may lead to IMFs that are steeper than those of Salpeter (1955)\ because, as dynamical work suggests, the less massive members of a core of massive clusters are preferentially ejected (\citealt{Clarke1992}; \cite{Pflamm2006}). In \cite{Kroupa2003}, the authors point toward $\alpha_{field} \gtrsim 2.8$, compatible with our results.  \cite{Scalo1986} obtain $\alpha=2.7$ for stars with $M>1M_\odot$ and in \cite{Kroupa2013} a slope of $\alpha=2.7 \pm 0.4 $ is reported for masses $M \gtrsim 1 M_\odot,$ both within $1\sigma$ of our result obtained with Stilism 3D extinction map. Also compatible with our results is the work of \cite{Rybizki2015}, that, from the population synthesis side, using Galaxia \citep{Sharma2011}, obtained a slope of $\alpha=3.02 \pm 0.06$ at the high mass range. Some studies found steeper values, such as for example \cite{MS79} who reported a galactic field IMF of $\alpha \approx 3.3,$ and later \cite{Massey1998} who found very steep values of the IMF of OB stars in the field for the SMC and LMC, with values around $\alpha \approx 4.5$ (while they found values similar to those of Salpeter (1955) for stars in OB associations). We note here that the value of the slope depends on the specific range of masses upon which it is based. Finally, we note that, as reported in \cite{Kroupa2003}, the mass function of the clusters themselves over the mass range of $10 M_\odot$ to $10^7 M_\odot$ can be represented with a power-law with slope $\beta$. Analysing the results presented in \cite{Kroupa2003} (see their Fig. 2) our results favour values of the slope of the clusters' mass function of $\beta \geq 2$.

The posterior PDF that we obtain for the inverse of the characteristic timescale of the thin disc SFH assuming a simple exponential law is not compatible with flat values. When we adopt the extinction map from \cite{Drimmel2001} we obtain $\gamma={0.09}_{-0.02}^{+0.03} Gyr^{-1}$. This is compatible with the value obtained by \cite{Aumer&Binney2009} (see their table 8: $\gamma \approx 0.09$) when considering the SFH of the thin and thick discs separately. The value obtained when using the Stilism extinction map is $\gamma={0.14}_{-0.02}^{+0.03} Gyr^-1$, that is, very  close to the results obtained in our case A ($\gamma={0.13}_{-0.03}^{+0.04} Gyr^-1$), and falls in the high range of the values discussed in \cite{Aumer&Binney2009}. Although the $\gamma$ values that we obtain depend on the assumption of the 3D extinction map, the value that we estimate for the present rate of star formation in the disc (averaged along the last 100 Myr) is very similar for both, being $1.2 \pm 0.2 M_\odot/yr$. This is possible because the total surface mass density at the position of the Sun is found to be higher for the case of $\gamma=0.14$. Our result is compatible with the values found by \cite{Robitaille2010} ($0.68-1.45 M_\odot/yr$) derived from \textit{Spitzer} young stellar objects. Additionally, \cite{Licquia2015} obtain a value for the current Milky Way rate of star formation of $1.65 \pm 0.19$ which accounts for both the disc and the bulge. Using chemical data, \cite{Snaith2015} found a higher value of the present rate of star formation, that is, approximately $2 M_\odot/yr$.

We focus now on our findings for the density laws. For the radial scale length ($h_R$) of the thin disc component we obtain a wide posterior PDF showing large uncertainties. Tycho-2 data up to a magnitude of $V_T=11$ cover heliocentric radii up to $1.5-2$ kpc. Therefore the constraints that can be obtained on the radial scale length counting the stars over all the longitude ranges, as we do, are weak. Nonetheless the most probable value obtained when using the Stilism extinction map ($h_R=2151pc$) is very close to the value obtained in \cite{Robin2012} ($h_R=2170pc$). For the total stellar volume mass density at the position of the Sun ($\rho_\odot$) we obtain a value of $\rho_\odot={0.051}_{-0.005}^{+0.002} M_\odot/pc^3 $ (see Table \ref{densvalues}). Excluding white and brown dwarfs, we obtain a value of $0.044_{-0.005}^{+0.002} M_\odot/pc^3$,  compatible within $1\sigma$ with the results in \cite{Boby2017} where he found a value for the giants + main sequence stars in the solar neighbourhood of $0.040 \pm {0.002} M_\odot/pc^3$ , and within $2\sigma$ with \cite{Mckee2015} who found a value of $0.036\pm {0.005} M_\odot/pc^3$. The results from \cite{Reid2002} ($0.034-0.037 M_\odot/pc^3$) and \cite{Flynn2006} ($0.0334 M_\odot/pc^3$) point towards lower values. We obtain a local dark matter volume density of $0.012_{-0.001}^{+0.001} M_\odot/pc^3$ when fitting the Galactic rotation curve in the process to ensure the LDSE. This value is compatible with \cite{CU10}, \cite{Salucci2010}, \cite{McMillan2011} and \cite{BT12}. Our value for the local total mass density ($0.113^{+0.002}_{-0.004} M_\odot/pc^3$) is compatible with the values of \cite{Kor2003} ($0.1-0.11 M_\odot/pc^3$), \cite{HF2000} ($0.102 \pm  0.1 M_\odot/pc^3$), and \cite{Leeuwen2007} ($0.112 \pm 0.019 M_\odot/pc^3$); all three were obtained with Hipparcos data.

We discuss below four points that could impact our parameter inference: (1) the fixed ingredients of BGM, (2) the Tycho-2 photometric error modelling, (3) the assumptions on the other Galactic components in the BGM (e.g. thick disc, halo, and bulge-bar) and (4) the choice of extinction map.

The assumptions for the fixed ingredients of the BGM, for example, atmosphere models, stellar evolutionary tracks, chemical distribution, or the mass limits of the IMF $x_1$ and $x_2$,   could produce degeneracies between the fixed ingredients and the explored ingredients or biases introduced by the fixed ingredients. One example of these biases could be the shift of the red peak between simulations and Tycho-2 data observed in Fig. \ref{bestfit1} that was already reported in \cite{Czekaj2014} and \cite{Mor2017}. This shift could be caused by several of the fixed BGM ingredients, for example the atmosphere models, the stellar evolutionary tracks, the photometric transformation between Johnson and Tycho bands, or the treatment of unresolved stellar multiple systems. In practice, a shift in the red peak implies that there is no combination of the six explored parameters that is able to exactly reproduce the data, and therefore the obtained posterior PDFs are affected by this impossibility. We suspect that this is the reason why the improvements on the fit comparing the MP variant with the DAV variant are mostly concentrated in the blue half of the CMD (see Fig. \ref{Bestfitlat3}). 

The Tycho-2 photometric error modelling could be an additional source of uncertainty for the derivation of the SFH and IMF parameters. Assumptions on the other galactic components considered in the BGM could also influence the results. The amount of stars from the inner region of the Galaxy and the halo in the solar neighbourhood is known to be negligible. We estimate that less than $0.5\%$ of the stars in our sample belong to the halo. The density of the bulge/bar stars is much smaller, even null. Therefore, our assumptions on the structure of the halo and the bulge/bar marginally affect our results. The case of the thick disc component is more complex. We assume the structure of the thick disc from \cite{Robin2014} with two star-formation episodes at 10 Gyr and 12 Gyr. With these assumptions we estimate that the amount of thick-disc stars in our sample is about $10\%$. This is small but not negligible. Most of these stars have ages of the order of 10 Gyr, and thus have a similar age to  the older stars from the thin disc. The consideration of other age distributions for the thick disc would imply that the thin and thick discs overlap in different age ranges. Thus the results on the thin disc, as shown here, could be impacted by our (fixed) hypotheses on the thick-disc SFH. This question will be considered in the near future. Other structures, such as the spiral arms, are not considered in the model, and therefore irregularities in the spatial density distribution are smoothed in the BGM. The simulated CMDs that fit the Tycho-2 CMDs are thus resulting from a smoothed simulated solar neighbourhood, while the Tycho-2 CMDs can result from inhomogeneous structures, such as clusters, associations, and resonant structures. This is why we emphasise that the IMF that we are deriving is the composite IMF (or integrated galactic IMF) as we are not simulating the inhomogeneities of the star formation in the Galactic disc.

As we can see throughout this paper the choice of the extinction map is a critical ingredient for our parameter inference. An overestimation or underestimation of the extinction generally leads to an underestimation or overestimation of the star counts that could bias our results. An example of the effects of the choice of the extinction map is the different IMFs (at high masses) and SFHs that we are obtaining for Drimmel and Stilism extinction maps. On average, the absorption of the Stilism map is higher than that of the Drimmel map, producing slightly smaller star counts, especially in the blue. As a consequence, to fit the observations when using the Drimmel map we need a steeper slope of the IMF, at high masses, because in the Drimmel map the bluest stars are less absorbed than in the Stilism map. The differences in the absorption of both maps is also responsible for the slightly different SFHs. For the moment we do not have enough evidence to decide which map is more precise. In a future paper, external tests of the Stilism map will be provided to answer this question.

\section{BGM FASt  in the context of Galaxy modelling}\label{Discussion}

We have presented the BGM Fast Approximate  Simulations (BGM FASt) which is a strategy to generate Milky Way simulations at low computational cost. The computational time of a BGM FASt simulation represents only $0.02\%$ of the equivalent simulation using the BGM Std strategy. We dedicate the following paragraphs to contextualising BGM FASt in the environment of the Galaxy models. We consider, for this contextualisation, the Galaxy models that could be fast enough to be used together with iterative processes like machine learning, ABC, or MCMC to study the structure and evolution of the Milky Way. 

The Galaxy model from \cite{Pasetto2016} is also a population synthesis model to deal with large amounts of data from surveys of the Milky Way. It allows for very quick computation of CMDs from the distribution function in a given line of sight avoiding star-by-star sampling. This is a very valuable property to work with iterative inference processes such as ABC or genetic algorithms. However, when making comparisons with observational data other than that from CMDs (e.g. parallax or proper motions),  star-by-star sampling (e.g. figure 15 in \citealt{Pasetto2016}) is required. Our strategy, on the other hand, starts with a Mother Simulation which is already sampled and therefore BGM FASt never needs to re-sample star-by-star. This allows for quick generation not only of CMDs but also distributions of other observables such as parallax or proper motions. Moreover, the CMDs built with BGM FASt have a rigorous treatment of the stellar multiple systems, where we consider if a stellar multiple system is merged or resolved according to the angular resolution of the observational catalogue to be compared with the simulations. This effect is not considered in the quick CMDs of \cite{Pasetto2016}. These authors paid special attention to the kinematic constraints and dynamical consistency of the model. As in BGM FASt, the authors consider stationary state but they emphasise the non-isothermality of the disc and the formulation for the mixed terms of the Jeans equation obtained with consistency from the potential; their method ends up with a dynamical consistency stronger than the one adopted in BGM FASt where isothermal state is considered and radial and vertical motions are assumed to be decoupled. However, as described in Sect. \ref{ldse}, the process that ensures LDSE in BGM FASt runs at a very cheap computational cost. Without doubt the stronger point of the Galaxy model in \cite{Pasetto2016}, in relation to BGM FASt, is the introduction of the spiral arms as a perturbation of the stellar distribution function. Although they have a perturbed disc, they still compute its dynamical constraints with the same Poisson solver as in \cite{Robin2003}, which makes the assumption of an axi-symmetric potential to solve the Poisson equation. In the future we may consider a similar technique (to that used in \cite{Robin2012}) for BGM FASt, such as the possibility of introducing spiral arms in BGM FASt using an axi-symmetric Mother Simulation. 

TRILEGAL has been used as a prior for a Bayesian study but its computational cost is probably too expensive for parameter exploration using iterative algorithms. Initially it provided only photometry of any field of the Galaxy but no kinematic parameters such as proper motions or radial velocities. Nowadays, TRILEGAL incorporates a kinematic module. Dynamical consistency is not analysed in \cite{Girardi2005}. Compared with TRILEGAL we consider BGM FASt to be more adapted to exploiting extremely large surveys.  

Galfast \citep{Juric2008} takes advantage of the use of graphics processing units (GPUs) instead of central processing units (CPU), and due to the computational cost reported in \cite{Juric2010}, it could be fast enough to be used, in some cases, together with Bayesian iterative techniques. However, as pointed out in \cite{Loebman2014} this galaxy model does not consider inputs such as SFH, age-metallicity relation, or the IMF. It is simply a sophisticated Monte Carlo generator designed to produce a snapshot of the current sky with the stellar content consistent with SDSS observations and where initially the luminosity function was taken from \cite{KTG93}. Therefore, nowadays Galfast is not ready to infer the IMF and SFH as BGM FASt is. 

Galaxia code from \cite{Sharma2011} is, for the moment, fast enough to work with modern Bayesian techniques as seen in \cite{Rybizki2015}. Galaxia can sample stars from an N-body simulation by splitting each N-body particle into several stellar populations of different ages and masses, imposing a given IMF and SFH.  On the one hand, the possibility to sample stars from an N-body simulation is a clear advantage from the point of view of dynamical consistency, as the N-body evolves naturally from the interaction of the particles. On the other hand, if one samples an N-body model with Equation (11) from \cite{Sharma2011},  for a given N-body particle, the kinematic behaviours of the young and old stars are going to be exactly the same, thus not following the Jeans equation. Another approach using Galaxia was taken by \cite{Rybizki2015}, where they  apply the SFH to build the N-body simulation, each particle
being representative of an age range. They subsequently use Galaxia to simply convert each particle to a single stellar population for a given IMF. They include the Jeans equation implicitly when calculating the local SFH for their volume complete sample (their Fig. 1). Recalculating the N-body input for different SFHs is computationally inexpensive but the use of Galaxia to explore the IMF is relatively slow. Here, BGM FASt is faster than Galaxia (Rybizki, J. private comm.). Alternatively, if one wants to study the SFH and the IMF simultaneously with Galaxia, one can use it in the analytical model mode based on the model by  \cite{Robin2003}. In Galaxia, the code can be edited to update the density distribution to the latest versions of the BGM (e.g. \citealt{Robin2012}, \citealt{Robin2014}, \citealt{Czekaj2014}). In BGM FASt, it is also straightforward to work with the most updated version of the BGM Std simply by changing the Mother Simulation. For the moment, Galaxia is not incorporating a rigorous treatment for the resolution of the stellar multiple systems according to the angular resolution of the pertinent observational catalogue as we do in BGM FASt. Therefore, CMDs coming from Galaxia do not consider this effect. Galaxia in its analytical mode uses the same dynamical constraints as \cite{Bienayme1987}, as we do in BGM FASt. The sampling of the distribution function used in Galaxia is known to be very efficient. It  incorporates, among others, a clever strategy that, given an imposed limiting apparent magnitude for the simulation, they computes the lowest stellar mass that can generate a visible star, and then they exclusively generate stars with masses above this limit. Their sampling strategy is one of the clues that makes Galaxia a tool fast enough to use it in an iterative processes.

In summary, several Galaxy models and codes have been built to explore the structure and evolution of the Milky Way. Their advantages and disadvantages depend on the specific scientific goals. They are certainly all extremely valuable tools for the study of the Galaxy but BGM FASt is especially suited to exploring more complex models with more free parameters, as is needed, for example, when working with Gaia data.

\section{Conclusions}\label{Conclusions}

We have developed, tested, and applied a new framework to perform BGM Fast Approximate Simulations (BGM FASt). We have shown BGM FASt to be a powerful tool to study the Milky Way using modern computational Bayesian techniques. This strategy allows one to explore large parameter spaces using huge observational surveys (e.g. Gaia DR2). We have demonstrated the robustness of BGM FASt through several validation tests. In the context of dynamical consistency, we have shown that the results obtained using the approximate method to ensure LDSE, implemented in BGM FASt, are totally compatible with the results obtained with the full LDSE, implemented in BGM Std. We have rigorously compared colour-magnitude diagrams and distributions of mass, age, and colour obtained from both BGM FASt and BGM Std, demonstrating the superior performance of BGM FASt. Thanks to the structure of BGM FASt, any improvement on the Galaxy model used as the Mother Simulation is naturally incorporated. 

As examples of application of BGM FASt, we have presented two science demonstration cases using solar neighbourhood data. Of special scientific interest is case B, where we have explored a 6D space of the thin disc component, including the SFH, the IMF, and the density laws, using Tycho-2 and assuming two 3D extinction maps, one from \cite{Drimmel2001} and another from Stilism \citep{Rosine2018}. We have seen that the values of the slopes of the IMF at the first and second mass ranges ($\alpha_1$ and $\alpha_2$), the values of the thin disc stellar mass density at the position of the Sun ($\rho_\odot$), and the values of the thin disc radial scale length ($h_R$) are almost independent of our choice of 3D extinction map. The  results obtained for the composite IMF (or IGIMF) for the low mass range ($\alpha_1={0.5}_{-0.3}^{+0.6}$) are compatible with a flat function and have large uncertainties. In the mass range between $0.5 M_\odot$ and $1.53M_\odot$ we obtain $\alpha_2={2.1}_{-0.3}^{+0.1}$. The value for the total stellar mass density in the solar neighbourhood is $0.051_{-0.005}^{+0.002} M_\odot/pc^3$ and the local dark matter density was found to be $0.012 \pm 0.001 M_\odot/pc^3$. We obtain a total mass volume density in the solar neighbourhood of $0.113_{-0.005}^{+0.002} M_\odot/pc^3$ (see Table \ref{densvalues}). The value of the radial scale length for the thin disc is obtained with very large uncertainties (e.g.  $2151_{-247}^{+421}pc$ using Stilism). Our results show that the determination of the slope of the IMF at the high mass range ($\alpha_3$) and the inverse of the characteristic time scale of the SFH ($\gamma$) is degenerated with the choice of the 3D extinction map. We have obtained a very steep slope of $\alpha_3={3.7}_{-0.2}^{+0.2}$ using the Drimmel extinction map and a more common slope $\alpha_3={2.9}_{-0.2}^{+0.2}$ using the Stilism extinction map. However, both results discard a Salpeter (1955) slope at the $\sim 3 \sigma$ level. Our result of the $\alpha_3$ at high mass ranges favour values of $\beta$ greater than 2.0 for the mass function of the clusters (see Fig. 2 in \citealt{Kroupa2003}). Finally, for the SFH we have obtained $\gamma={0.09}_{-0.2}^{+0.3}$ and $\gamma={0.14}_{-0.2}^{+0.3}$ when using the maps of Drimmel and Stilism, respectively. Despite the differences in the shape of the two SFHs that we found, the present rate of star formation that we derived from them is almost identical, being $1.2 \pm 0.2 M_\odot/yr$. With the obtained results we have notably improved the fit of the BGM with the solar neighbourhood stellar content. We conclude that the BGM with these new parameters will provide a better tool to simulate new Milky Way surveys. 

BGM FASt framework will also allow us, in the future, to constrain the kinematics, the age-metallicity, and chemo-dynamics, among others. It is also possible the use of more complex expressions for the IMF, as for example including dependences of the IMF with the metallicity, as done in \cite{Jerabkova2018}. Time variation of the thin disc structure can also be  introduced as in \cite{Amores2017}. Moreover,  BGM FASt can also be used to constrain the SFH (and the age of the thin disc) considering a more flexible, non-parametric distribution. The structure of the BGM FASt code is built in order to work efficiently with the ABC algorithms but it can be used with other techniques, such as for example machine learning tools. Although BGM FASt is designed to work with a BGM simulation, as a Mother Simulation it can be used with simulations from other Galaxy models that can be described with Equation (\ref{split}) even assuming different functional forms of the fundamental functions. In this case, Equation (\ref{wii}) should be used to compute the weights. The next step in this work will be the use of the Gaia data together with a non-parametric SFH in BGM FASt. 
 
To conclude, we want to stress that BGM FASt also constitutes a large technical step in population synthesis galaxy modelling, as for the first time a simulation of the Milky Way is performed using the Apache Hadoop and Apache Spark environments \citep{ApacheSpark}. The appropriate big data platform and the efficient ABC algorithm that we use together with the BGM FASt allow us to address fundamental questions of the Milky Way structure and evolution using extremely large surveys.

\begin{acknowledgements}
Substantial thanks go to J. Rybizki, the referee, for the constructive report which helped to improve the quality of the work. This work was supported by the MINECO (Spanish Ministry of Economy) - FEDER through grant ESP2014-55996-C2-1-R and MDM-2014-0369 of ICCUB (Unidad de Excelencia 'Mar\'ia de Maeztu'),  the French Agence Nationale de la Recherche under contract  ANR-2010-BLAN-0508-01OTP and the  European Community's Seventh Framework Programme (FP7/2007-2013) under grant agreement GENIUS FP7 - 606740. We also acknowledge Dr. X. Luri and the team of engineers (GaiaUB-ICCUB) in charge of setting up and maintaining the big data platform (GDAF) at University of Barcelona. We also acknowledge E. Jennings for some clarifications about the python package astroABC. 
\end{acknowledgements}

\bibliographystyle{aa}
\bibliography{ref2}

\begin{thebibliography}{70}
\expandafter\ifx\csname natexlab\endcsname\relax\def\natexlab#1{#1}\fi

\bibitem[{{Am{\^o}res} {et~al.}(2017){Am{\^o}res}, {Robin}, \&
  {Reyl{\'e}}}]{Amores2017}
{Am{\^o}res}, E.~B., {Robin}, A.~C., \& {Reyl{\'e}}, C. 2017, \aap, 602, A67

\bibitem[{{Arenou}(2011)}]{Arenou2011}
{Arenou}, F. 2011, in American Institute of Physics Conference Series, Vol.
  1346, American Institute of Physics Conference Series, ed. J.~A. {Docobo},
  V.~S. {Tamazian}, \& Y.~Y. {Balega}, 107--121

\bibitem[{{Arenou} {et~al.}(2017){Arenou}, {Luri}, {Babusiaux}, {Fabricius},
  {Helmi}, {Robin}, {Vallenari}, {Blanco-Cuaresma}, {Cantat-Gaudin},
  {Findeisen}, {Reyl{\'e}}, {Ruiz-Dern}, {Sordo}, {Turon}, {Walton}, {Shih},
  {Antiche}, {Barache}, {Barros}, {Breddels}, {Carrasco}, {Costigan},
  {Diakit{\'e}}, {Eyer}, {Figueras}, {Galluccio}, {Heu}, {Jordi},
  {Krone-Martins}, {Lallement}, {Lambert}, {Leclerc}, {Marrese}, {Moitinho},
  {Mor}, {Romero-G{\'o}mez}, {Sartoretti}, {Soria}, {Soubiran}, {Souchay},
  {Veljanoski}, {Ziaeepour}, {Giuffrida}, {Pancino}, \&
  {Bragaglia}}]{Arenou2017}
{Arenou}, F., {Luri}, X., {Babusiaux}, C., {et~al.} 2017, \aap, 599, A50

\bibitem[{{Aumer} \& {Binney}(2009)}]{Aumer&Binney2009}
{Aumer}, M. \& {Binney}, J.~J. 2009, \mnras, 397, 1286

\bibitem[{{Awiphan} {et~al.}(2016){Awiphan}, {Kerins}, \&
  {Robin}}]{Awiphan2016}
{Awiphan}, S., {Kerins}, E., \& {Robin}, A.~C. 2016, \mnras, 456, 1666

\bibitem[{{Beaumont} {et~al.}(2008){Beaumont}, {Cornuet}, {Marin}, \&
  {Robert}}]{Beaumont2008}
{Beaumont}, M.~A., {Cornuet}, J.-M., {Marin}, J.-M., \& {Robert}, C.~P. 2008,
  ArXiv e-prints [\eprint[arXiv]{0805.2256}]

\bibitem[{{Bertelli} {et~al.}(2008){Bertelli}, {Girardi}, {Marigo}, \&
  {Nasi}}]{Bertelli2008}
{Bertelli}, G., {Girardi}, L., {Marigo}, P., \& {Nasi}, E. 2008, \aap, 484, 815

\bibitem[{{Bertelli} {et~al.}(2009){Bertelli}, {Nasi}, {Girardi}, \&
  {Marigo}}]{Bertelli2009}
{Bertelli}, G., {Nasi}, E., {Girardi}, L., \& {Marigo}, P. 2009, \aap, 508, 355

\bibitem[{{Bienaym{\'e}} {et~al.}(1987){Bienaym{\'e}}, {Robin}, \&
  {Creze}}]{Bienayme1987}
{Bienaym{\'e}}, O., {Robin}, A.~C., \& {Creze}, M. 1987, \aap, 180, 94

\bibitem[{{Bienaym{\'e}} {et~al.}(2015){Bienaym{\'e}}, {Robin}, \&
  {Famaey}}]{Bienayme2015}
{Bienaym{\'e}}, O., {Robin}, A.~C., \& {Famaey}, B. 2015, \aap, 581, A123

\bibitem[{{Bonnell} {et~al.}(2007){Bonnell}, {Larson}, \&
  {Zinnecker}}]{Bonnell2007}
{Bonnell}, I.~A., {Larson}, R.~B., \& {Zinnecker}, H. 2007, Protostars and
  Planets V, 149

\bibitem[{{Bovy}(2017)}]{Boby2017}
{Bovy}, J. 2017, \mnras, 470, 1360

\bibitem[{{Bovy} \& {Tremaine}(2012)}]{BT12}
{Bovy}, J. \& {Tremaine}, S. 2012, \apj, 756, 89

\bibitem[{{Caldwell} \& {Ostriker}(1981)}]{Caldwell1981}
{Caldwell}, J.~A.~R. \& {Ostriker}, J.~P. 1981, \apj, 251, 61

\bibitem[{{Capitanio} {et~al.}(2017){Capitanio}, {Lallement}, {Vergely},
  {Elyajouri}, \& {Monreal-Ibero}}]{Capitanio2017}
{Capitanio}, L., {Lallement}, R., {Vergely}, J.~L., {Elyajouri}, M., \&
  {Monreal-Ibero}, A. 2017, \aap, 606, A65

\bibitem[{{Catena} \& {Ullio}(2010)}]{CU10}
{Catena}, R. \& {Ullio}, P. 2010, \jcap, 8, 004

\bibitem[{{Chabrier} \& {Baraffe}(1997)}]{Chabrier1997}
{Chabrier}, G. \& {Baraffe}, I. 1997, \aap, 327, 1039

\bibitem[{{Clarke} \& {Pringle}(1992)}]{Clarke1992}
{Clarke}, C.~J. \& {Pringle}, J.~E. 1992, \mnras, 255, 423

\bibitem[{{Czekaj} {et~al.}(2014){Czekaj}, {Robin}, {Figueras}, {Luri}, \&
  {Haywood}}]{Czekaj2014}
{Czekaj}, M.~A., {Robin}, A.~C., {Figueras}, F., {Luri}, X., \& {Haywood}, M.
  2014, \aap, 564, A102

\bibitem[{{Drimmel} \& {Spergel}(2001)}]{Drimmel2001}
{Drimmel}, R. \& {Spergel}, D.~N. 2001, \apj, 556, 181

\bibitem[{{Elmegreen} \& {Scalo}(2006)}]{Elmegreen2006}
{Elmegreen}, B.~G. \& {Scalo}, J. 2006, \apj, 636, 149

\bibitem[{{Flynn} {et~al.}(2006){Flynn}, {Holmberg}, {Portinari}, {Fuchs}, \&
  {Jahrei{\ss}}}]{Flynn2006}
{Flynn}, C., {Holmberg}, J., {Portinari}, L., {Fuchs}, B., \& {Jahrei{\ss}}, H.
  2006, \mnras, 372, 1149

\bibitem[{{Gaia Collaboration} {et~al.}(2018){Gaia Collaboration}, {Brown},
  {Vallenari}, {Prusti}, {de Bruijne}, {Babusiaux}, {Bailer-Jones}, {Biermann},
  {Evans}, {Eyer}, \& et~al.}]{GaiaDR2}
{Gaia Collaboration}, {Brown}, A.~G.~A., {Vallenari}, A., {et~al.} 2018, \aap,
  616, A1

\bibitem[{{Girardi} {et~al.}(2005){Girardi}, {Groenewegen}, {Hatziminaoglou},
  \& {da Costa}}]{Girardi2005}
{Girardi}, L., {Groenewegen}, M.~A.~T., {Hatziminaoglou}, E., \& {da Costa}, L.
  2005, \aap, 436, 895

\bibitem[{{Haywood} {et~al.}(1997){Haywood}, {Robin}, \& {Creze}}]{Haywood1997}
{Haywood}, M., {Robin}, A.~C., \& {Creze}, M. 1997, \aap, 320, 428

\bibitem[{{Holmberg} \& {Flynn}(2000)}]{HF2000}
{Holmberg}, J. \& {Flynn}, C. 2000, \mnras, 313, 209

\bibitem[{{Jennings} \& {Madigan}(2017)}]{Jennings2017}
{Jennings}, E. \& {Madigan}, M. 2017, Astronomy and Computing, 19, 16

\bibitem[{{Jerabkova} {et~al.}(2018){Jerabkova}, {Zonoozi}, {Kroupa},
  {Beccari}, {Yan}, {Vazdekis}, \& {Zhang}}]{Jerabkova2018}
{Jerabkova}, T., {Zonoozi}, A.~H., {Kroupa}, P., {et~al.} 2018, ArXiv e-prints
  [\eprint[arXiv]{1809.04603}]

\bibitem[{{Juric} {et~al.}(2010){Juric}, {Cosic}, {Vinkovic}, \&
  {Ivezic}}]{Juric2010}
{Juric}, M., {Cosic}, K., {Vinkovic}, D., \& {Ivezic}, Z. 2010, in Bulletin of
  the American Astronomical Society, Vol.~42, American Astronomical Society
  Meeting Abstracts \#215, 222

\bibitem[{{Juri{\'c}} {et~al.}(2008){Juri{\'c}}, {Ivezi{\'c}}, {Brooks},
  {Lupton}, {Schlegel}, {Finkbeiner}, {Padmanabhan}, {Bond}, {Sesar},
  {Rockosi}, {Knapp}, {Gunn}, {Sumi}, {Schneider}, {Barentine}, {Brewington},
  {Brinkmann}, {Fukugita}, {Harvanek}, {Kleinman}, {Krzesinski}, {Long},
  {Neilsen}, {Nitta}, {Snedden}, \& {York}}]{Juric2008}
{Juri{\'c}}, M., {Ivezi{\'c}}, {\v Z}., {Brooks}, A., {et~al.} 2008, \apj, 673,
  864

\bibitem[{{Kendall} \& {Stuart}(1973)}]{Kendall1973}
{Kendall}, M.~G. \& {Stuart}, A. 1973, {The Advanced Theory of Statistics. Vol.
  2, Ch. 18, Ed. Griffin (London)}

\bibitem[{{Korchagin} {et~al.}(2003){Korchagin}, {Girard}, {Borkova},
  {Dinescu}, \& {van Altena}}]{Kor2003}
{Korchagin}, V.~I., {Girard}, T.~M., {Borkova}, T.~V., {Dinescu}, D.~I., \&
  {van Altena}, W.~F. 2003, \aj, 126, 2896

\bibitem[{{Kroupa}(2008)}]{Kroupa2008}
{Kroupa}, P. 2008, in Astronomical Society of the Pacific Conference Series,
  Vol. 390, Pathways Through an Eclectic Universe, ed. J.~H. {Knapen}, T.~J.
  {Mahoney}, \& A.~{Vazdekis}, 3

\bibitem[{{Kroupa} {et~al.}(1993){Kroupa}, {Tout}, \& {Gilmore}}]{KTG93}
{Kroupa}, P., {Tout}, C.~A., \& {Gilmore}, G. 1993, \mnras, 262, 545

\bibitem[{{Kroupa} \& {Weidner}(2003)}]{Kroupa2003}
{Kroupa}, P. \& {Weidner}, C. 2003, \apj, 598, 1076

\bibitem[{{Kroupa} {et~al.}(2013){Kroupa}, {Weidner}, {Pflamm-Altenburg},
  {Thies}, {Dabringhausen}, {Marks}, \& {Maschberger}}]{Kroupa2013}
{Kroupa}, P., {Weidner}, C., {Pflamm-Altenburg}, J., {et~al.} 2013, {The
  Stellar and Sub-Stellar Initial Mass Function of Simple and Composite
  Populations}, ed. T.~D. {Oswalt} \& G.~{Gilmore}, 115

\bibitem[{{Lagarde} {et~al.}(2017){Lagarde}, {Robin}, {Reyl{\'e}}, \&
  {Nasello}}]{Lagarde2017}
{Lagarde}, N., {Robin}, A.~C., {Reyl{\'e}}, C., \& {Nasello}, G. 2017, \aap,
  601, A27

\bibitem[{{Lallement} {et~al.}(2018){Lallement}, {Capitanio}, {Ruiz-Dern},
  {Danielski}, {Babusiaux}, {Vergely}, {Elyajouri}, {Arenou}, \&
  {Leclerc}}]{Rosine2018}
{Lallement}, R., {Capitanio}, L., {Ruiz-Dern}, L., {et~al.} 2018, ArXiv
  e-prints [\eprint[arXiv]{1804.06060}]

\bibitem[{{Licquia} \& {Newman}(2015)}]{Licquia2015}
{Licquia}, T.~C. \& {Newman}, J.~A. 2015, \apj, 806, 96

\bibitem[{{Loebman} {et~al.}(2014){Loebman}, {Ivezi{\'c}}, {Quinn}, {Bovy},
  {Christensen}, {Juri{\'c}}, {Ro{\v s}kar}, {Brooks}, \&
  {Governato}}]{Loebman2014}
{Loebman}, S.~R., {Ivezi{\'c}}, {\v Z}., {Quinn}, T.~R., {et~al.} 2014, \apj,
  794, 151

\bibitem[{{Marin} {et~al.}(2011){Marin}, {Pudlo}, {Robert}, \&
  {Ryder}}]{Marin2011}
{Marin}, J.-M., {Pudlo}, P., {Robert}, C.~P., \& {Ryder}, R. 2011, ArXiv
  e-prints [\eprint[arXiv]{1101.0955}]

\bibitem[{Marjoram {et~al.}(2003)Marjoram, Molitor, Plagnol, \&
  Tavar{\'e}}]{Marjoram2003}
Marjoram, P., Molitor, J., Plagnol, V., \& Tavar{\'e}, S. 2003, Proceedings of
  the National Academy of Sciences, 100, 15324

\bibitem[{{Marshall} {et~al.}(2006){Marshall}, {Robin}, {Reyl{\'e}},
  {Schultheis}, \& {Picaud}}]{Marshall2006}
{Marshall}, D.~J., {Robin}, A.~C., {Reyl{\'e}}, C., {Schultheis}, M., \&
  {Picaud}, S. 2006, \aap, 453, 635

\bibitem[{{Massey}(1998)}]{Massey1998}
{Massey}, P. 1998, in Astronomical Society of the Pacific Conference Series,
  Vol. 142, The Stellar Initial Mass Function (38th Herstmonceux Conference),
  ed. G.~{Gilmore} \& D.~{Howell}, 17

\bibitem[{{McKee} {et~al.}(2015){McKee}, {Parravano}, \&
  {Hollenbach}}]{Mckee2015}
{McKee}, C.~F., {Parravano}, A., \& {Hollenbach}, D.~J. 2015, \apj, 814, 13

\bibitem[{{McMillan}(2011)}]{McMillan2011}
{McMillan}, P.~J. 2011, \mnras, 414, 2446

\bibitem[{{Miller} \& {Scalo}(1979)}]{MS79}
{Miller}, G.~E. \& {Scalo}, J.~M. 1979, \apjs, 41, 513

\bibitem[{{Mor} {et~al.}(2017){Mor}, {Robin}, {Figueras}, \&
  {Lemasle}}]{Mor2017}
{Mor}, R., {Robin}, A.~C., {Figueras}, F., \& {Lemasle}, B. 2017, \aap, 599,
  A17

\bibitem[{{Pasetto} {et~al.}(2012){Pasetto}, {Chiosi}, \&
  {Kawata}}]{Pasetto2012}
{Pasetto}, S., {Chiosi}, C., \& {Kawata}, D. 2012, \aap, 545, A14

\bibitem[{{Pasetto} {et~al.}(2016){Pasetto}, {Natale}, {Kawata}, {Chiosi},
  {Hunt}, \& {Brogliato}}]{Pasetto2016}
{Pasetto}, S., {Natale}, G., {Kawata}, D., {et~al.} 2016, \mnras, 461, 2383

\bibitem[{{Pflamm-Altenburg} \& {Kroupa}(2006)}]{Pflamm2006}
{Pflamm-Altenburg}, J. \& {Kroupa}, P. 2006, \mnras, 373, 295

\bibitem[{{Reid} {et~al.}(2002){Reid}, {Gizis}, \& {Hawley}}]{Reid2002}
{Reid}, I.~N., {Gizis}, J.~E., \& {Hawley}, S.~L. 2002, \aj, 124, 2721

\bibitem[{{Robin} {et~al.}(2017){Robin}, {Bienaym{\'e}},
  {Fern{\'a}ndez-Trincado}, \& {Reyl{\'e}}}]{Robin2017}
{Robin}, A.~C., {Bienaym{\'e}}, O., {Fern{\'a}ndez-Trincado}, J.~G., \&
  {Reyl{\'e}}, C. 2017, \aap, 605, A1

\bibitem[{{Robin} {et~al.}(2012){Robin}, {Marshall}, {Schultheis}, \&
  {Reyl{\'e}}}]{Robin2012}
{Robin}, A.~C., {Marshall}, D.~J., {Schultheis}, M., \& {Reyl{\'e}}, C. 2012,
  \aap, 538, A106

\bibitem[{{Robin} {et~al.}(2003){Robin}, {Reyl{\'e}}, {Derri{\`e}re}, \&
  {Picaud}}]{Robin2003}
{Robin}, A.~C., {Reyl{\'e}}, C., {Derri{\`e}re}, S., \& {Picaud}, S. 2003,
  \aap, 409, 523

\bibitem[{{Robin} {et~al.}(2014){Robin}, {Reyl{\'e}}, {Fliri}, {Czekaj},
  {Robert}, \& {Martins}}]{Robin2014}
{Robin}, A.~C., {Reyl{\'e}}, C., {Fliri}, J., {et~al.} 2014, \aap, 569, A13

\bibitem[{Robitaille \& Whitney(2010)}]{Robitaille2010}
Robitaille, T.~P. \& Whitney, B.~A. 2010, The Astrophysical Journal Letters,
  710, L11

\bibitem[{{Rybizki} \& {Just}(2015)}]{Rybizki2015}
{Rybizki}, J. \& {Just}, A. 2015, \mnras, 447, 3880

\bibitem[{{Salpeter}(1955)}]{Salpeter1955}
{Salpeter}, E.~E. 1955, \apj, 121, 161

\bibitem[{{Salucci} {et~al.}(2010){Salucci}, {Nesti}, {Gentile}, \& {Frigerio
  Martins}}]{Salucci2010}
{Salucci}, P., {Nesti}, F., {Gentile}, G., \& {Frigerio Martins}, C. 2010,
  \aap, 523, A83

\bibitem[{{Scalo}(1986)}]{Scalo1986}
{Scalo}, J.~M. 1986, \fcp, 11, 1

\bibitem[{{Sharma} {et~al.}(2011){Sharma}, {Bland-Hawthorn}, {Johnston}, \&
  {Binney}}]{Sharma2011}
{Sharma}, S., {Bland-Hawthorn}, J., {Johnston}, K.~V., \& {Binney}, J. 2011,
  \apj, 730, 3

\bibitem[{{Simion} {et~al.}(2017){Simion}, {Belokurov}, {Irwin}, {Koposov},
  {Gonzalez-Fernandez}, {Robin}, {Shen}, \& {Li}}]{simion2017}
{Simion}, I.~T., {Belokurov}, V., {Irwin}, M., {et~al.} 2017, \mnras, 471, 4323

\bibitem[{{Sisson} \& {Fan}(2010)}]{Sisson2010}
{Sisson}, S. \& {Fan}, Y. 2010, ArXiv e-prints [\eprint[arXiv]{1001.2058}]

\bibitem[{{Snaith} {et~al.}(2015){Snaith}, {Haywood}, {Di Matteo}, {Lehnert},
  {Combes}, {Katz}, \& {G{\'o}mez}}]{Snaith2015}
{Snaith}, O., {Haywood}, M., {Di Matteo}, P., {et~al.} 2015, \aap, 578, A87

\bibitem[{{Sofue}(2015)}]{Sofue2015}
{Sofue}, Y. 2015, \pasj, 67, 75

\bibitem[{{Tapiador} {et~al.}(2017){Tapiador}, {Berihuete}, {Sarro}, {Julbe},
  \& {Huedo}}]{Tapiador2017}
{Tapiador}, D., {Berihuete}, A., {Sarro}, L.~M., {Julbe}, F., \& {Huedo}, E.
  2017, Astronomy and Computing, 19, 1

\bibitem[{{Tinsley}(1980)}]{Tinsley1980}
{Tinsley}, B.~M. 1980, \fcp, 5, 287

\bibitem[{{van Leeuwen}(2007)}]{Leeuwen2007}
{van Leeuwen}, F., ed. 2007, Astrophysics and Space Science Library, Vol. 350,
  {Hipparcos, the New Reduction of the Raw Data}

\bibitem[{Zaharia {et~al.}(2012)Zaharia, Chowdhury, Das, Dave, Ma, McCauly,
  Franklin, Shenker, \& Stoica}]{ApacheSpark}
Zaharia, M., Chowdhury, M., Das, T., {et~al.} 2012, in Presented as part of the
  9th {USENIX} Symposium on Networked Systems Design and Implementation ({NSDI}
  12) (San Jose, CA: {USENIX}), 15--28

\end{thebibliography}


\begin{appendix}

\section{Testing BGM FASt vs. Std simulations}\label{ApA}

In this appendix we aim to present the detailed figures of the tests BGM FASt versus BGM Std reported in Sect. 6 when comparing the behaviour of colour, mass, and age distributions. Additionally we provide a comparison of colour, mass, and age distribution between the MP variant (using the combination of the six most probable of the explored parameters) simulated with BGM FASt and the MP variant (defined in Sect. \ref{CaseB}) simulated using BGM Std.

\subsection{Testing the BGM FASt performance}\label{ApA1}

\begin{figure*}
 \begin{subfigure}{.5\textwidth}
  \centering
  \includegraphics[width=.8\linewidth]{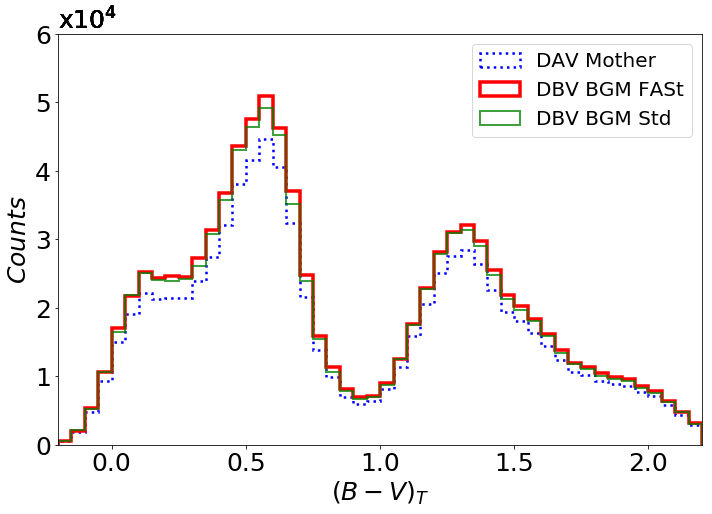}
 
\end{subfigure}%
\begin{subfigure}{.5\textwidth}
  \centering
  \includegraphics[width=.8\linewidth]{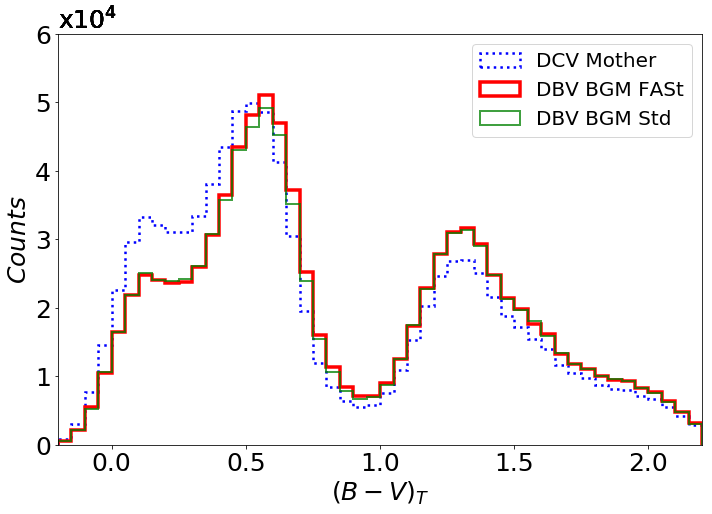}
  
\end{subfigure}
\begin{subfigure}{.5\textwidth}
  \centering
  \includegraphics[width=.8\linewidth]{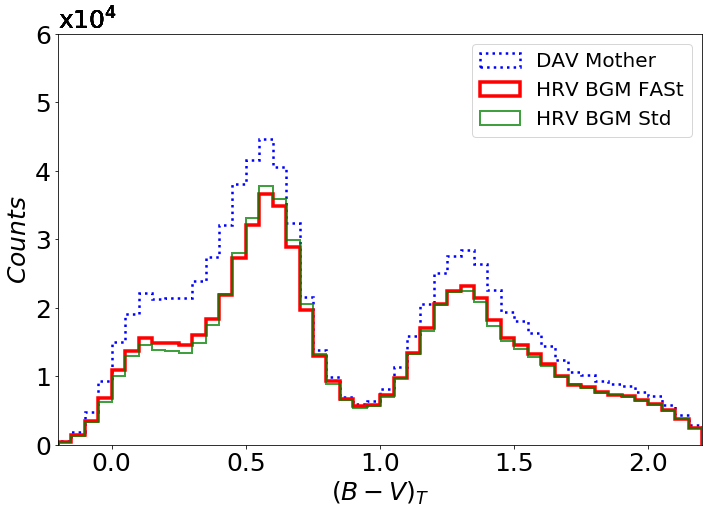}
  \
\end{subfigure}
\begin{subfigure}{.5\textwidth}
  \centering
  \includegraphics[width=.8\linewidth]{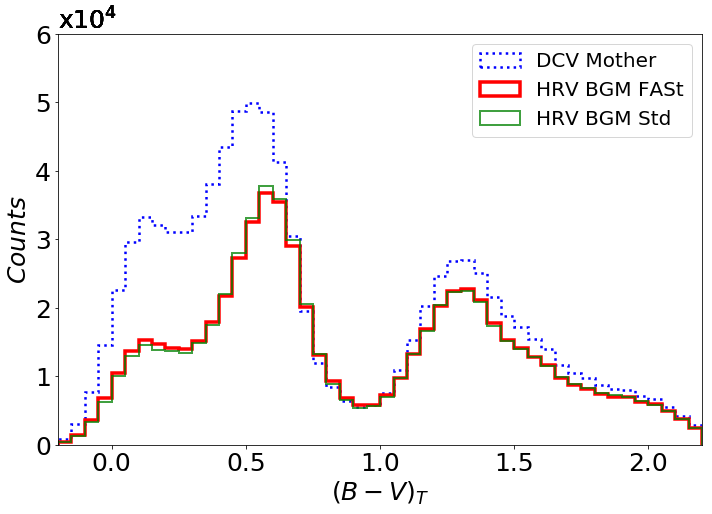}
  
\end{subfigure}

 \begin{subfigure}{.5\textwidth}
  \centering
  \includegraphics[width=.8\linewidth]{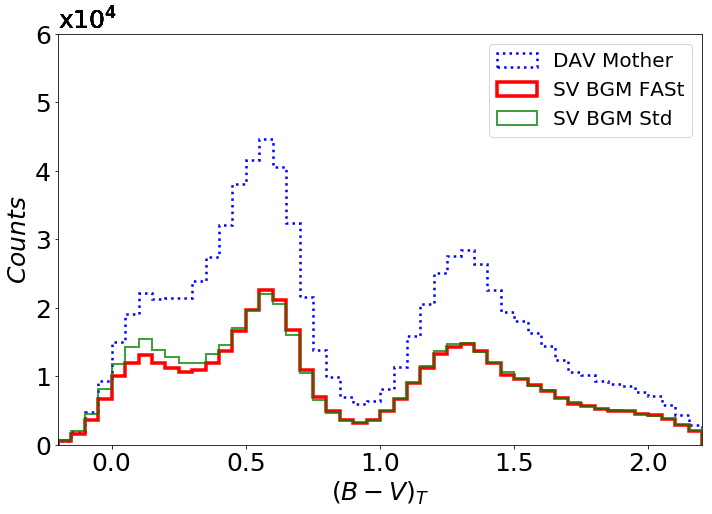}
\end{subfigure}%
\begin{subfigure}{.5\textwidth}
  \centering
  \includegraphics[width=.8\linewidth]{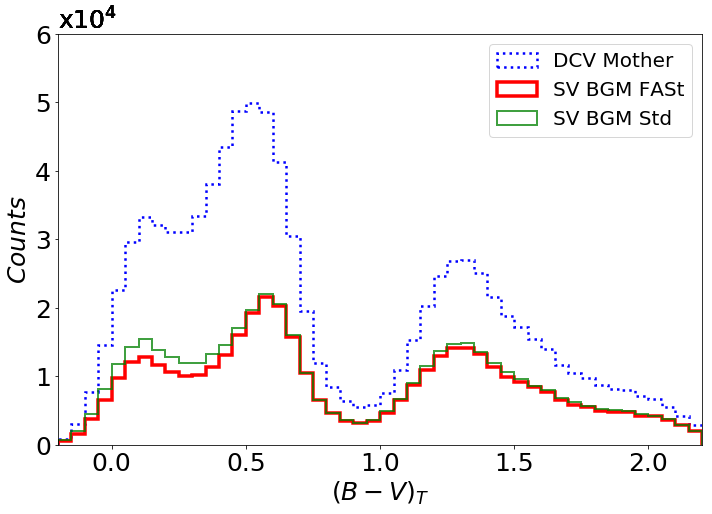}
\end{subfigure}
\begin{subfigure}{.5\textwidth}
  \centering
  \includegraphics[width=.8\linewidth]{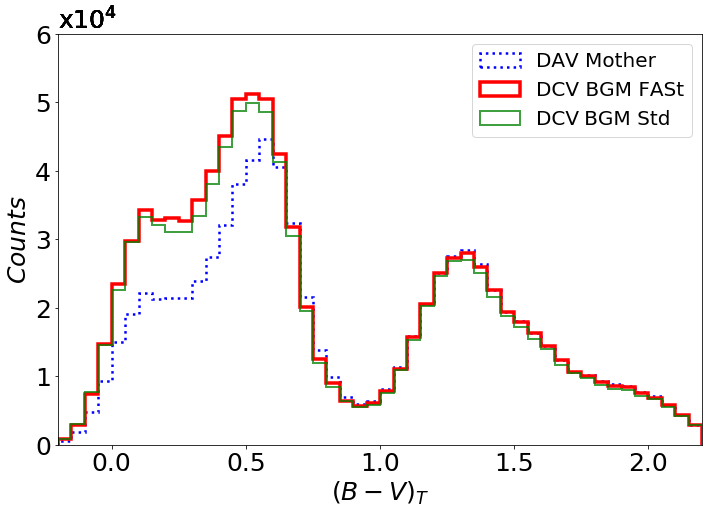}
  
\end{subfigure}
\begin{subfigure}{.5\textwidth}
  \centering
  \includegraphics[width=.8\linewidth]{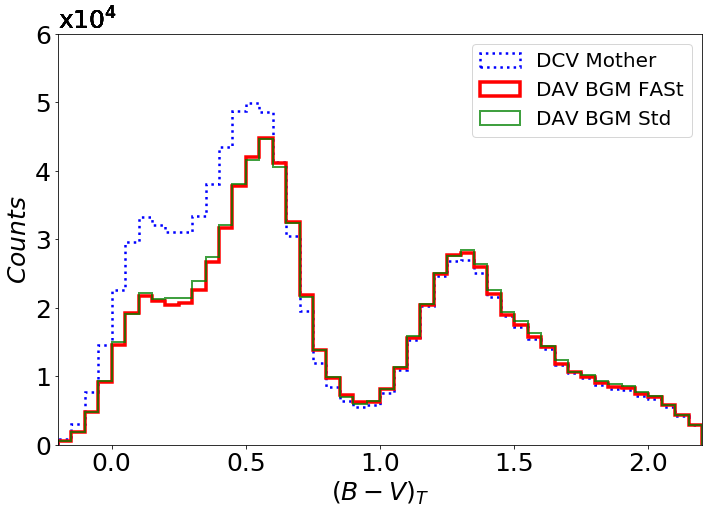}
  
\end{subfigure}
     \caption{Colour distribution,$(B-V)_T$ , for the BGM FASt vs. BGM Std tests presented in  Table \ref{table:2}. All the plotted simulations  use the \cite{Drimmel2001} extinction map. The dotted blue line is for the Mother Simulation (DAV variant for the first column and DCV variant for the second column). The thin green line and the thick red line signify BGM Std and BGM FASt simulations, respectively.}
     \label{A1}
        \end{figure*}

\begin{figure*}
  \begin{subfigure}{.5\textwidth}
  \centering
  \includegraphics[width=.78\linewidth]{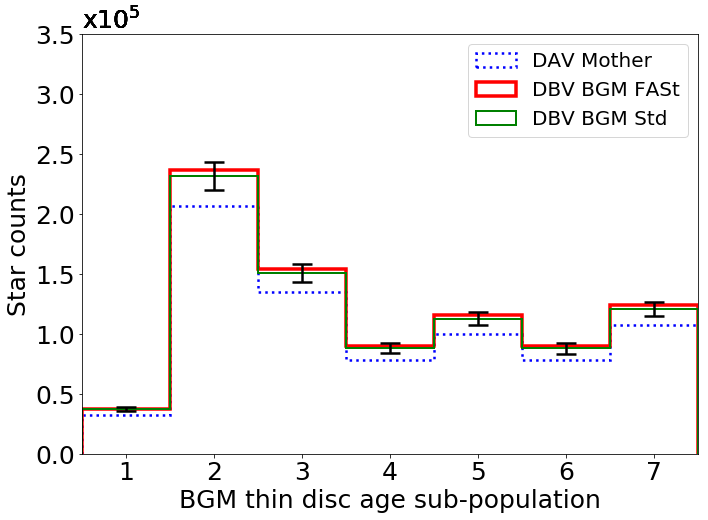}
 
\end{subfigure}%
\begin{subfigure}{.5\textwidth}
  \centering
  \includegraphics[width=.78\linewidth]{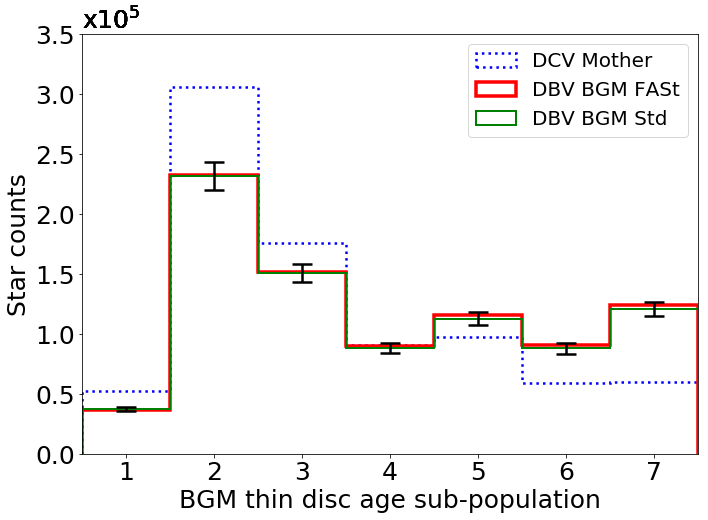}
  
\end{subfigure}
\begin{subfigure}{.5\textwidth}
  \centering
  \includegraphics[width=.78\linewidth]{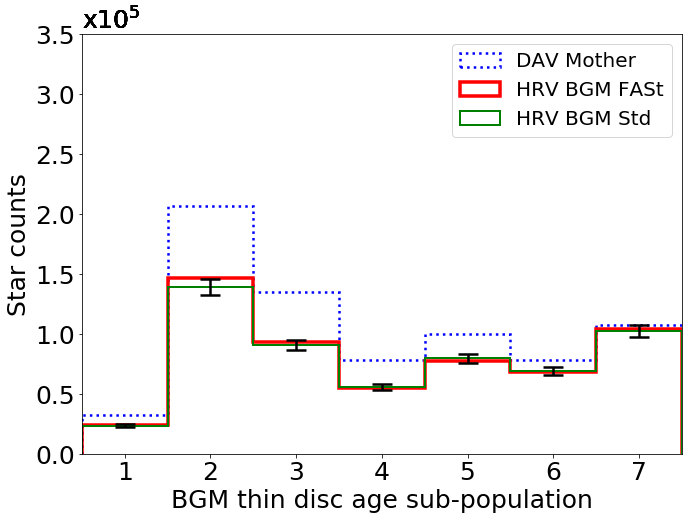}
  \
\end{subfigure}
\begin{subfigure}{.5\textwidth}
  \centering
  \includegraphics[width=.78\linewidth]{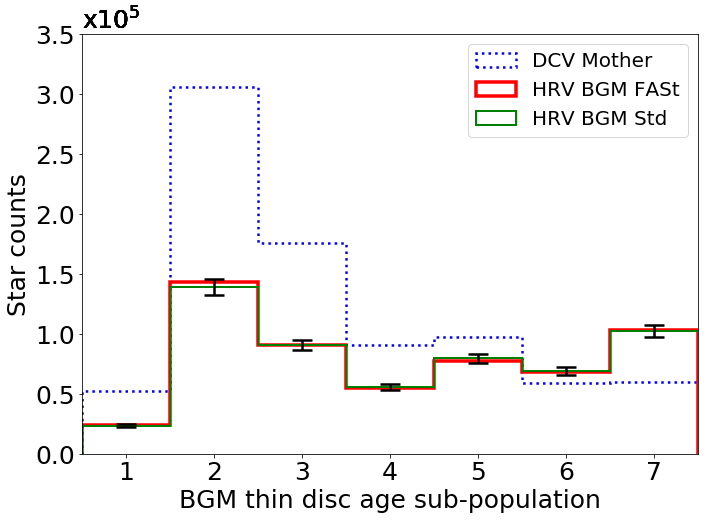}
  
\end{subfigure}

 \begin{subfigure}{.5\textwidth}
  \centering
  \includegraphics[width=.78\linewidth]{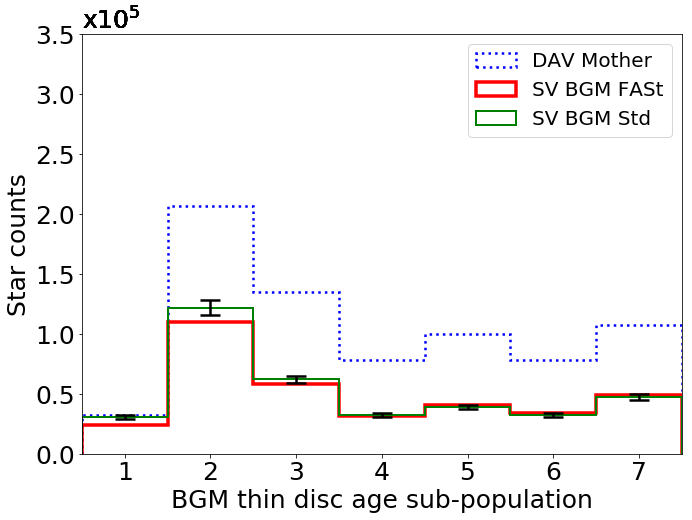}
\end{subfigure}%
\begin{subfigure}{.5\textwidth}
  \centering
  \includegraphics[width=.78\linewidth]{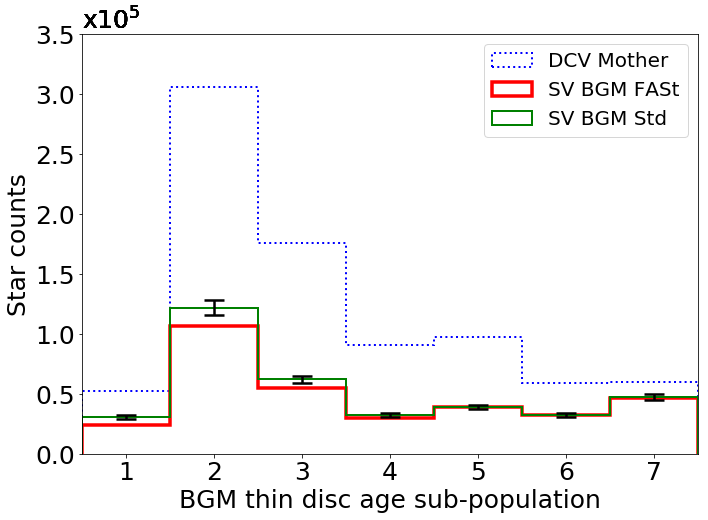}
\end{subfigure}
\begin{subfigure}{.5\textwidth}
  \centering
  \includegraphics[width=.78\linewidth]{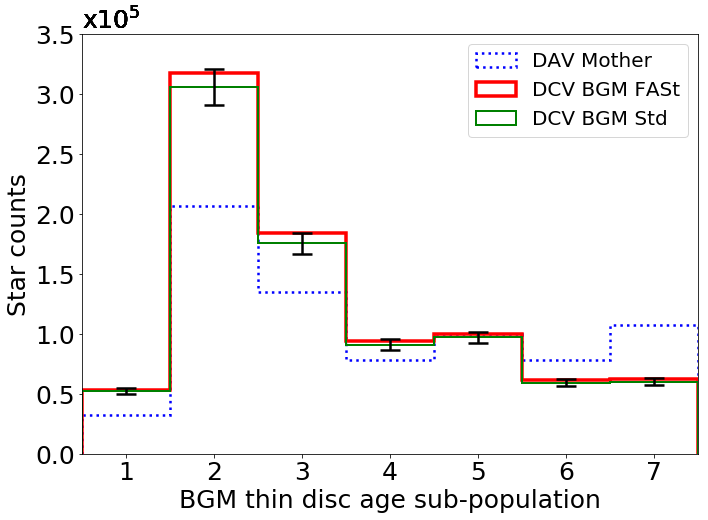}
  
\end{subfigure}
\begin{subfigure}{.5\textwidth}
  \centering
  \includegraphics[width=.78\linewidth]{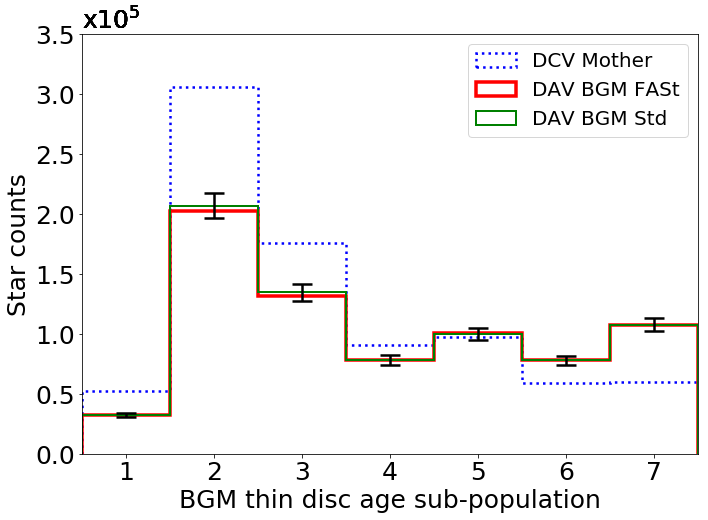}
  
\end{subfigure}
     \caption{Age sub-population distribution for the BGM FASt vs. BGM Std tests presented in  Table \ref{table:2}. All the plotted simulations  use the \cite{Drimmel2001} extinction map. The  blue dotted line is for the Mother Simulation (DAV variant for the first column and DCV variant for the second column). The thin green line and the thick red line signify BGM Std and BGM FASt simulations, respectively. The error bars are set to be $5\%$ of the stars in the bin to visualise if the differences are below or above it.}
     \label{A2}
        \end{figure*}

\begin{figure*}
 \begin{subfigure}{.5\textwidth}
  \centering
  \includegraphics[width=.8\linewidth]{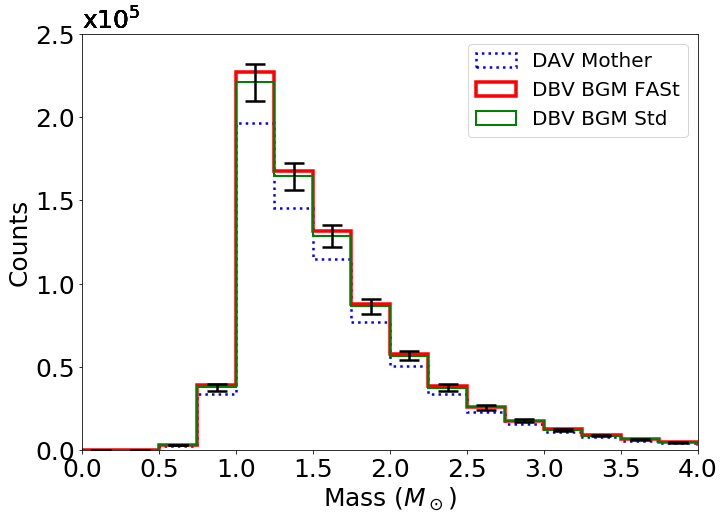}
 
\end{subfigure}%
\begin{subfigure}{.5\textwidth}
  \centering
  \includegraphics[width=.8\linewidth]{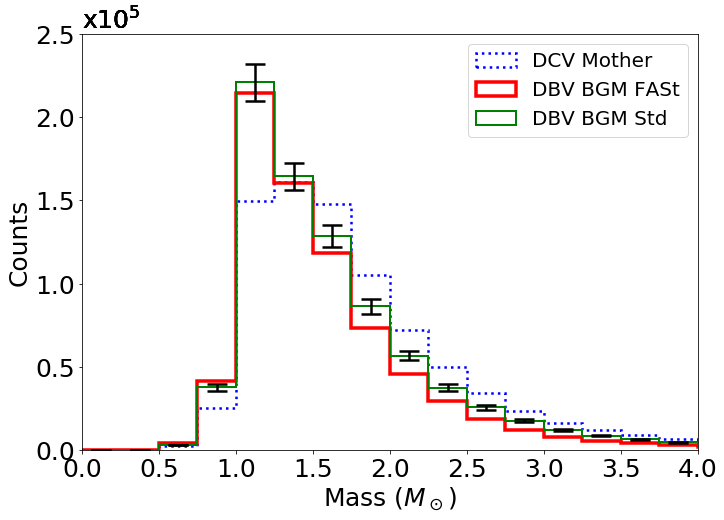}
  
\end{subfigure}
\begin{subfigure}{.5\textwidth}
  \centering
  \includegraphics[width=.8\linewidth]{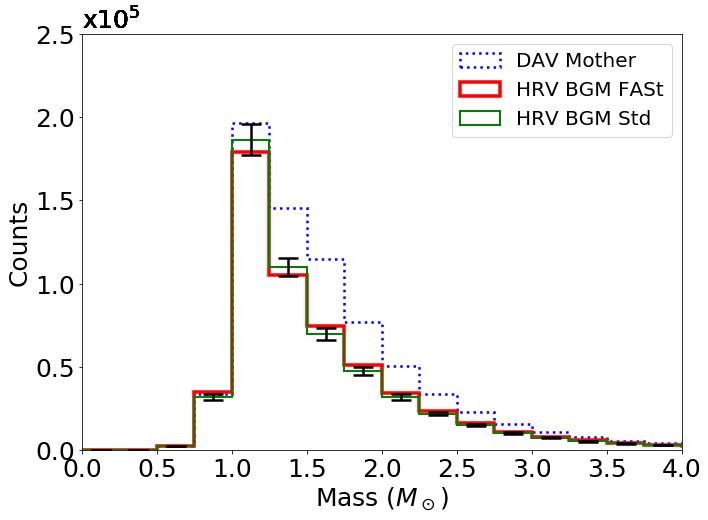}
  \
\end{subfigure}
\begin{subfigure}{.5\textwidth}
  \centering
  \includegraphics[width=.8\linewidth]{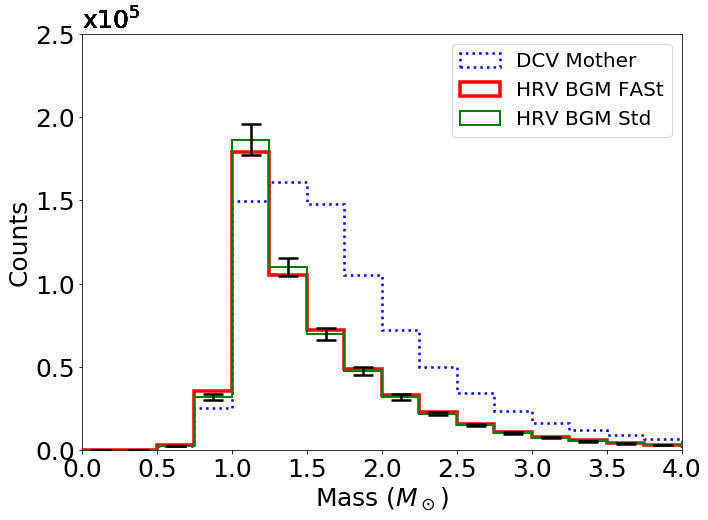}
  
\end{subfigure}
 
 \begin{subfigure}{.5\textwidth}
  \centering
  \includegraphics[width=.8\linewidth]{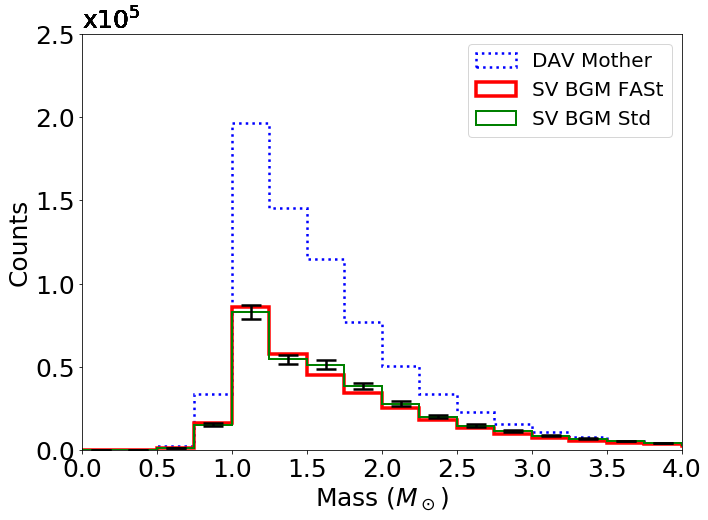}
\end{subfigure}%
\begin{subfigure}{.5\textwidth}
  \centering
  \includegraphics[width=.8\linewidth]{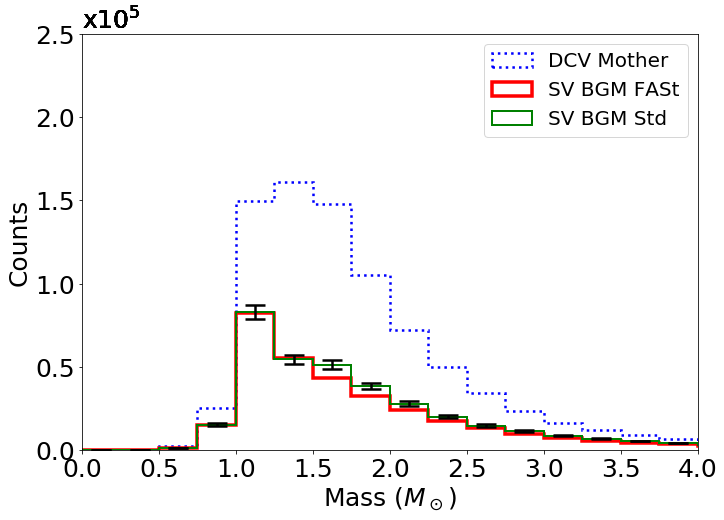}
\end{subfigure}
\begin{subfigure}{.5\textwidth}
  \centering
  \includegraphics[width=.8\linewidth]{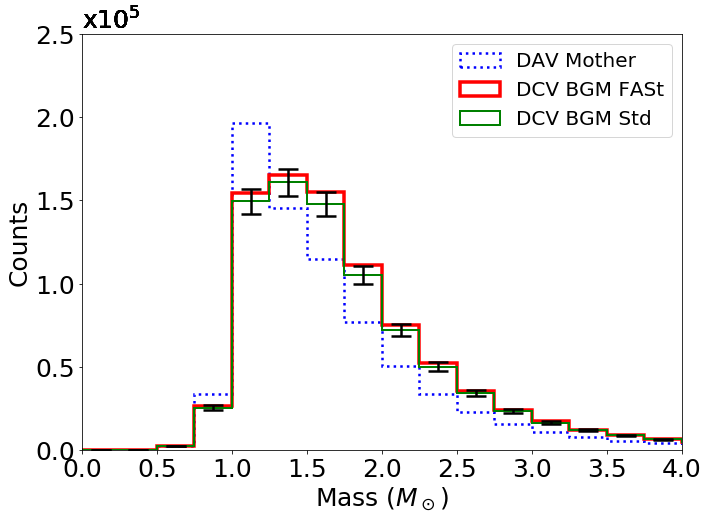}
  
\end{subfigure}
\begin{subfigure}{.5\textwidth}
  \centering
  \includegraphics[width=.8\linewidth]{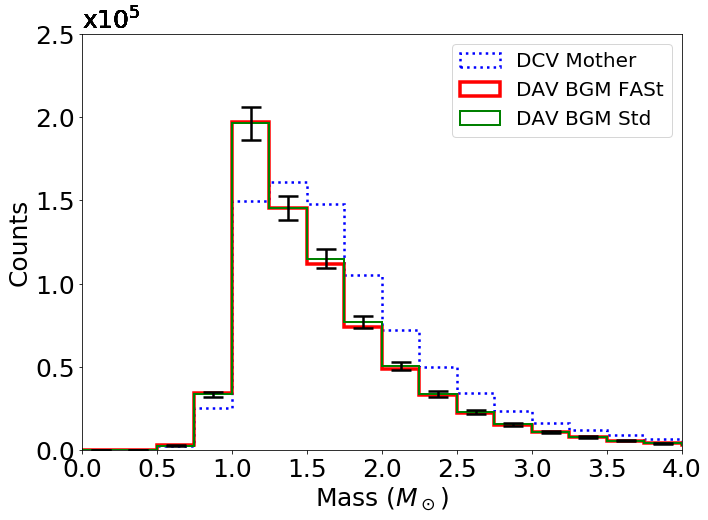}
  
\end{subfigure}
    \caption{ Mass distribution for the BGM FASt vs. BGM Std tests presented in  Table \ref{table:2}. All the plotted simulations  use \cite{Drimmel2001} extinction map. The dotted blue line is for the Mother Simulation (DAV variant for the first column and DCV variant for the second column). The thin green line and the  thick red line signify BGM Std and BGM FASt simulations, respectively. The error bars are set to be $5\%$ of the stars in the bin to visualise if the differences are below or above it.}
    \label{A3}
        \end{figure*}

In Figs. \ref{A1}, \ref{A2}, and \ref{A3} we present the colour, age, and mass distributions of the tests BGM FASt versus BGM Std presented and discussed in Sect. \ref{SolarN}. The age distribution is grouped by age-subpopulation (see Sect. \ref{stdthin}). Comments on these data are presented in Sect. \ref{fastvsstd}.

\subsection{Testing the MP variant}\label{ApA2}   

To test if the obtained BGM FASt MP variant is equivalent to the BGM Std MP variant we have repeated the tests of Sects. \ref{ldsetest} and \ref{fastvsstd} for this new variant. We have compared the eccentricities of the Einasto density profiles and the stellar volume mass density at the position of the Sun obtained from both the approximate local dynamical statistical equilibrium used in BGM FASt and the full LDSE used in the standard BGM. We find differences of the eccentricities of the Einasto density profiles always smaller than $1\%$ while we find differences in the stellar volume densities at the position of the Sun smaller than $3\%$. Additionally we have checked that the difference in the local dark matter density is about $3.5\%$ with negligible effects on the rotation curve as demonstrated in Sect. \ref{ldsetest}. These differences are inside the margins reported in Sect. \ref{ldsetest}. 

\begin{figure}
   \centering
   \includegraphics[width=\hsize]{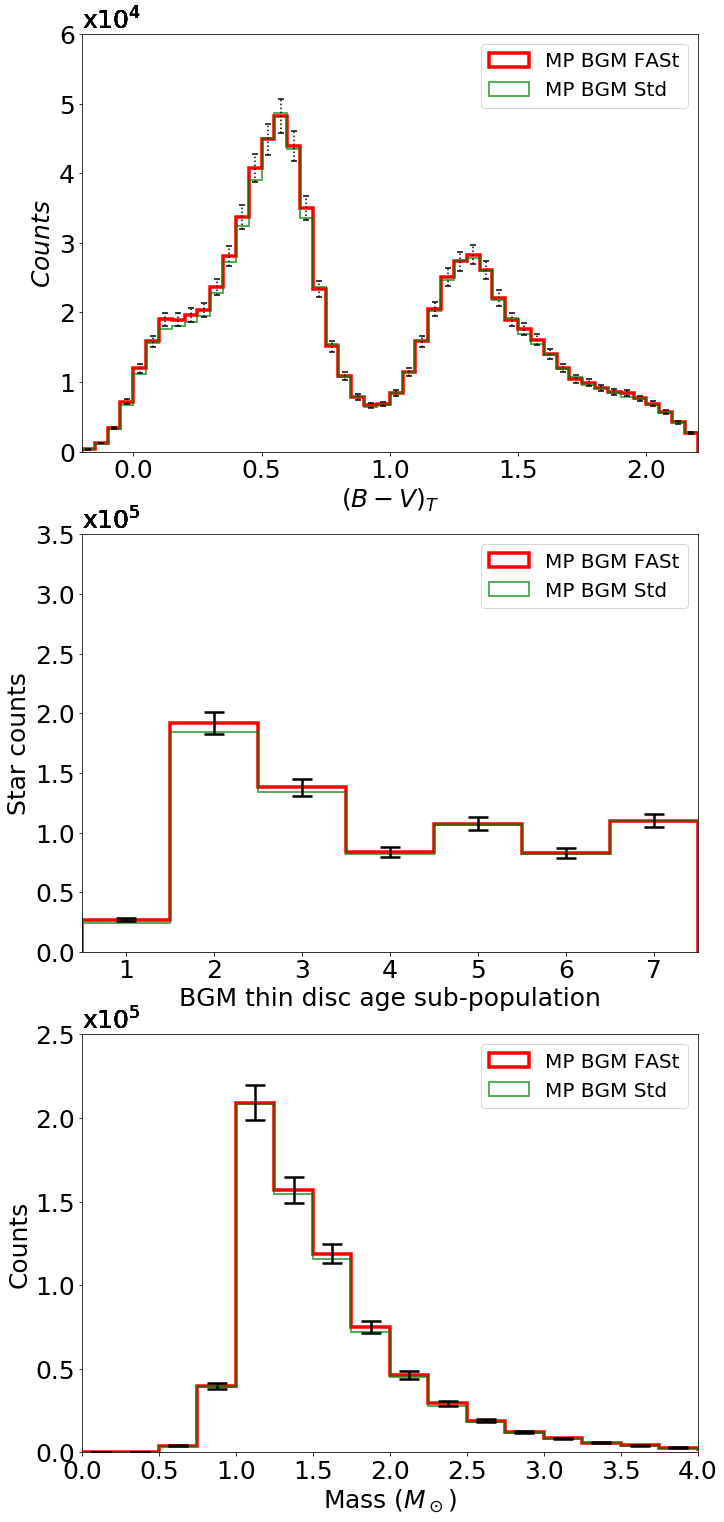}
      \caption{\textbf{Top:}$(B-V)_T$ distribution. \textbf{Middle:} Star counts for each of the seven age sub-populations of the thin disc assumed in the BGM. \textbf{Bottom:} Mass distribution. The simulations are done using the combination of the six most probable parameters (MP variant) with the \cite{Drimmel2001} extinction map; in red the simulation is done with BGM FASt while in black it is done with BGM Std.  The error bars are set to be $5\%$ of the stars in the bin to visualise that the differences between BGM FASt and BGM Std are below it. The simulations are samples limited in visual apparent magnitude $V_T < 11$ and photometric errors are not considered.}
         \label{bestfit2}
   \end{figure}

In Fig. \ref{bestfit2} we present the colour, age, and mass distribution of the MP variant using the \cite{Drimmel2001} extinction map. The age distribution is grouped by age-subpopulation (see Sect. \ref{stdthin}). From Sect. \ref{SolarN} we conclude that differences between BGM FASt and BGM Std are usually below $5\%$. Therefore, the error bars in the plots are set to be $5\%$ of the star counts in the bin to visualise that the differences are smaller. We note that within the error bars both distributions match each other very well, demonstrating that BGM FASt provides a good approximation to the BGM standard for the best model variant fitting Tycho-2 data.

\section{Evaluating the sampling noise in BGM}\label{ApB}
\subsection{Sampling noise in BGM Standard}\label{ApB1}

In this section we analyse the star-generation strategy of BGM Std to demonstrate that the discrepancies larger than $5\%$ that we find occasionally when comparing BGM FASt and BGM Std (see Sect, 6 and Appendix A) can be explained by the sampling noise for very-low-mass reservoirs in BGM Std.

As described in Equation (1) of \cite{Czekaj2014}, for each thin-disc age sub-population the mass available to be spent on star production in a given volume element is quoted as the mass reservoir.

\begin{figure*}\label{B1}
\begin{subfigure}{.33\textwidth}
  \centering
  \includegraphics[width=0.95\linewidth]{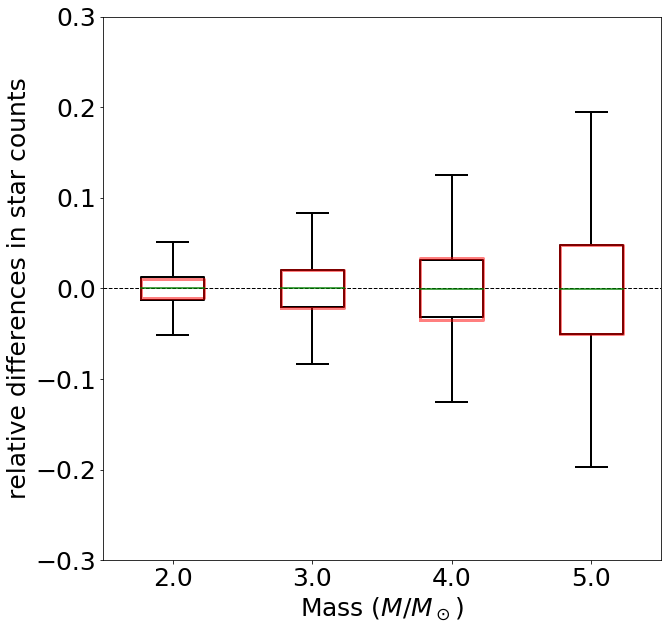}
\end{subfigure}%
 \begin{subfigure}{.33\textwidth}
  \centering
  \includegraphics[width=0.95\linewidth]{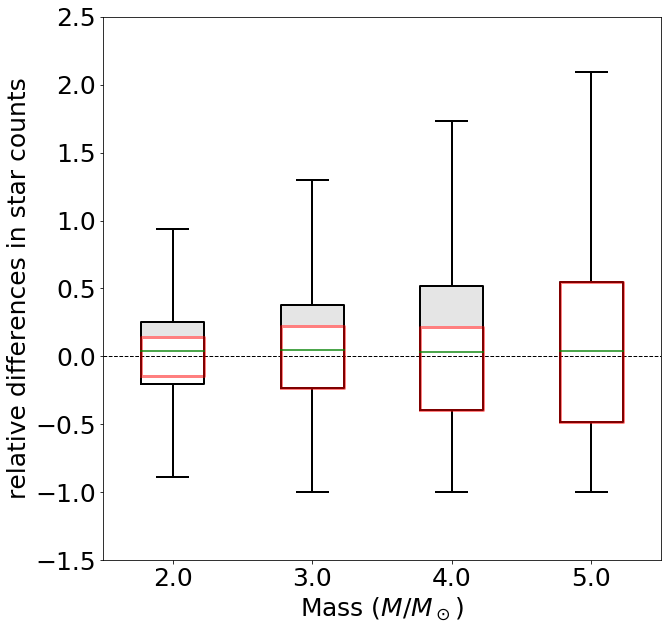}
\end{subfigure}%
 \begin{subfigure}{.33\textwidth}
  \centering
  \includegraphics[width=0.95\linewidth]{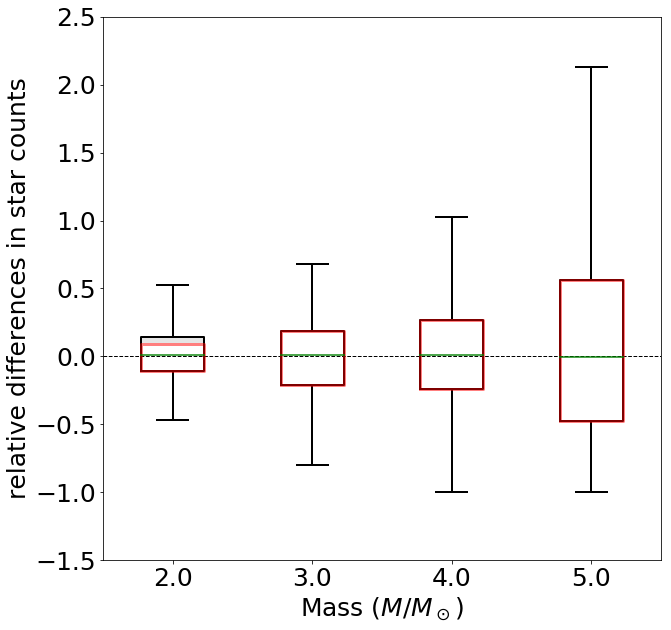}
\end{subfigure}%

    \caption{\textbf{Left panel: }Behaviour of the noise in BGM Std when the mass reservoir is large. This behaviour is obtained by reproducing a full star generation process of a mass reservoir of $10^4 M_\odot$ for the youngest age sub-population $10^4$
times. The limits of the boxes show the position of the first and the third quartile. The limits of the bars show the position of $-1.5 \cdot IQR$ and $+1.5 \cdot IQR$, where IQR is the interquartile range. Everything beyond the limits of the bar is considered an outlier. In black we show the relative differences in star counts between the expected number of stars in a given mass bin and the number of stars obtained with the standard BGM generation strategy. The red boxes show what these relative differences would be if the generation were to precisely follow a Poisson distribution centred in the expected value. The noise behaves approximately as a Poisson distribution, as expected. \textbf{Middle and right panels:} Behaviour of the noise in BGM Std when the mass reservoir is small. This is obtained by by reproducing a full star generation process of a mass reservoir of $150 M_\odot$ for the youngest age sub-population $10^4$
times . This small mass reservoir only appears occasionally. The middle panel is for very flat slopes of the IMF at high mass range (in this case $\alpha_3=2.35$). The right panel is for an IMF slope of $\alpha_3=3.2$ closer to the best slopes fitting the data. In black we show the relative differences in star counts between the expected number of stars in a given mass bin and the number of stars obtained with the standard BGM generation strategy. The red boxes show what these relative differences would be  if the generation were to precisely follow a Poisson distribution centred in the expected value. The grey shadow emphasises the differences between the distribution obtained with BGM Std and the one that would follow exactly a Poisson distribution. We note that the effect is very small for the right panel. We show the results only for masses up to $5.5 M_\odot$ as about $99\%$ of the stars in the simulated samples (limited at $V_T=11$) have masses smaller than $5.5 M_\odot$.}
         \label{anoise}
   \end{figure*}

In Fig. \ref{anoise}  we present, for the youngest sub-population of the thin disc component in BGM and for masses bigger than $1.53M_\odot$, the relative differences in star counts per mass bin between the sampled stars and the stars predicted by the imposed distribution function $\mathcal{G}(\tau,M,Z,\bar{x},\bar{v},\alpha)$ along $10^4$ realisations. We expect that the Poisson distribution describes the distribution of the number of sampled stars of a given interval (e.g. mass or age) along the $10^4$ realisations. The red boxes in the figure represent the relative differences between the sampled stars and the stars predicted by $\mathcal{G}(\tau,M,Z,\bar{x},\bar{v},\alpha)$ if the sampling distribution would follow exactly a Poisson distribution centred on the predicted theoretical value. The black boxes are the relative differences in star counts per mass bin between the stars generated by BGM Std and the stars predicted by the imposed $\mathcal{G}(\tau,M,Z,\bar{x},\bar{v},\alpha)$ along $10^4$ realisations. In the left panel we show the result for a mass reservoir of $10^4 M_\odot$ for a Salpeter IMF while in the middle and right panels we show the results for a mass reservoir of $150M_\odot$ for a Salpeter IMF (middle) and Kroupa-like IMF (right). We note that for the mass reservoir of $10^4 M_\odot$ the BGM standard generation behaves as expected and approximately follows the Poisson distribution.  We detect small discrepancies with the Poisson distribution that are caused by the fact that the mass in the mass reservoir runs out. It is important to notice that the mean value of the distribution coincides with the expected theoretical value, differing from each other by less than 0.01\%.  Regarding the middle panel (Salpeter IMF), we note that for the first three mass bins the noise is not Poissonian (red boxes are for Poisson distribution centred in the predicted value) and is clearly biased towards higher values, thus producing and oversampling the stars with masses $1.53M_\odot<M<4.5M_\odot$; the grey shadow emphasises this bias towards higher values. In the right plot (Kroupa-like IMF) we note that the deviation from the Poisson noise is much smaller, marginally affecting the distribution.

We can conclude that when the mass reservoir is large enough ($M>10^4M_\odot$), we can consider as a first approximation that the probability of an occurrence of a star generation event is not affecting the probability of the occurrence of the following star generation event (necessary condition for a Poisson distribution). The results therefore approximately follow a Poisson distribution. When the mass reservoir is small (e.g. $M<500M_\odot$) the approximation that the probability of the occurrence of two star generation events is conditionally independent is no longer valid  and as a consequence the obtained distribution is slightly biased and does not follow the Poisson distribution.

We have performed the test for all populations and masses. We only find remarkable effects for the youngest population and for masses larger than $M>1.53M_\odot$ when the slope of the IMF in the high mass range is flat. It is important to emphasise that we roughly estimate that only about $10\%$ of the stars in the samples that we use in this paper (up to $V_T=11$) are generated in mass reservoirs small enough to be affected by the effects discussed in this appendix. 

The discrepancies bigger than $5\%$ found in colour, mass, and age distributions (Figs. \ref{A1}, \ref{A2} and \ref{A3}) between BGM FASt and BGM Std simulations that we reported in Sect. \ref{SolarN} can be explained by the non-Poissonian sampling noise that we found in BGM Std for the youngest sub-populations at the high mass range when the mass reservoir is small. The comparisons in Sect. \ref{ldsetest} of the densities in a sphere around the Sun are only minorly affected by this as BGM Std, when working in the sphere mode, has very large mass reservoirs to generate the stars at the position of the Sun. It is important to note that the non-Poissonian noise has a minor effect on the BGM simulations that better fit the thin disc at the solar neighbourhood. Work is in progress to further diminish the effects of the noise in the small mass reservoirs.

\subsection{Weighting the sampling noise in BGM FASt}\label{ApB2}

The BGM Std simulation that we use as a Mother Simulation is a random realisation of an imposed distribution function for the generated stars in the Galaxy ($\mathcal{G}(\tau,M,Z,\bar{x},\bar{v},\alpha)$). As a consequence when we weight the stars in BGM FASt strategy we also weight the noise. This implies, as we demonstrate in this appendix, that the noise in a BGM FASt simulation is approximately a factor $\sqrt{\bar{w}}$ of the noise in the Mother Simulation (where $\bar{w}$ is the mean weight applied to the Mother Simulation).  

If we perform $n$ realizations of a BGM Std simulation with an imposed $\mathcal{G}$, the distribution of the number of stars $X_i$ in a given interval of a given parameter, along  the $n$ realisations, can be approximated by a Poisson distribution. If the number of stars is large enough, the Poisson distribution can be approximated by a Gaussian distribution with $\mu=E[X_i]$ and $\sigma^2=E[X_i]$, where $E[X_i]$ is the expected number of stars in the interval given by the imposed distribution function of the generated stars in the Galaxy ($\mathcal{G}(\tau,M,Z,\bar{x},\bar{v},\alpha)$). We discuss below which is, for $n$ realisations of a BGM FASt simulation, the distribution of the number of stars $X_i$ in a given interval of a given parameter. We start from a Mother Simulation with an imposed distribution function $\mathcal{G}_{MSt}$ such as that in a given interval $E[X^{MSt}_i]=N_{MSt}$, where $N_{MSt}$ is the expected number of stars. Performing $n$ realisations, we obtain, as discussed above, an approximately Gaussian distribution with $\mu_{MSt}=E[X^{MSt}_i]=N_{MSt}$ and  $\sigma_{MSt}^2=E[X^{MSt}_i]=N_{MSt}$. The sample mean, for a given interval, of the $n$ realisations of the Mother Simulation is known to be an unbiased estimator for $\mu$, and is computed as follows.

\begin{equation}\label{mean}
\bar{X}_{MSt}=\frac{1}{n}\sum_{i=1}^n X^{MSt}_i
.\end{equation}

\noindent The sample variance of the $n$ realisations is known to be  an unbiased estimator for
$\sigma^2$, and is computed as follows.

\begin{equation}\label{variance}
S^2_{MSt}=\frac{1}{n-1} \cdot \sum_{i=1}^n (X^{MSt}_i-\bar{X}_{MSt})^2 
.\end{equation}

\noindent 

Next, we build $n$ realisations of a BGM FASt simulation with an imposed distribution function, $\mathcal{G}_{FASt}$, applying the pertinent weights $w_i$ to the stars of the $n$ realisations of the Mother Simulation. We assume here that the mean weight applied to a given interval of a given parameter is approximately the same for the n realisations, that is $\bar{w}$. The expected number of stars in the interval for the BGM FASt is therefore $E[X_i]=\bar{w} \cdot N_{MSt}$. If the distribution were Poissonian we would expect, as discussed above, a Gaussian with  $\mu=\bar{w} \cdot N_{MSt}$ and $\sigma^2=\bar{w} \cdot N_{MSt}$. As the $n$ realisations of the BGM FASt are built from the $n$ realisations of the Mother Simulation then we can write $X^{FASt}_i=\bar{w} \cdot X_i^{MSt}$ and the sample mean $\bar{X}_{FASt}$ as follows.

\begin{equation}\label{mean2}
\bar{X}_{FASt}=\frac{1}{n}\sum_{i=1}^n X^{FASt}_i=\frac{1}{n}\sum_{i=1}^n \bar{w} \cdot X^{MSt}_i = \bar{w} \cdot \bar{X}^{MSt}
,\end{equation}
 
\noindent For $n \to \infty$ we can write $\mu^{FASt}=\bar{w} \cdot \mu^{MSt}$ and that is exactly the expected value of stars in the given range for BGM FASt as mentioned above, $E[X^{FASt}_i]=\bar{w} \cdot N_{MSt}$.  

\begin{figure}\label{B2}

  \centering
  \includegraphics[width=\hsize]{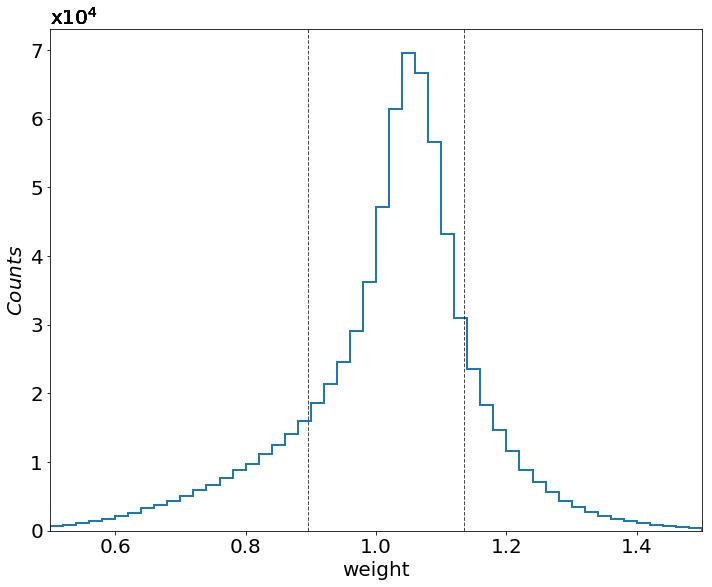}
  \caption{Distribution of weights applied to the Mother Simulation to obtain the MP variant when using the \cite{Drimmel2001} extinction map. The vertical dashed lines indicate the 0.16 and 0.84 quantiles.}
         \label{weightdist}
   \end{figure}

Now if we compute the sample variance:

\begin{equation*}\label{variance2}
S^2_{FASt}=\frac{1}{n-1} \cdot \sum_{i=1}^n (X^{FASt}_i-\bar{X}^{FASt})^2=
\end{equation*}
\begin{equation}\label{variance2}
= \frac{1}{n-1} \cdot \sum_{i=1}^n (\bar{w} \cdot X_i^{MSt}-\bar{w} \cdot \bar{X}^{MSt})^2=\bar{w}^2 \cdot S^2_{MSt}
,\end{equation}

\noindent for $n \to \infty$ we can write $\sigma_ {FASt}^2=\bar{w}^2 \cdot \sigma_ {MSt}^2=\bar{w}^2 \cdot N_{MSt}$ but we note that the variance for the Poisson distribution with $E[X_i]=\bar{w} \cdot N_{MSt}$ would be $\sigma^2=\bar{w} \cdot N_{MSt}$. As a consequence, the noise in BGM FASt simulation is a factor $\sqrt{\bar{w}}$ of the noise that a BGM Std simulation would have in its place. Therefore, $\sigma_{FASt}=\sqrt{w} \cdot \sigma_{Std}$. This effect causes an increase or reduction of the variance proportional to the value of the weight. In the range of the parameter space that we are exploring, the mean values of the weights used go from about 0.8 to about  1.20. In Fig. \ref{weightdist} we present the distribution of the weights needed to generate the MP variant. Its mean value is $1.023$ while the quantiles $0.16$ and $0.84$ take the values $0.90$ and $1.14,$ respectively. In this case the distribution of the weights is such that the effects discussed in this section are very small, as can be seen in Fig. \ref{bestfit2} when comparing the colour distribution of the MP variant simulated with BGM FASt and BGM Full. If for future studies we need weights much further away from 1 to explain the observational data, it will be necessary to introduce one or more intermediate steps to run BGM Std simulations closer to the data to perform the final parameter exploration.

\end{appendix}

\end{document}